\documentclass{article}

\usepackage[english]{babel}
\usepackage[T1]{fontenc}
\usepackage[utf8]{inputenc}
\usepackage{subfig}
\usepackage{graphicx}
\usepackage{xcolor}
\usepackage{float}
\usepackage{varwidth}
\usepackage{amsmath,amssymb,amsthm,textcomp}
\usepackage{enumerate}
\usepackage{multicol}
\usepackage{tikz}
\usepackage{hyperref}
\usepackage{fancyhdr}

\usepackage{geometry}
\usepackage{amsmath}
\usepackage{amsfonts}
\usepackage{color}
\usepackage{amssymb,amsthm}
\usepackage{mathtools}
\usepackage{mathrsfs}
\usepackage{dsfont}
\usepackage{hyperref}    
\usepackage[linesnumbered,vlined]{algorithm2e} 
\usepackage{cleveref}    

\usepackage{natbib}
    \bibpunct[, ]{(}{)}{,}{a}{}{,}%

\newcommand{\NN}{\mathbb{N}}

\newcommand{\RR}{\mathbb{R}}

\newtheorem{example}{Example}[section]

\newcommand{\ignore}[1]{}
\def\defi{\vcentcolon=}
\def\diff{\mathop{}\!\mathrm{d}}

\author{V. Guigues\\EMAp\\ FGV \and 
A. J. Kleywegt\\Industrial and\\ Systems Engineering\\ Georgia Institute\\ of Technology \and
G. Amorim\\EMAp\\FGV \and
A. Krauss\\Department of\\ Informatics\\ PUC-RJ \and
V. H. Nascimento\\ PESC\\ UFRJ
}

\title{LASPATED: A Library for the Analysis of Spatio-Temporal Discrete Data}

\date{}

\begin{document}

\maketitle

\begin{abstract}
We describe methods, tools, and a software library called LASPATED
to fit nonhomogeneous spatio-temporal Poisson process models using spatio-temporal data and space-time discretization.
The methods approximate the arrival intensity function of the Poisson process by discretizing space and time, and estimating arrival intensity as a function of subregion and time interval.
With such methods, it is typical that the dimension of the estimator is large relative to the amount of data, and therefore the performance of the estimator can be improved by using additional data.
The first method considered uses additional data to add a regularization term to the likelihood function for estimating the intensity of the Poisson process.
The second method considered uses additional data to estimate arrival intensity as a function of covariates.
We describe how the LASPATED Python package performs various types of space and time discretization, and how it calibrates Poisson models, with options to use regularization or covariates.
We demonstrate the use of our methods using both simulated and real data, and compare the results of regularized estimators with the results of basic maximum likelihood estimators.
The experiments with real data calibrate models of the arrival process of emergencies to be handled by the Rio de Janeiro emergency medical service.\\
\end{abstract}


\par {\textbf{Keywords:}} Poisson process, spatio-temporal discretization, mathematical software, library, discrete data, emergency health services








\section{Introduction}

In many applications, one would like to use data to fit a function such as $f : \mathcal{S} \times [0,T] \mapsto \RR$, where $\mathcal{S}$ is a bounded subset of $\RR^{n}$.
Often, the domain $\mathcal{S}$ is called ``space'' and the domain $[0,T]$ is called ``time''.
For example, $\mathcal{S}$ may represent locations on a $2$-dimensional map (in which case $n = 2$), or locations in a $3$-dimensional space (in which case $n = 3$), or origin-destination pairs on a map (in which case $n = 4$), and $[0,T]$ may represent time of the day (in which case $T = 24$~hours), or time of the week (in which case $T = 168$~hours).

A widely used nonparametric approach for fitting such a function~$f$ with data is to discretize space and time in a way that may depend on the data, and to fit a function from a chosen class on each discrete subset of the domain.
There are many ways to construct such a discretization.
Also, there are benefits to allowing the discretization to depend on the data.
For example, a finer discretization may be chosen in regions of space and time with a higher concentration of data.
There are also many function classes that can be used on each discrete subset of the domain.
For example, the fitted function may be piecewise constant or piecewise linear, that is, it may be constant or linear on each discrete subset of the domain.
Furthermore, sometimes one has data of covariates on the same space-time domain, with the covariates being correlated with the dependent variable.
Even if the main purpose of the study is not to model the relation between the dependent variable and the covariates, these covariate data can be used in various ways to improve the estimates.
For example, the covariates can be used in the choice of discretization, for example, by choosing the subsets of the discretization to be uniform in terms of the covariate values.
The covariates can also be used as independent variables in a fitted parametric or semiparametric model.
The covariates can also be used to regularize the model estimates, especially when the data are sparse.

Software tools for managing the choice of discretization, the regularization, and the model estimation, can be of great use.
This paper describes such software tools for function estimation with spatio-temporal data.

\subsection{Application}

Next, we describe the application that motivated the development of these software tools, and that will also be used in this paper to demonstrate the use of these tools with real data.
Phone calls about medical emergencies arrive at a call center for emergency medical services (EMS's).
Each emergency is characterized by a type, a location, and the arrival time of the call.
The type of emergency is determined by the classification system used by the EMS, such as the Medical Priority Dispatch System (MPDS) or the Association of Public-Safety Communications Officials (APCO) system.
Typical classification systems classify emergencies by the anatomical region (e.g.,\ chest) affected, the cause of the emergency (e.g.,\ animal bites), the importance of response time (e.g.,\ ``hot'' versus ``cold''), and the level of emergency service needed (e.g.,\ basic life support versus advanced life support).
The location is typically specified by an address, but can be converted to a latitude-longitude coordinate.
The arrival time of the call is specified by a date-time combination.

We would like to estimate the arrival rate of emergencies as a function of emergency type, location, and time.
Such an arrival rate function can be used to optimize the types and number of ambulances needed, the types and number of crew members needed, the crew schedules, and the positioning and dispatch of ambulances, and to develop a simulation of the emergency response system.
We consider methods that discretize space and time and then estimate the arrival rate as a function of discrete space and time.
While working on such a project (see \cite{guiklevhn2022}, \cite{websiteambrouting24}, \cite{ourheuristic241}), we identified the lack of easy-to-use tools for this type of analysis.
Thus, the LASPATED library was created as a general tool to fill this gap.

\subsection{Overview of the LASPATED Library}

Various discretization methods to partition space and time were implemented in LASPATED, summarized next.
LASPATED provides four types of space discretization: in rectangles, in hexagons, based on the Voronoi diagram/Dirichlet tessellation for a set of points in $\mathcal{S}$, and customized discretizations.
LASPATED also provides functions to combine information from two different space discretizations, such as functions to calculate the areas of intersection of pairs of subregions from two discretizations, or to calculate location data such as population in the intersection of pairs of subregions from two discretizations.

LASPATED also provides various types of time discretization.
For example, a simple partitioning of time could be a partition of the week in time intervals of $1$~hour each.
Then the corresponding time intervals are 
$$\left\{\mbox{Monday [0:00,1:00)},\mbox{Monday [1:00,2:00)},\mbox{Monday [2:00,3:00)},\ldots,\mbox{Sunday [23:00,24:00)}\right\}.$$
Time intervals of different types can also be distinguished.
For example, there could be different time intervals such as Monday [0:00,1:00) for different holidays, or there could be different time intervals such as Friday [20:00,21:00) depending on the scheduling of a sport event at the time.
For instance, for the time window Friday [20:30,21:00), one could have
\begin{itemize}
\item
a time interval for Friday [20:30,21:00) on days which are holidays and a major sport event is scheduled;
\item
a time interval for Friday [20:30,21:00) on days which are holidays and a major sport event is not scheduled;
\item
a time interval for Friday [20:30,21:00) on days which are not holidays and a major sport event is scheduled; and
\item
a time interval for Friday [20:30,21:00) on days which are not holidays and a major sport event is not scheduled.
\end{itemize}
Given a point in space $\ell_{0} \in \mathcal{S}$ and time $t_{0} \in [0,T]$, possibly with additional characteristics such as holiday or event type, LASPATED determines the combination of spatial subregion and time interval for the chosen discretization that contains the given point $(\ell_{0},t_{0})$.
Furthermore, given a data set of observations consisting of several points, LASPATED computes the number of points in the data set in each combination of spatial subregion and time interval for the chosen discretization.
For example, given a data set of medical emergencies, with a week of data associated with each observation, LASPATED computes the number of emergencies in the data set for each combination of type of emergency, spatial subregion, and time interval, for the chosen discretization, for each week of data.

LASPATED also facilitates calibration of Poisson models using data prepared with the discretization subroutines.
For example, suppose that the number of points of a spatio-temporal Poisson process is a function of type, spatial location, and time, and that a piecewise constant arrival rate model $\lambda_{c,i,t}$ for type~$c$, spatial subregion~$i$, and time interval~$t$, is estimated.
LASPATED provides two approaches discussed in Section~\ref{sec:modelcalib} to calibrate such models.
Both approaches allow one to estimate with sparse data, for example when there are many combinations of subregion and time window in the discretization relative to the amount of data.
In the first approach, the intensities are the solutions of an optimization problem which optimizes a linear combination of the log-likelihood and a regularization term that may include covariates or penalize non-smoothness of intensities regarding space and time.
In the second approach, covariates are used as auxiliary explanatory variables in the model. 

To the best of our knowledge, LASPATED is the first software that:
\begin{enumerate}
\item implements the calibration of the models we propose for spatio-temporal data;
\item provides a complete analyis and calibration of statistical models for spatio-temporal
data taking only historical data as inputs and providing as outputs a discretized process,
Poisson intensities, and scenarios for future arrivals;
\item offers such a large set of options for spatial discretization;
\item offers such a large set of options for time discretization;
\item handles arrival types for discretization and provides outputs using information from
several discretizations of several types of data.
\end{enumerate}

The outline of the paper is as follows. 
In Section~\ref{sec:modelcalib}, we describe models to be estimated with discretized spatio-temporal data, and explain how the parameters of these models can be calibrated, making provision for cases in which relatively little data are available.
Time discretization functionalities are described in Section~\ref{sec:timedis}.
Space discretization is discussed in Sections~\ref{sec:spaced} and~\ref{sec:adddisc}.
A more detailed description of each function is provided in the LASPATED user manual \citep{laspatedmanual}.
The user can also learn how to use LASPATED with an online video tutorial\footnote{available at https://www.youtube.com/watch?v=tHjhEkySn4E\&list=PLJOegoo5cBB3cAQUAv05iCBExDCPvr4Ve}.
In Section~\ref{sec:examples}, we demonstrate various models and results obtained with LASPATED, and we compare the results with basic maximum likelihood estimates, using simulated data.
We also demonstrate the use of LASPATED to calibrate models of the arrival process of emergency calls to the Rio de Janeiro emergency medical service. Finally, in Section \ref{sec:rep}, we explain how to run our Python repliction script which allows us to run all examples and experiments presented in this paper.

\section{Related Literature}
\label{sec:literature}

Spatio-temporal point processes consisting of data with a location and a timestamp arise in many applications such as the epicenters of earthquakes \citep{ogata1,ogata2}, sudden crimes such as robberies and assaults \citep{paya:23}, epidemiology \citep{elli:01,wall:04}, and medical emergencies \citep{schm:12}.
For the theory of point processes, see \cite{dale:03,dale:08}.
\cite{moller03,scha:05,gelf:10,cres:11,digg:14}; and \cite{gonz:16} provide overviews of statistical models and estimation for spatial and spatio-temporal point processes.
\cite{biva:13} and \cite{wikl:19} describe the use of \texttt{R} for processing spatial and spatio-temporal data and for model estimation.
\cite{pebe:12} describes the \texttt{R} package \texttt{spacetime} for converting spatio-temporal data among different formats, aggregating spatio-temporal data, and creating various types of plots.

The references above consider two types of models of the intensity function $\lambda : X \mapsto \RR_{+}$ of a nonhomogeneous spatial or spatio-temporal Poisson process.
One type of model is a nonparametric model such as the following.
Let $\{x_{n}\}_{n=1}^{N} \subset X$ denote the observed points.
A kernel function $\kappa : \RR_{+} \mapsto \RR_{+}$, and a smoothing parameter~$h > 0$ is chosen.
Then the intensity model is given by
\begin{equation}
\lambda(x) \ \ = \ \ \frac{1}{h^2} \sum_{n=1}^{N} \kappa\left(\frac{\left\|x - x_{n}\right\|}{h}\right).
\label{eqn:kernel nonparametric intensity}
\end{equation}
One challenge is that the effort to compute $\lambda(x)$ for a given~$x$ increases in the size~$N$ of the dataset.
Another challenge is to select a good value for the smoothing parameter~$h$.
Several software packages make provision for the estimation of such kernel-based intensity models, including \texttt{splancs} \citep{rowl:93} and \texttt{spatstat} \citep{badd:05a}.
In contrast, LASPATED makes provision for the estimation of nonparametric models based on discretization of~$X$.
For example, let $\{X_{i}\}_{i=1}^{I}$ denote a partition of~$X$.
Then the intensity model is given by
\begin{equation}
\lambda(x) \ \ = \ \ \sum_{i=1}^{I} \mathds{1}_{\{x \in X_{i}\}} \lambda_{i}
\label{eqn:discretized nonparametric intensity}
\end{equation}
where $\lambda_{i}$, $i = 1,\ldots,I$, is estimated with data.
LASPATED facilitates both the discretization as well as the parameter estimation.

Another type of model is a parametric model such as the following.
Let $z_{k} : X \mapsto \RR$, $k = 1,\ldots,K$, denote a chosen collection of covariate functions, and let $\beta_{k}$, $k = 1,\ldots,K$, denote corresponding parameters to be estimated.
Then the intensity model is given by
\begin{equation}
\lambda(x) \ \ = \ \ \exp\left(\sum_{k=1}^{K} \beta_{k} z_{k}(x)\right).
\label{eqn:unconstrained parametric intensity}
\end{equation}
An advantage of the $\exp\left(\cdot\right)$ on the right side is that $\lambda(x) > 0$ for all~$\beta$ and all~$x$.
A disadvantage is that $\exp\left(\cdot\right)$ grows much faster when its argument is large than when its argument is small, which often results in poor fit and numerical instability.
In contrast, LASPATED makes provision for the estimation of constrained parametric models such as
\begin{equation}
\lambda(x) \ \ = \ \ \sum_{k=1}^{K} \beta_{k} z_{k}(x)
\qquad \mbox{subject to} \qquad \beta_{k} z_{k}(x) \ > \ 0 \ \ \forall \ x \in X
\label{eqn:constrained parametric intensity}
\end{equation}
where typically $z(X)$ is bounded.
Several software packages make provision for the estimation of parametric intensity models of the form~\eqref{eqn:unconstrained parametric intensity}, including \texttt{spatstat} \citep{badd:05a} and \texttt{NHPoisson} \citep{cebr:15}.
More specifically, \texttt{NHPoisson} is an \texttt{R} package for the estimation and testing of parametric intensity functions of the form $\lambda(t) = \exp\big(\beta^{\top} z(t)\big)$ in one dimension, say time~$t$.
It is assumed that the covariate values are piecewise constant over integer intervals, that is, $z(t)$ is constant for all $t \in (k, k+1]$ for integers~$k$.
\texttt{NHPoisson} can use two optimization routines, \texttt{optim} and \texttt{nlminb}, to compute the maximum likelihood estimates $\hat{\beta}$ of $\beta$.
\texttt{NHPoisson} also computes the inverse of the negative of the Hessian of the log-likelihood function as an asymptotic estimate of the covariance matrix of $\hat{\beta}$.
\texttt{NHPoisson} computes the residuals and correlation metrics such as the Pearson correlation coefficient, and performs distribution tests such as the Kolmogorov-Smirnov test.
\cite{cebr:15} also specifically consider Poisson process approximations of peak over threshold models.

\cite{badd:13} proposed hybrid point process models with densities given by the normalized product of densities of a number of point processes, and described the use of \texttt{spatstat} for
statistical inference for the hybrid models.


\cite{hans:13} describes the ppstat-package for analyzing data from multivariate point processes in time or one-dimensional space based on a specification of the conditional intensity process.

\ignore{
In this paper, we consider spatio-temporal point processes that produce data with continuous locations and timestamps.
As explained in the introduction, our software LASPATED has two main functionalities: (i) space and time discretization of the observations and (ii) calibration of two statistical models for these discretized observations: one model that minimizes a regularized likelihood and another model that uses covariates.
To the best of our knowledge, this is the first software that offers a large set of functionalities for both (i) and (ii).
More precisely, for spatial discretization, we are only aware of Uber Python library \textit{H3}, see \cite{uber} and some functionalities offered by Python GeoPandas library available at {url{https://geopandas.org/en/stable/}}.
Our spatial discretization tools use and enrich these libraries, see Section~\ref{sec:spaced} for details.
Moreover, we are not aware of time discretization tools that are as generic as the ones we provide in LASPATED, see Section~\ref{sec:timedis} for details.
Finally, we are not aware of a software providing the calibration of the models described in Section \ref{sec:modelcalib} for stochastic spatio-temporal data.
The discretization functionalities are implemented in Python.
The model calibration functions were implemented in C++ and Matlab and can be run from
C++, Matlab, and Python.
}

\section{Poisson Models with Discretized Spatio-temporal Data}
\label{sec:modelcalib}

In this section, we describe the models that motivated the work on LASPATED, and we explain how LASPATED solves the optimization problems for the calibration of these models.
We will use the medical emergency application as a running example to explain the models.

Let $i \in \mathcal{I}$ index the subsets of the space discretization forming a partition of the region $\mathcal{S} \subset \RR^{n}$; the elements of $\mathcal{I}$ will be called zones.
Let $t \in \mathcal{T}$ index the subsets of the time discretization forming a partition of all times of interest; the elements of $\mathcal{T}$ will be called time intervals.
Let $\mathcal{C}$ denote the set of point types; $\mathcal{C}$ can be any finite set that forms a partition of all space-time points of interest; the elements of $\mathcal{C}$ will be called types.

\subsection{Model without Covariates}
\label{sec:model1}

Each time interval $t \in \mathcal{T}$ has a duration $\mathcal{D}_{t}$ (in time units).
It is assumed that data are observed for multiple occurrences of the same interval~$t$, and that each time data are observed for a type~$c \in \mathcal{C}$, zone~$i \in \mathcal{I}$, and interval~$t \in \mathcal{T}$, all such space-time points in the same time interval are recorded in the data.
For example, suppose that $\mathcal{T}$ forms a partition of the week in time intervals of $1$~hour each.
Then it is assumed that the data contain observations of emergencies of type~$c$ in zone~$i$ during hour~$t$ for multiple weeks, and that each time emergency arrivals of type~$c$ in zone~$i$ are observed during a particular hour~$t$ of a particular week, all the emergency arrivals of the same type in the same zone during the same hour~$t$ of that week are recorded.
(It is planned to make provision for censored data in future work.)
All the emergency arrivals of a particular type~$c$ in a particular zone~$i$ during a particular hour~$t$ of a week together are called an observation.
For each $c \in \mathcal{C}$, $i \in \mathcal{I}$, and $t \in \mathcal{T}$, let $N_{c,i,t}$ denote the number of observations for type~$c$, zone~$i$, and time interval~$t$, and let these observations be indexed by $n \in \mathcal{N}_{c,i,t} \defi \{1,\ldots,N_{c,i,t}\}$.
For each $c \in \mathcal{C}$, $i \in \mathcal{I}$, $t \in \mathcal{T}$, and $n \in \mathcal{N}_{c,i,t}$, let $M_{c,i,t,n}$ denote the number of points (arrivals) for observation~$n$ of type~$c$, zone~$i$, and time interval~$t$, and let $M_{c,i,t} \defi \sum_{n=1}^{N_{c,i,t}} M_{c,i,t,n}$ denote the total number of points over all observations for type~$c$, zone~$i$, and time interval~$t$.

Assume that $\big\{M_{c,i,t,n} \, : \, c \in \mathcal{C}, i \in \mathcal{I}, t \in \mathcal{T}, n \in \mathcal{N}_{c,i,t}\big\}$ are independent (but not necessarily identical) Poisson distributed random variables.
Let $\lambda_{c,i,t}$ denote the mean number of points per length of time (such as per hour) for type~$c$, zone~$i$, and time interval~$t$.
Then random variable $M_{c,i,t,n}$ is Poisson distributed with mean $\lambda_{c,i,t} \mathcal{D}_{t}$.
Let $\lambda \defi \big(\lambda_{c,i,t}, c \in \mathcal{C}, i \in \mathcal{I}, t \in \mathcal{T}\big)$.
Then the likelihood function is
\[
L(\lambda) \ \ = \ \ \prod_{c \in \mathcal{C}} \prod_{i \in \mathcal{I}} \prod_{t \in \mathcal{T}} \prod_{n \in \mathcal{N}_{c,i,t}} e^{-\lambda_{c,i,t} \mathcal{D}_{t}} \frac{(\lambda_{c,i,t}\mathcal{D}_{t})^{M_{c,i,t,n}}}{M_{c,i,t,n}!}
\]
and intensities $\lambda$ that maximize the log-likelihood are the same intensities that solve
\begin{equation}
\label{eqn:maxlikelihood1}
\min_{\lambda} \left\{\mathscr{L}(\lambda) \ \ \defi \ \ \sum_{c \in \mathcal{C}} \sum_{i \in \mathcal{I}} \sum_{t \in \mathcal{T}} \left[N_{c,i,t} \lambda_{c,i,t}\mathcal{D}_{t} - M_{c,i,t} \log\left(\lambda_{c,i,t}\right)\right]\right\}.
\end{equation}

A typical issue with such applications is that the distribution of data is far from uniform --- there are a few combinations of type~$c$, zone~$i$, and time interval~$t$ with many observations, but for most combinations of type~$c$, zone~$i$, and time interval~$t$ there are very few observations.
In such cases, it may be advantageous to use data from ``neighboring'' observations, or to bring additional data to bear on the estimation problem.

For example, suppose that $\mathcal{T}$ forms a partition of the week in time intervals of $1$~hour each, but that it is expected that many hours of the week are similar to other hours of the week in terms of arrival rates.
A simple approach to incorporate such an idea is to partition $\mathcal{T}$ into a collection $\mathcal{G}$ of subsets of $\mathcal{T}$ in such a way that it may be reasonable to expect that, for each $c \in \mathcal{C}$, $i \in \mathcal{I}$, and $G \in \mathcal{G}$, the values of $\lambda_{c,i,t}$ for different $t \in G$ will be close to each other (but not necessarily the same).
Let $W_{G} \ge 0$ denote a similarity weight for $G \in \mathcal{G}$.
Another approach to incorporate such an idea is to specify a similarity weight $w_{t,t'} \ge 0$ for each (unordered) pair $t,t' \in \mathcal{T}$.
For example, $w_{t,t'} = w > 0$ if $t$ and $t'$ are neighboring time intervals, and $w_{t,t'} = 0$ otherwise.
Similarly, it may be expected that many zones are similar to other zones in terms of arrival rates.
Such an idea can also be incorporated by partitioning the set $\mathcal{I}$ of zones into subsets, each with its own similarity weight, or by specifying for each pair $i,j \in \mathcal{I}$, a similarity weight $w_{i,j} \ge 0$.
An example loss function with similarity regularization that uses the first approach for time intervals and the second approach for zones is given by
\begin{align}
\ell(\lambda) \ \ = \ \ &
\displaystyle \sum_{c \in \mathcal{C}} \sum_{i \in \mathcal{I}} \sum_{G \in \mathcal{G}} \sum_{t \in G} \left[N_{c,i,t} \lambda_{c,i,t} \mathcal{D}_{t} - M_{c,i,t} \log\left(\lambda_{c,i,t}\right) + \frac{W_{G}}{2} \sum_{t' \in G} N_{c,i,t} N_{c,i,t'} \left(\lambda_{c,i,t} - \lambda_{c,i,t'}\right)^2\right] \vspace{2mm} \nonumber \\
& \displaystyle {} + \sum_{c \in \mathcal{C}} \sum_{i,j \in \mathcal{I}} \sum_{t \in \mathcal{T}} \frac{w_{i,j}}{2} N_{c,i,t} N_{c,j,t} \left(\lambda_{c,i,t} - \lambda_{c,j,t}\right)^2.
\label{eqn:regularization1}
\end{align}
Intensity estimates~$\lambda$ are then obtained by solving 
\begin{equation}
\label{eqn:model0}
\min_{\lambda > 0} \; \ell(\lambda).
\end{equation}

\subsection{Model with Covariates}
\label{sec:modelcov}

Each type~$c$, zone~$i$, and time interval~$t$ may have covariates that are correlated with the arrival rates~$\lambda_{c,i,t}$, and data of these covariates can be used to partly compensate for sparse data.
For example, emergency arrival rates in different zones and time intervals can be expected to be correlated with the population and other measures of economic activity in the zones, as well as with festivals and other events during the time intervals.
For each $c \in \mathcal{C}$, $i \in \mathcal{I}$, and $t \in \mathcal{T}$, let $x_{c,i,t} \defi (x_{c,i,t,1},\ldots,x_{c,i,t,K})$ denote the covariate values of type~$c$, zone~$i$, and time interval~$t$.
For example, $x_{c,i,t,1}$ may be the population count with home addresses in a zone~$i$, and $x_{c,i,t,2}$ may be an indicator that a major sports event is scheduled in a zone~$i$ during a time interval~$t$.
Then consider the model
\begin{equation}
\label{eqn:covariate model}
\lambda(x_{c,i,t}) \mathcal{D}_{t} \ \ = \ \ \beta^{\top} x_{c,i,t}
\end{equation}
where $\beta = (\beta_{1},\ldots,\beta_{K})$ are the model parameters.
Let $\mathcal{X}_{c,i,t}$ denote the set of all possible values of $x_{c,i,t}$.
Often $\mathcal{X}_{c,i,t}$ can be chosen to be a polyhedron.
Note that it should hold that
\begin{equation}
\beta^{\top} x_{c,i,t} \ \ \ge \ \ 0 \qquad \forall \ x_{c,i,t} \in \mathcal{X}_{c,i,t}, \ \forall \ c \in \mathcal{C}, i \in \mathcal{I}, t \in \mathcal{T}.
\label{eqn:feassetref1}
\end{equation}

To facilitate such a model, let $N = \sum_{c \in \mathcal{C}} \sum_{i \in \mathcal{I}} \sum_{t \in \mathcal{T}} N_{c,i,t}$ denote the total number of observations, and let these observations be indexed $n = 1,\ldots,N$.
For each observation $n \in \{1,\ldots,N\}$, let $M_{n}$ denote the number of arrival points for observation~$n$ and let $x^{n} \defi (x^{n}_{1},\ldots,x^{n}_{K})$ denote the covariate values of observation~$n$.
Then the negative log-likelihood function is given by
\begin{equation}
\label{eqn:maxlikelihood3}
\mathscr{L}(\beta) \ \ = \ \ \sum_{n=1}^{N} \left[\beta^{\top} x^{n} - M_{n} \log\left(\beta^{\top} x^{n}\right)\right].
\end{equation}

Next, we provide two examples of such models.

\begin{example}
\label{excov1}
Index $\mathcal{C} = \{1,\ldots,|\mathcal{C}|\}$, $\mathcal{T} = \{1,\ldots,|\mathcal{T}|\}$, and let $K_{1} \defi |\mathcal{C} \times \mathcal{T}|$.
For each $c \in \mathcal{C}$, $t \in \mathcal{T}$, and $k = (c - 1)|\mathcal{T}| + t$, let $x^{n}_{k}$ be the number of people resident (population count) in the zone of observation~$n$ if observation~$n$ is for type~$c$ and time interval~$t$, and $x^{n}_{k} = 0$ otherwise.
The next covariate is a set of occupational land use areas (in km$^2$), for instance the areas of commercial activities and public facilities, of industrial activities, and of undeveloped land.
Index the occupational land uses by $\mathcal{O} = \{1,\ldots,|\mathcal{O}|\}$.
For each $c \in \mathcal{C}$, $t \in \mathcal{T}$, $m \in \mathcal{O}$, and $k = K_{1} + (c - 1) |\mathcal{T}| |\mathcal{O}| + (t-1) |\mathcal{O}| + m$, let $x^{n}_{k}$ be the area of occupational land use~$m$ in the zone of observation~$n$ if observation~$n$ is for type~$c$ and time interval~$t$, and $x^{n}_{k} = 0$ otherwise.
Then $\sum_{k=1}^{K_{1}} \beta_{k} x^{n}_{k}$ denotes the forecasted number of arrivals for the type of observation~$n$ in the zone of observation~$n$ during the time interval of observation~$n$ due to people being in the residential area, which is modeled as proportional to the number of people resident in the zone of observation~$n$ with a proportionality coefficient that depends on the type and the time interval.
Similarly, $\sum_{k = K_{1} + 1}^{K_{1}(1+|\mathcal{O}|)} \beta_{k} x^{n}_{k}$ denotes the forecasted number of arrivals for the type of observation~$n$ in the zone of observation~$n$ during the time interval of observation~$n$ due to people being in the different occupational areas which is modeled as proportional to the areas of occupational land use in the zone of observation~$n$ with a proportionality coefficient that depends on the type and the time interval.
If $K_{1} |\mathcal{O}|$ is large, the number of parameters $\beta_{k}$ of the model specified in this example is also large.
Similar to~\eqref{eqn:regularization1}, similarity regularization can be used to estimate the parameters, for example, by minimizing a loss function such as
\[
\ell(\beta) \ \ = \ \ \sum_{n=1}^{N} \left[\beta^{\top} x^{n} - M_{n} \log\left(\beta^{\top} x^{n}\right)\right] + \sum_{k,k'=1}^{K_{1}(1+|\mathcal{O}|)} \frac{w_{k,k'}}{2} \left(\beta_{k} - \beta_{k'}\right)^2.
\]
\end{example}

\begin{example}
\label{excov2}
This example will be used to demonstrate the modeling of arrivals of emergency calls to an emergency medical service.
The sets $\mathcal{C}$, $\mathcal{I}$, and $\mathcal{O}$ are the same as in Example~\ref{excov1}.
Let $\mathcal{T}$ denote the indices of discrete time periods during a day, for example, $\mathcal{T} = \{1,\ldots,48\}$ if each day is discretized into $30$-minute intervals.
Let $\mathcal{D}$ denote the set of indices of the (normal) days of the week, as well as indices for special days such as holidays.
Thus, the cardinality of $\mathcal{D}$ is 7 plus the number of special days.
A pair $(d,t) \in \mathcal{D} \times \mathcal{T}$ specifies a time interval.
For each zone $i \in \mathcal{I}$, consider covariates $x_{i} \in \mathbb{R}^{1+|\mathcal{O}|}$, where $x_{i}(1)$ is the population count in zone~$i$, and $x_{i}(j+1)$ is the area (in km$^2$) of occupational land use~$j$ in zone~$i$ for $j=1,\ldots,|\mathcal{O}|$.
Here, we assume that these data (population and land type areas) do not depend on time, which is reasonable for moderate time periods.
Then, for each type~$c$, zone~$i$, day~$d$, and time period~$t$, the arrival intensity is given by $\lambda_{c,i,d,t} \mathcal{D}_{t} = \beta_{c,d,t}^{\top} x_{i}$ for some vector $\beta_{c,d,t} \in \mathbb{R}^{1+|\mathcal{O}|}$).
Furthermore, for each type~$c$, zone~$i$, day~$d$, and time period~$t$, let $N_{c,i,d,t}$ denote the number of observations, let $M_{c,i,d,t,n}$ denote the number of arrival points for observation~$n$, and let $M_{c,i,d,t} \defi \sum_{n=1}^{N_{c,i,d,t}} M_{c,i,d,t,n}$.
Then the negative of the log-likelihood function is given by
\begin{align}
\mathscr{L}(\beta) \ \ & = \ \ \displaystyle \sum_{c \in \mathcal{C}} \sum_{i \in \mathcal{I}} \sum_{d \in \mathcal{D}} \sum_{t \in \mathcal{T}} \sum_{n = 1}^{N_{c,i,d,t}} \left[\beta_{c,d,t}^{\top} x_{i} - M_{c,i,d,t,n} \log(\beta_{c,d,t}^{\top} x_{i})\right] \nonumber \\
& = \ \ \displaystyle \sum_{c \in \mathcal{C}} \sum_{i \in \mathcal{I}} \sum_{d \in \mathcal{D}} \sum_{t \in \mathcal{T}} \left[N_{c,i,d,t} \beta_{c,d,t}^{\top} x_{i} - M_{c,i,d,t} \log(\beta_{c,d,t}^{\top} x_{i})\right].
\label{eqn:covariateloglike}
\end{align}
Note that call rates $\beta_{c,d,t}^{\top} x_{i}$ should be positive.
Additionally, for each type~$c$, day~$d$, and time period~$t$, the coefficients $\beta_{c,d,t}(1)$ that represent the ratio of emergencies to population should be small (say less than 1), which implies the constraints $0 \leq \beta_{c,d,t}(1) \leq 1$.
Therefore, the estimation problem is to solve the optimization problem
\begin{equation}
\label{model2}
\begin{array}{rl}
\min & \ \ \mathscr{L}(\beta) \\
\mbox{s.t.} & \ \ \beta_{c,d,t}^{\top} x_{i} \ > \ 0,
\quad 0 \ \leq \ \beta_{c,d,t}(1) \ \leq \ 1,
\qquad \forall \ c \in \mathcal{C}, d \in \mathcal{D}, t \in \mathcal{T}.
\end{array}
\end{equation}

\if{
\begin{equation}\label{eq:lossf}
\begin{array}{lcl}
\ell(\beta)&=&
\displaystyle 
\sum_{c \in \mathcal{C}}
\sum_{d \in \mathcal{D}}
\sum_{t \in \mathcal{T}}
\sum_{i \in \mathcal{I}}
N_{c,d,t,i}\beta_{c,d,t}^{\top} x_i -
M_{c,d,t,i}
\log(\beta_{c,d,t}^{\top} x_i)\\
&& \displaystyle + \frac{1}{2}
\sum_{c \in \mathcal{C}}
\sum_{G \in \mathcal{G}}
\sum_{(d,t) \in G}
\sum_{(d',t') \in G}W_G 
 \left\|\frac{\beta_{c,d,t}}{\mathcal{D}_t}-\frac{\beta_{c,d',t'}}{\mathcal{D}_{t'}}\right\|_{2}^2.
\end{array}
\end{equation}
}\fi

Depending on the solver, better numerical performance may be obtained by replacing the constraints of~\eqref{model2} with
\begin{equation}
\label{feassetref1b}
\beta_{c,d,t}^{\top} x_{i} \ \geq \ \varepsilon,
\quad 0 \ \leq \ \beta_{c,d,t}(1) \ \leq \ 1,
\qquad \forall \ c \in \mathcal{C}, d \in \mathcal{D}, t \in \mathcal{T},
\end{equation}
for some $\varepsilon > 0$ sufficiently small.
\end{example}

\subsection{Solving the Optimization Problems}
\label{sec:opt_sol}

The calibration of models such as those of Section~\ref{sec:model1} requires solving an optimization problem~\eqref{eqn:model0} of the form $\min\{\ell(\lambda) \, : \, \lambda > 0\}$.
The problem can also be formulated as
\begin{equation}
\label{eqn:reformmodel1}
\min \; \{\ell(\lambda) \; : \; \lambda \in C\}
\end{equation}
where $C$ is a closed convex set such as
\begin{equation}
\label{eqn:Ceps}
C \ \ = \ \ \{\lambda \; : \; \lambda \geq \varepsilon \textbf{e}\}
\end{equation}
for a chosen $\varepsilon > 0$ and \textbf{e} a vector of ones.
The objective function~$\ell$ of~\eqref{eqn:reformmodel1} is differentiable on such $C$.

The calibration of models such as those of Section~\ref{sec:modelcov} requires solving an optimization problem~\eqref{model2} of the form
\begin{equation}
\label{optlikelihbeta}
\min \; \{\mathscr{L}(\beta) \; : \; \beta \in B\}
\end{equation}
where $B$ is given by constraints such as~\eqref{feassetref1b}.
In both cases, the problem is a convex optimization problem with a differentiable objective function on the feasible set.

The algorithms described below for solving problems such as~\eqref{eqn:reformmodel1} take regularization weights such as $W_{G}$ and $w_{i,j}$ as input.
LASPATED also provides cross validation functions, that facilitate use of the data to select regularization weights.

\paragraph{Projected gradient with line search.}
LASPATED provides the following two methods for solving problems~\eqref{eqn:reformmodel1} and~\eqref{optlikelihbeta}: (1)~a projected gradient method with line search along a feasible direction, and (2)~a projected gradient method with line search along the boundary.
These methods follow \cite{iusem2003} except for the update of $\Delta$.
\if{
\footnote{In \cite{iusem2003}, $\Delta$ is any parameter in a given compact interval $[\tilde{\Delta}, \hat{\Delta}] \subset (0,\infty)$.
The choice $\Delta = \bar{\Delta} / 2^{j}$ will satisfy $\Delta \in [\tilde{\Delta}, \hat{\Delta}] \subset (0,\infty)$ when a finite number of outer iterations~$k$ is performed since it can be shown that each line search will terminate after a finite number of iterations.}
}\fi
Here, we provide the pseudocode for these methods for a problem of form~\eqref{eqn:reformmodel1}.
Let $\Pi_{C}(\lambda)$ denote the projection of the point $\lambda$ onto a closed convex set~$C$.
The pseudocode presents methods that are run for a chosen number of iterations and that start from any feasible $\lambda_{0} \in C$.
For alternative stopping criteria, see the discussion at the end of this section. \\

\hrule
\vspace*{0.2cm}
\par {\textbf{Projected gradient method with Armijo line search along a feasible direction for convex problem \eqref{eqn:reformmodel1}.}}
\vspace*{0.2cm}
\hrule
\vspace*{0.2cm}
\par Initialization: Choose $\sigma \in (0,1)$, $\overline{\Delta} > 0$, MaxNumberIteration $> 0$, and an initial feasible point $\lambda_{0} \in C$
\par $\lambda = \lambda_{0}$, $k = 0$, 
$\Delta = \overline{\Delta}$
\par While $k <$ MaxNumberIteration
\par \hspace*{1cm} Compute $\ell(\lambda)$ and $\nabla \ell(\lambda)$
\par \hspace*{1cm} $z = \Pi_{C}(\lambda - \Delta \nabla \ell(\lambda))$
\par \hspace*{1cm} \mbox{zAux} = z
\par \hspace*{1cm} Compute $\ell(\mbox{zAux})$
\par \hspace*{1cm} $j = 0$
\par \hspace*{1cm} While $\left(\displaystyle{\ell(\mbox{zAux}) \ > \ \ell(\lambda) + \frac{\sigma}{2^{j}} \nabla \ell(\lambda)^{\top} (z - \lambda)}\right)$
\par \hspace*{2cm} $j \leftarrow j + 1$
\par \hspace*{2cm} $\displaystyle{\mbox{zAux} = \lambda + \frac{1}{2^{j}} (z - \lambda)}$
\par \hspace*{2cm} Compute $\ell(\mbox{zAux})$
\par \hspace*{1cm} End While
\par \hspace*{1cm} $\lambda = \mbox{zAux}$
\par \hspace*{1cm} $\displaystyle{\Delta \leftarrow \frac{2 \Delta}{2^{j}}}$
\par \hspace*{1cm} $k \leftarrow k+1$
\par End While
\vspace*{0.2cm}
\hrule
\vspace*{0.5cm}

\hrule
\vspace*{0.2cm}
\par {\textbf{Projected gradient method with Armijo line search along the boundary for convex problem \eqref{eqn:reformmodel1}.}}
\vspace*{0.2cm}
\hrule
\vspace*{0.2cm}
\par Initialization: Choose $\sigma \in (0,1)$, $\bar{\Delta} > 0$, MaxNumberIteration $> 0$, and an initial feasible point $\lambda_{0} \in C$
\par $\lambda = \lambda_{0}$, $k = 0$, $\Delta = \overline{\Delta}$
\par While $k <$ MaxNumberIteration
\par \hspace*{1cm} Compute $\ell(\lambda)$ and $\nabla \ell(\lambda)$
\par \hspace*{1cm} $\displaystyle z = \Pi_{C}\left(\lambda - \Delta \nabla \ell(\lambda)\right)$
\par \hspace*{1cm} Compute $\ell(\mbox{z})$
\par \hspace*{1cm} $j = 0$
\par \hspace*{1cm} While $\left(\displaystyle \ell(z) > \ell(\lambda) + \sigma \nabla \ell(\lambda)^{\top} (z - \lambda)\right)$
\par \hspace*{2cm} $j \leftarrow j + 1$
\par \hspace*{2cm} $\displaystyle z = \Pi_{C}\left(\lambda - \frac{\Delta}{2^{j}} \nabla \ell(\lambda)\right)$
\par \hspace*{2cm} Compute $\ell(\mbox{z})$
\par \hspace*{1cm} End While
\par \hspace*{1cm} $\lambda = z$
\par \hspace*{1cm} $\displaystyle{\Delta \leftarrow \frac{2 \Delta}{2^{j}}}$
\par \hspace*{1cm} $k \leftarrow k+1$
\par End While
\vspace*{0.2cm}
\hrule
\vspace*{0.2cm}

Note that for problem~\eqref{eqn:reformmodel1} with $C = \{\lambda \, : \, \lambda \geq \varepsilon \textbf{e}\}$, the projection $\Pi_{C}(\lambda) = \max\{\lambda, \varepsilon \textbf{e}\}$ onto $C$ is easy to compute, where the max is taken componentwise.
Furthermore, the first derivatives of $\ell$ are given by
\[
\frac{\partial \ell}{\partial \lambda_{c,i,t}}(\lambda) \ \ = \ \ N_{c,i,t} \mathcal{D}_{t} - \frac{M_{c,i,t}}{\lambda_{c,i,t}} + W_{G(t)} \sum_{t' \in G(t)} 
N_{c,i,t} N_{c,i,t'} (\lambda_{c,i,t} - \lambda_{c,i,t'}) + \sum_{j \in \mathcal{I}} w_{i,j} N_{c,i,t} N_{c,j,t} (\lambda_{c,i,t} - \lambda_{c,j,t}),
\]
where $G(t)$ denotes the group of time intervals to which $t$ belongs.
For problem \eqref{optlikelihbeta}, the first derivatives of $\mathscr{L}$ are given by
\[
\frac{\partial \mathscr{L}}{\partial \beta_{c,d,t}}(\beta) \ \ = \ \ \sum_{i \in \mathcal{I}} \left[N_{c,i,d,t} x_{i} - \frac{M_{c,i,d,t} x_{i}}{\beta_{c,d,t}^{\top} x_{i}}\right].
\]
\if{
$$
\frac{\partial \ell(\beta)}{\partial \beta_{c,d,t}}
=
\sum_{i} N_{c,d,t,i}x_i -
\sum_{i} \frac{M_{c,d,t,i}x_i}{\beta_{c,d,t}^{\top} x_i}
+ 2 \frac{W_G}{\mathcal{D}_{t}}
\displaystyle 
\sum_{(d',t') \in G((d,t))}
\left(\frac{\beta_{c,d,t}==========================}{\mathcal{D}_{t}} 
-\frac{\beta_{c,d',t'}}{\mathcal{D}_{t'}}\right),
$$
where 
$G((d,t))$
is the group in $\mathcal{G}$
which contains the pair
$(d,t)$.
}\fi
For problem~\eqref{optlikelihbeta} with $B = \left\{\beta \, : \, \beta_{c,d,t}^{\top} x_{i} \geq \varepsilon, \, 0 \leq \beta_{c,d,t}(1) \leq 1,
\, \forall \; c \in \mathcal{C}, d \in \mathcal{D}, t \in \mathcal{T}\right\}$,
the projection $\Pi_{B}(\beta)$ onto $B$ is computed by solving a convex quadratic problem.

\paragraph{Stopping test for projected gradient method with line search.}
LASPATED makes provision for several stopping criteria such as convergence of the objective function values, a maximum number of iterations, or stopping when optimality conditions are approximately satisfied.
We give more details about this latter criterion.
Consider an optimization problem
\[
(\mbox{P}) \qquad \ell_{*} \ \ = \ \ \min \; \{\ell(\lambda) \; : \; \lambda \in C\}
\]
where $\ell$ and $C$ are convex, and $C$ is compact.
At the end of iteration~$k$, points $\lambda_{0},\ldots,\lambda_{k}$ have been generated, and the upper bound
\begin{equation}
\label{uppbd}
u_{k} \ \ = \ \ \min\big\{\ell(\lambda_{0}),\ldots,\ell(\lambda_{k})\big\}
\end{equation}
on $\ell_{*}$ can be computed.
One can minimize the linear approximation of $\ell$ at $\lambda_{k}$ to compute the lower bound
\begin{equation}
\label{lowbd}
\ell_{k} \ \ = \ \ \min \, \left\{\ell(\lambda_{k}) + \nabla \ell(\lambda_{k})^{\top} (\lambda - \lambda_{k}) \; : \; \lambda \in C\right\}
\end{equation}
or one can minimize the piecewise linear approximation of $\ell$ at $\lambda_{0},\ldots,\lambda_{k}$ to compute the lower bound
\begin{align}
\ell_{k} \ \ & = \ \ \min \left\{\max \left\{\ell(\lambda_{i}) + \nabla \ell(\lambda_{i})^{\top} (\lambda - \lambda_{i}) \; : \; i \in \{0,\ldots,k\}\right\} \; : \; \lambda \in C\right\} \nonumber \\
& = \ \ \min \left\{\theta \; : \; \theta \ge \ell(\lambda_{i}) + \nabla \ell(\lambda_{i})^{\top} (\lambda - \lambda_{i}) \; \forall \; i \in \{0,\ldots,k\}, \; \lambda \in C\right\}
\label{lowbd2}
\end{align}
The algorithm is stopped when
\begin{equation}
u_{k} - \ell_{k} \ \ \leq \ \ \varepsilon,
\end{equation}
thus obtaining an $\varepsilon$-optimal solution.
\if{
and $C$ is compact with diameter $D(C) \defi \max\{\|x - y\|_{2} \, : \, x,y \in C\}$.
Let $\delta_{C}(\lambda) \defi 0$ if $\lambda \in C$ and $\delta_{C}(\lambda) \defi \infty$ if $\lambda \notin C$.
Consider $(\lambda_{k},u_{k})$ such that $\lambda_{k} \in C$,
\begin{equation}
\label{eqn:uk}
u_{k} \ \ \in \ \ \partial(\ell + \delta_{C})(\lambda_{k}),
\end{equation}
and $\|u_{k}\|_{2} \leq \epsilon$ for specified $\epsilon > 0$ (meaning that the optimality conditions are approximately satisfied at $\lambda_{k}$).
Then it holds for all $\lambda \in C$ that
\[
\ell(\lambda) \ \ \geq \ \ \ell(\lambda_{k}) + \langle u_{k}, \lambda - \lambda_{k} \rangle \ \ \geq \ \ \ell(\lambda_{k}) - \|u_{k}\|_{2} D(C) \ \ \geq \ \ \ell(\lambda_{k}) - \epsilon D(C)
\]
and thus, for $\varepsilon$ replaced by $\epsilon D(C)$, $\ell(\lambda_{k}) \leq \ell_{*} + \varepsilon$, that is, $\lambda_{k}$ is an $\varepsilon$-optimal solution of (P).

Next, we describe the computation of subgradients $u_{k} \in \partial(\ell + \delta_{C})(\lambda_{k})$ and that become small, so that $\|u_{k}\|_{2}$ can be used as a stopping criterion.
Consider the case with $C = \{\lambda : A \lambda \leq b\}$.
Then condition~\eqref{eqn:uk} can be written 
\[
u_{k} \ \ \in \ \ \partial \ell(\lambda_{k}) + N_{C}(\lambda_{k})
\]
where $N_{C}(\lambda_{k})$ denotes the normal cone to $C$ at $\lambda_{k}$.
If $\ell$ is differentiable, then it follows from
\[
N_{\{\lambda : A \lambda \leq b\}}(\lambda_{k}) \ \ = \ \ \{A^{\top} \gamma \; : \; \gamma \geq 0, \; \gamma^{\top} (b - A \lambda_{k}) = 0\}
\]
that
\begin{equation}
\label{eqn:subdifferential}
u_{k} \ \ \in \ \ \left\{\nabla \ell(\lambda_{k}) + A^{\top} \gamma \; : \; \gamma \geq 0, \; \gamma^{\top} (b - A \lambda_{k}) = 0\right\}.
\end{equation}
If some constraints $A \lambda \leq b$ are active at $\lambda_{k}$, then the subdifferential on the right of~\eqref{eqn:subdifferential} is unbounded, whereas it is desirable to compute small subgradients $u_{k}$ to serve as stopping criterion.
One could solve auxiliary optimization problems such as
\begin{eqnarray*}
{\mbox{Opt}}_{k} \ \ = \ \ \min_{u, \gamma} & & \|u\|_{1} \\
\mbox{s.t.} & & u \ \ = \ \ \nabla \ell(\lambda_{k}) + A^{\top} \gamma \\
& & \gamma \ \ \geq \ \ 0, \quad \gamma^{\top} (b - A \lambda_{k}) \ \ = \ \ 0.
\end{eqnarray*}
or
\begin{eqnarray*}
{\mbox{Opt}}_{k} \ \ = \ \ \min_{u,\gamma} & & \|u\|_{2} \\
\mbox{s.t.} & & u \ \ = \ \ \nabla \ell(\lambda_{k}) + A^{\top} \gamma \\
& & \gamma \ \ \geq \ \ 0, \quad \gamma^{\top} (b - A \lambda_{k}) \ \ = \ \ 0
\end{eqnarray*}
and stop the algorithm when ${\mbox{Opt}}_{k} \leq \varepsilon / D(C)$.
\ignore{
In practice we can stop the algorithm when Opt$_{k}$ is small but if $C$ is not compact there is no guarantee that $\lambda_{k}$ will be an $\varepsilon$-optimal solution with $\varepsilon$ small as can be seen from this simple one dimensional example:
Consider $\ell(x) = 10^{-k} |x|$ with $k$ a large positive integer.
Then it is clear that the condition $u \in \nabla \ell(x)$ with $u$ of small norm does not imply that $\ell(x)$ is close to 0 since $|\ell'(x)| = 10^{-k}$ even if $|x| > 10^{k} \varepsilon$.
}
Next, we show how to compute subgradients $u_{k}$ that go to~$0$ without solving auxiliary optimization problems.
First, consider the projected gradient algorithm with Armijo search along the boundary for convex problem~\eqref{eqn:reformmodel1} with $C = \{\lambda : A \lambda \leq b\}$.
The projection steps are given by $\lambda_{k+1} = \pi_{C}\big(\lambda_{k} - [\bar{\Delta} / 2^{s(k)}] \nabla \ell(\lambda_{k})\Big)$.
The algorithm that computes the projection $\pi_{C}\big(\lambda_{k} - [\bar{\Delta} / 2^{s(k)}] \nabla \ell(\lambda_{k})\Big)$ returns an optimal dual solution $\gamma_{k+1}$ corresponding to the constraint $A \lambda \leq b$.
It follows from the optimality conditions for the projection problem at iteration~$k-1$ that
\begin{eqnarray}
\label{eqn:projection opt 1}
\lambda_{k} - \lambda_{k-1} + \frac{\bar{\Delta}}{2^{s(k-1)}} \nabla \ell(\lambda_{k-1}) + A^{\top} \gamma_{k} \ \ = \ \ 0 \\
\gamma_{k} \ \ \geq \ \ 0, \qquad \gamma_{k}^{\top} (b - A \lambda_{k}) \ \ = \ \ 0
\label{eqn:projection opt 2}
\end{eqnarray}
and at iteration~$k$ that
\begin{eqnarray}
\label{eqn:projection opt 3}
\lambda_{k+1} - \lambda_{k} + \frac{\bar{\Delta}}{2^{s(k)}} \nabla \ell(\lambda_{k}) + A^{\top} \gamma_{k+1} \ \ = \ \ 0 \\
\gamma_{k+1} \ \ \geq \ \ 0, \qquad \gamma_{k+1}^{\top} (b - A \lambda_{k+1}) \ \ = \ \ 0.
\label{eqn:projection opt 4}
\end{eqnarray}
It follows from~\eqref{eqn:projection opt 2} that $[2^{s(k)}/{\bar{\Delta}}] \gamma_{k}$ is a valid choice for $\gamma$ in~\eqref{eqn:subdifferential}, and thus we can choose
\begin{eqnarray}
\frac{\bar{\Delta}}{2^{s(k)}} u_{k} & = & \frac{\bar{\Delta}}{2^{s(k)}} \nabla \ell(\lambda_{k}) + A^{\top} \gamma_{k} \nonumber \\
\Rightarrow \qquad u_{k} & = & \frac{2^{s(k)}}{\bar{\Delta}} \big[\lambda_{k} - \lambda_{k+1} + A^{\top} (\gamma_{k} - \gamma_{k+1})\big]
\label{eqn:subdifferential2}
\end{eqnarray}
where~\eqref{eqn:subdifferential2} follows from~\eqref{eqn:projection opt 3}.
Thus, the stopping criterion $\|u_{k}\|_{2} \leq \varepsilon / D(C)$ becomes
\[
\frac{2^{s(k)}}{\bar{\Delta}} \Big\|\lambda_{k} - \lambda_{k+1} + A^{\top} (\gamma_{k} - \gamma_{k+1})\Big\|_{2} \ \ \leq \ \ \frac{\varepsilon}{D(C)}
\]
which implies that $\lambda_{k}$ is an $\varepsilon$-optimal solution.
It was shown in \cite{iusem2003} that $\{\lambda_{k}\}$ converges, and thus $\|\lambda_{k} - \lambda_{k+1}\| \to 0$ as $k \to \infty$.
Furthermore, denoting $v_k=\|\lambda_k - \lambda_{k-1}\|$ 
it follows from~\eqref{eqn:projection opt 1} and~\eqref{eqn:projection opt 3} that
\begin{eqnarray*}
A^{\top}(\gamma_{k} - \gamma_{k+1}) & = & \lambda_{k+1} - \lambda_{k} + \frac{\bar{\Delta}}{2^{s(k)}} \nabla \ell(\lambda_{k}) - \lambda_{k} + \lambda_{k-1} - \frac{\bar{\Delta}}{2^{s(k-1)}} \nabla \ell(\lambda_{k-1}) \\
\Rightarrow \qquad \left\|A^{\top}(\gamma_{k} - \gamma_{k+1})\right\| & \le & v_{k+1} + v_k + \left\|\frac{\bar{\Delta}}{2^{s(k)}} \nabla \ell(\lambda_{k}) - \frac{\bar{\Delta}}{2^{s(k-1)}} \nabla \ell(\lambda_{k-1})\right\|.
\end{eqnarray*}

We will assume:
\begin{itemize}
\item[(A1)] $\ell$ is Lipschitz continuously differentiable:
there is a constant~$L$ such that
\[
\left\|\nabla \ell(\lambda_{1}) - \nabla \ell(\lambda_{2})\right\|_{2} \ \ \le \ \ L \left\|\lambda_{1} - \lambda_{2}\right\|_{2}
\]
for all $\lambda_{1}, \lambda_{2} \in C$.
\end{itemize}
Since  the sequence $(\lambda_k)$ converges
it is bounded: there is some compact set $D \subset C$ to which it belongs and therefore 
by Assumption (A1) there
is some constant $0 \leq \kappa(D)<+\infty$ such that
for all $k$ we have
$\|\nabla \ell(\lambda_k)\| \leq \kappa(D)$. Setting $w_k={\bar \Delta}\|\nabla \ell(\lambda_k)-\nabla \ell(\lambda_{k-1})\|$, it follows that
\begin{eqnarray*}
\|A^{\top}(\gamma_{k} - \gamma_{k+1})\| & \leq &  v_{k+1} + v_{k}+ w_k + \kappa(D) {\bar \Delta}\left| 
\frac{1}{2^{s(k)}}-\frac{1}{2^{s(k-1)}}
\right|.
\\
\end{eqnarray*}

From the convergence of the sequence
$(\lambda_k)$ and the continuity of
$\nabla \ell$, we have that
$$
\displaystyle \lim_{k \rightarrow +\infty} v_k=\lim_{k \rightarrow +\infty} w_k= 0.
$$

For any $\lambda \in C$, let $N_{C}(\lambda)$ denote the normal cone of $C$ at $\lambda$, let $T_{C}(\lambda)$ denote the tangent cone of $C$ at $\lambda$. 
Then for any $x$ we have
\begin{equation}\label{projnt}
x-\lambda=\Pi_{N_{C}(\lambda)}\big(x-\lambda \big)
+
\Pi_{T_{C}(\lambda)}\big(x-\lambda \big).
\end{equation}
Taking $x=\lambda-\Delta \nabla \ell(\lambda)$ 
in the relation above and setting
\[
v(\lambda) \ \ \defi \ \ \Pi_{N_{C}(\lambda)}\big(-\Delta \nabla \ell(\lambda)\big)
\qquad \mbox{and} \qquad
w(\lambda) \ \ \defi \ \ \Pi_{T_{C}(\lambda)}\big(-\Delta \nabla \ell(\lambda)\big),
\]
we get
$$
-\Delta \nabla \ell(\lambda) 
= v(\lambda) + w(\lambda)
$$
where
$$
v(\lambda)^T w(\lambda)=0.
$$
Furthermore, let
\[
z(\lambda) \ \ \defi \ \ \Pi_{C}\big(\lambda - \Delta \nabla \ell(\lambda)\big).
\]
Next, it is shown in \cite{iusem2003}  that $\lambda_{k} \to \lambda^*$ as $k \to \infty$.
By optimality conditions, we also have
that $-\nabla \ell(\lambda^*) \in N_{C}(\lambda^*)$, and thus $w(\lambda^*) = 0$.
Then it follows from the continuity of $\nabla \ell$ and projection that $w(\lambda_{k}) \to 0$.
Next, taking $x=z(\lambda_k)$ and
$\lambda=\lambda_k$
in relation \eqref{projnt}, we get
$$
z(\lambda_k)-\lambda_k =\Pi_{N_{C}(\lambda_k)}\big(z(\lambda_k)-\lambda_k \big)
+
\Pi_{T_{C}(\lambda_k)}\big(z(\lambda_k)-\lambda_k \big).
$$
Moreover, it follows from $C$ being a polyhedron and $w(\lambda_{k}) \to 0$ that $z(\lambda_{k}) = \lambda_{k} + w(\lambda_{k})$ for all~$k$ sufficiently large.

Also, it follows from the Lipschitz continuity of $\nabla \ell$ that
\[
\ell\big(z(\lambda_{k})\big) \ \ \le \ \ \ell\big(\lambda_{k}\big) + \nabla \ell\big(\lambda_{k}\big)^{\top} [z(\lambda_{k}) - \lambda_{k}] + \frac{L}{2} \|z(\lambda_{k}) - \lambda_{k}\|_{2}^2.
\]
Thus, if $\Delta < 2 (1 - \sigma) / L$, for all $k$ large enough we have
\begin{eqnarray*}
\ell\big(z(\lambda_{k})\big) & \le & \ell\big(\lambda_{k}\big) + \sigma \nabla \ell\big(\lambda_{k}\big)^{\top} [z(\lambda_{k}) - \lambda_{k}] + (1 - \sigma) \nabla \ell\big(\lambda_{k}\big)^{\top} [z(\lambda_{k}) - \lambda_{k}] + \frac{L}{2} \|z(\lambda_{k}) - \lambda_{k}\|_{2}^2 \\
& = & \ell\big(\lambda_{k}\big) + \sigma \nabla \ell\big(\lambda_{k}\big)^{\top} [z(\lambda_{k}) - \lambda_{k}] - \frac{1 - \sigma}{\Delta} \big(v(\lambda_{k}) + w(\lambda_{k})\big)^{\top} w(\lambda_{k}) + \frac{L}{2} \|w(\lambda_{k})\|_{2}^2 \\
& = & \ell\big(\lambda_{k}\big) + \sigma \nabla \ell\big(\lambda_{k}\big)^{\top} [z(\lambda_{k}) - \lambda_{k}] - \frac{1 - \sigma}{\Delta} w(\lambda_{k})^{\top} w(\lambda_{k}) + \frac{L}{2} \|w(\lambda_{k})\|_{2}^2 \\
& < & \ell\big(\lambda_{k}\big) + \sigma \nabla \ell\big(\lambda_{k}\big)^{\top} [z(\lambda_{k}) - \lambda_{k}],
\end{eqnarray*}
and thus $s(k) = 0$ for all $k$ large enough.

The computations are similar for the
other variant of projected gradient:
for the projected gradient method with Armijo search along a feasible direction when
$C=\{\lambda: A \lambda \leq b\}$, we have
$$ \lambda_{k+1}=\lambda_{k} + \frac{1}{2^{s(k)}}(z_{k} - \lambda_{k})
$$
and 
$$
z_{k}=\pi_{C}\Big(\lambda_{k}-\frac{\bar \Delta}{2^{s(k-1)}}\nabla \ell(\lambda_{k})\Big),
$$
where again ${\bar \Delta}/2^{s(k)}$
is the notation used for
the step for iteration $k+1$.
By the optimality conditions we get
that there is $\gamma_{k}$ such that
$$
\begin{array}{l}
\displaystyle 
z_{k} - \lambda_{k} + ({\bar \Delta}/2^{s(k-1)})
\nabla \ell(\lambda_{k}) + A^{\top} \gamma_{k} = 0,\\
\displaystyle \gamma_{k} \geq 0,\;
\gamma_{k}^{\top}(b-A z_{k})=0
\end{array}
$$
which 
is of form
\eqref{eqn:subdifferential}
with $\lambda_{k}$ replaced by
$z_{k}$,
$$
u_{k}=\nabla \ell(z_{k})-\nabla \ell(\lambda_{k})
+(2^{s(k-1)+s(k)}/{\bar \Delta})(\lambda_{k}-\lambda_{k+1}),
$$
and $\gamma:=(2^{s(k-1)}/{\bar \Delta}) \gamma_{k}$. From our previous discussion,
we can use as a stopping criterion
the condition
$\|u_{k}\|_{2} \leq \frac{\varepsilon}{D(C)}$
which implies that
$z_{k}$ is an approximate
$\varepsilon$-primal solution. This stopping criterion can be conservative if the diameter of $C$ is large.

}\fi

\section{Time Discretization}
\label{sec:timedis}

LASPATED provides multiple types of time discretization.
Some time discretization methods are based on a periodic pattern.
The duration of the periodic pattern can be chosen, for example, $7$~days or $10$~days or $3$~months.
To make provision for holidays and special events, LASPATED also facilitates time discretization methods that are not based on a periodic pattern.
The subsets of the time discretization can be chosen to be time intervals, or finite unions of time intervals.
For example, if the duration of the periodic pattern is $7$~days, then Monday [08:00,09:00] $\cup$ Friday [17:00,18:00] can be chosen to be one subset of the time discretization.
Next, we mention some special cases of time discretization facilitated by LASPATED.

\subsection{Periodic with Equal Length Time Intervals}

The simplest time discretization in LASPATED uses a periodic pattern, in which the duration of the periodic pattern is partitioned into time intervals of equal length.
That is, each subset of the time discretization is a single time interval, and all these time intervals have the same length.

\ignore{
The duration of a period, in the time unit of the first argument of the method, is specified in the third argument of the method.
Therefore, the third argument needs to be a multiple of the second argument: it is the duration of an elementary time window times the number of elementary time windows in a period.
In Example \ref{list:1}, we consider two such discretizations.
In the first one, we want to count the number of arrivals for every day of the week.
Therefore, the first argument of add\_time\_discretization is 'D' (the time unit is a day), the second argument is 1 (an elementary time window lasts 1 day), and the third argument is 7 (the one week period has 7 days).
In the second discretization, we want to count the number of arrivals in time windows of 30 mins every day.
Therefore, the first argument of add\_time\_discretization is 'm' (the time unit is a minute), the second argument is 30 (an elementary time window lasts 30 minutes), and the third argument is 30*48 (the one day period has 30*48 minutes).

\begin{lstlisting}[label={list:1},caption=Elementary time windows repetition.]
# 7 days in a week
app.add_time_discretization('D', 1, 7)

# 48 30-minutes slots in a day
app.add_time_discretization('m', 30, 30*48)
\end{lstlisting}
}

\subsection{Periodic with Unequal Length Time Intervals}

Another time discretization in LASPATED also uses a periodic pattern, in which the duration of the periodic pattern is also partitioned into time intervals, but the time intervals may have unequal lengths.
As in the previous method, each subset of the time discretization is a single time interval, but unlike the previous method, all these time intervals do not have the same length.

\subsection{Customized Subsets}

LASPATED makes provision for customized subsets to facilitate holidays and special events.
Each customized subset is assigned a unique index number, and consists of one or more time intervals.
Each time interval is specified by its start time and its end time, as well as the index of the subset that the time interval belongs to.
Time points that do not belong to any customized interval, belong to the customized subset with index~$0$.
LASPATED also allows time intervals to repeat.
For example, a time interval that starts and ends on date 2016-01-01 (New Year's Day) may be specified to repeat each year.
If the time interval is specified as repeating yearly, then for all observations on the same day of the year (such as January 1), the time intervals' start time during the day, end time during the day, as well as the index of the subset that the time intervals belong to, will be the same for all years.

\section{Space Discretization}
\label{sec:spaced}

\subsection{Defining Borders}

The first step for space discretization is to choose a coordinate system and to specify the border of a region that contains the locations of all the points in the dataset.
The border specifies the region that will be discretized in space.
The border can be specified with LASPATED using different methods.

\subsubsection{Custom map}

A custom map can be provided to LASPATED by specifying the coordinates of a sequence of vertices on the border of the region.
Typically, a Shapefile is provided for this purpose; the user manual and the video tutorial contain details and examples.

\ignore{
LASPATED method add\_max\_borders adds a max\_border attribute to object app.
Parameter {\tt{data}} of this method is a GeoDataFrame object\footnote{see \url{https://geopandas.org/en/stable/docs/reference/api/geopandas.GeoDataFrame.html}}.
This GeoDataFrame object has an attribute geometry encoding the vertices of the border and this object is obtained calling function \textbf{gpd.read\_file} which takes as argument a shapefile (with .shp extension) of the custom map (region under consideration).
In Example \ref{list:4} below, we show how to use add\_max\_borders on two examples with data lying in Rio de Janeiro for the first example and data in New-York for the second.
The respective shape files rj.shp and ny.shp can be found on internet and also in folders Data/rj and Data/ny of the github project.

\begin{lstlisting}[label={list:4},caption=Adding borders from a custom map.]
import geopandas as gpd
import matplotlib as plt

# GeoDataFrame object read from Rio de janeiro shape file
custom_map = gpd.read_file(r'../Data/rj/rj.shp')
# adding as border
app.add_max_borders(data=custom_map)

app.max_borders.plot()
plt.show()

# GeoDataFrame object read from New-York shape file
custom_map = gpd.read_file(r'../Data/ny/ny.shp')
# adding as border
app.add_max_borders(data=custom_map)

# Plotting the border
app.max_borders.plot()
plt.show()
\end{lstlisting}
}

\subsubsection{Rectangular border and convex hull}

LASPATED can be instructed to determine various regions that contain the locations of all the points in a dataset, such as the smallest rectangle that contains all the points in the dataset, or the approximate convex hull of all the points in the dataset.
\ignore{
If a customized border is not available, we can use a rectangular border containing all events or the convex hull of the space observations.
In the first case, the argument of parameter {\tt{method}} of LASPATED method add\_max\_borders is "rectangle" while in the second case this parameter {\tt{method}} is "convex".
The following Example \ref{list:5} shows how to define these borders.
Only one of these two options must be chosen.

\begin{lstlisting}[label={list:5},caption=Automatically defined borders.]
# adding rectangular borders
app.add_max_borders(method="rectangle")

# adding as border the convex hull of observations 
app.add_max_borders(method="convex")

# plotting borders and a sample of the events
import matplotlib as plt
fig, ax = plt.subplots()
app.max_borders.plot(ax=ax)
app.events_data.sample(10).plot(markersize=10, color='red', ax=ax)
plt.show()
\end{lstlisting}
}

Once borders are specified, the space discretization step partitions the region inside the specified border into subregions.
Similar to the subsets of the time discretization, the subregions of the space discretization can be chosen to be simple shapes such as rectangles or hexagons, or unions of simple shapes.
Next we mention some special cases of space discretization facilitated by LASPATED.

\subsection{Equal Sized Rectangular Space Discretization}

\begin{figure}
\centering
\begin{tabular}{c}
\includegraphics[scale=0.6]{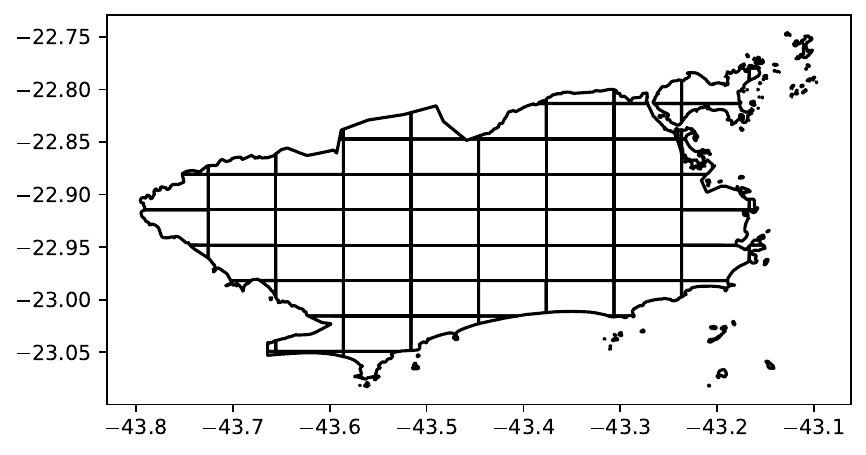}
\end{tabular}
\caption{\label{figurerect1010}
Space discretization of a region containing the city of Rio de Janeiro into $10 \times 10 = 100$ rectangles, 76 of which have nonempty intersection with the region and are shown in the figure.}
\end{figure}

\begin{figure}
\centering
\begin{tabular}{c}
\includegraphics[scale=0.2]{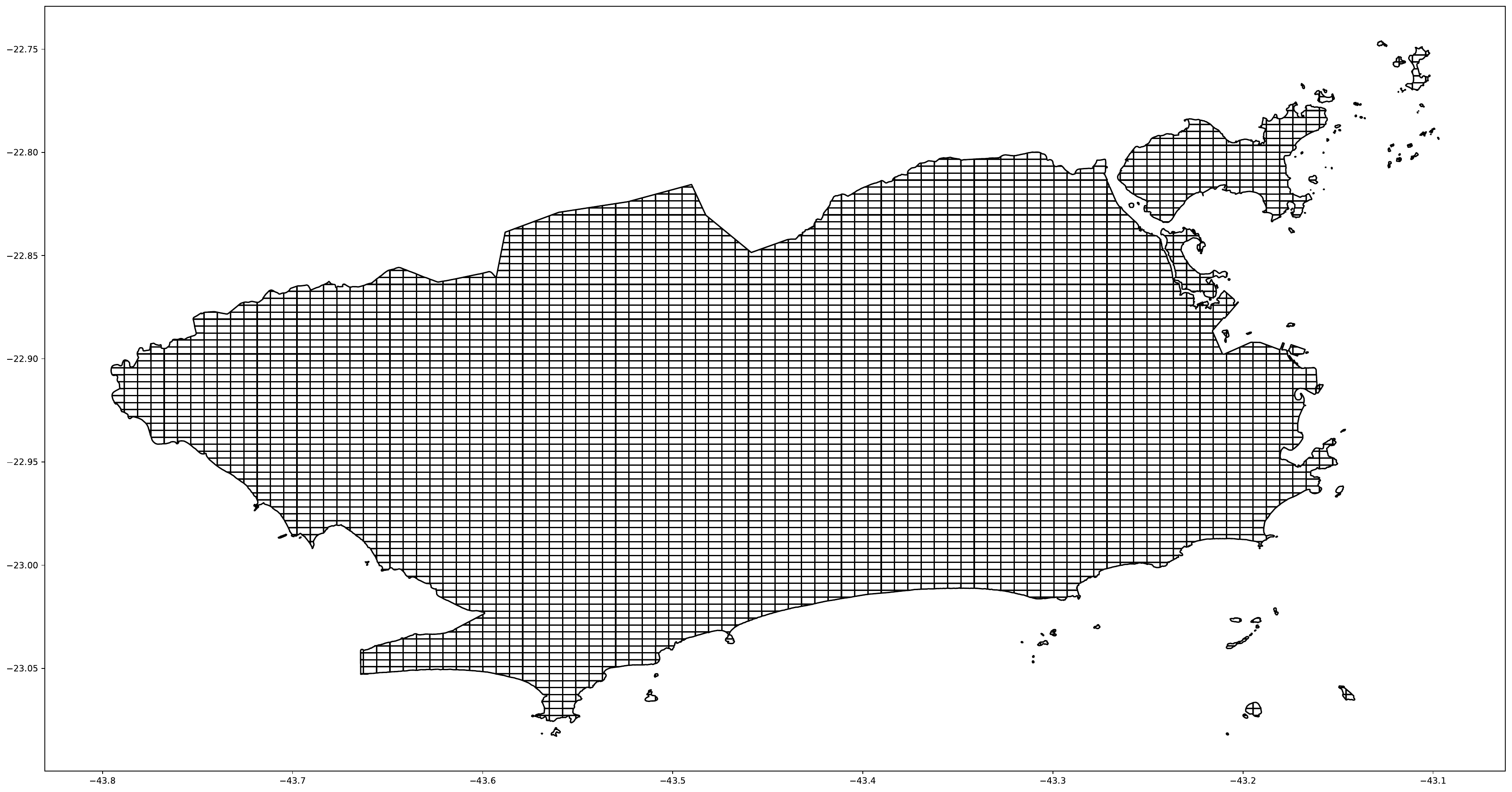}
\end{tabular}
\caption{\label{figurerect100100}
Space discretization of a region containing the city of Rio de Janeiro into $100 \times 100 = 10000$ rectangles, 4916 of which have nonempty intersection with the region and are shown in the figure.}
\end{figure}

The simplest space discretization in LASPATED partitions the region into equal sized rectangles, in such a way that adjacent rectangles share a common face, that is, if the boundaries of rectangle~$A$ and rectangle~$B$ intersect, then either rectangle~$A$ and rectangle~$B$ intersect in one (corner) vertex, or rectangle~$A$ and rectangle~$B$ share an edge between $2$~vertices.
For example, the rectangles $A = [0,2] \times [0,2]$ and $B = [2,4] \times [1,3]$ are adjacent but do not share a common face.
If the region is not a union of these equal sized rectangles, then the intersections of the rectangles with the region may not be equal sized.
For example, Figure~\ref{figurerect1010} shows a discretization of a custom region containing the city of Rio de Janeiro into $10 \times 10 = 100$ rectangles, and Figure~\ref{figurerect100100} shows a discretization of the same region into $100 \times 100 = 10000$ rectangles.
Both discretizations were obtained with LASPATED.

\subsection{Equal Sized Hexagonal Space Discretization}

\begin{figure}
\centering
\begin{tabular}{c}
\includegraphics[scale=0.9]{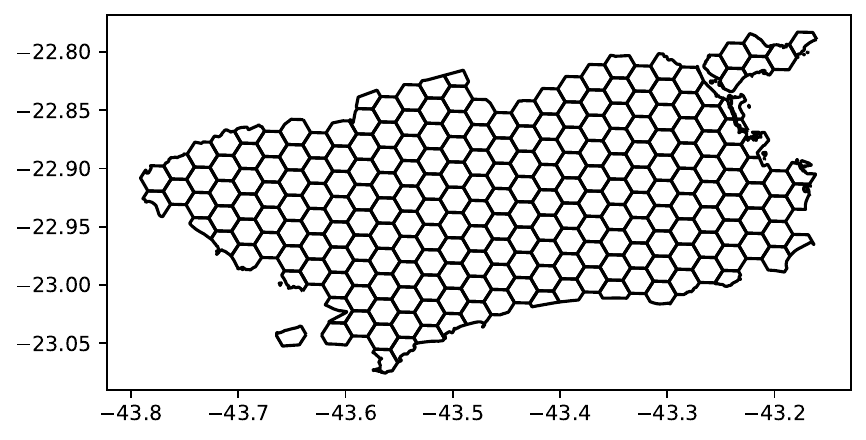}
\end{tabular}
\caption{\label{fig:uber7}
Space discretization of a region containing the city of Rio de Janeiro into hexagons using the Uber library H3 with scale parameter equal to $7$.}
\end{figure}

\begin{figure}
\centering
\begin{tabular}{c}
\includegraphics[scale=0.6]{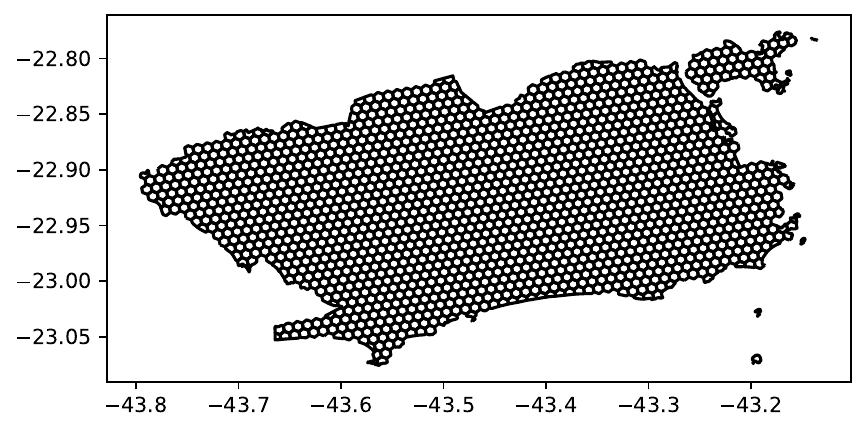}
\end{tabular}
\caption{\label{fig:uber8}
Space discretization of a region containing the city of Rio de Janeiro into hexagons using the Uber library H3 with scale parameter equal to $8$.}
\end{figure}

Another simple space discretization in LASPATED partitions the region into equal sized hexagons, in such a way that adjacent hexagons share a common edge.
(The Uber Python package \textit{H3} was used for the discretization.)
Figure~\ref{fig:uber7} shows a discretization of a region containing the city of Rio de Janeiro into hexagons using a scale parameter equal to 7, and Figure~\ref{fig:uber8} shows a discretization of the same region into hexagons using a scale parameter equal to 8.
Both discretizations were obtained with LASPATED.

\ignore{
LASPATED also provides an hexagonal discretization of the studied area.
This discretization is again obtained with LASPATED method {\textbf{add\_geo\_discretization}} now specifying 'H' for parameter \textbf{discr\_type} and an additional scale integer parameter \textbf{hex\_discr\_param} taking values between one and 16, where the smaller the integer the coarser the discretization.

\begin{lstlisting}[label={list:7},caption=Hexagonal discretization.]

# Hexagonal discretization with scale parameter 7
# The corresponding discretization for Rio de Janeiro city is given in  Figure 3
app.add_geo_discretization(
    discr_type='H',
    hex_discr_param=7
)

import matplotlib as plt
app.geo_discretization.boundary.plot()
plt.show()

# Hexagonal discretization with scale parameter 8
# The corresponding discretization for Rio de Janeiro city is given in  Figure 4
app.add_geo_discretization(
    discr_type='H',
    hex_discr_param=8
)

# Plotting the regions
app.geo_discretization.boundary.plot()
plt.show()
\end{lstlisting}
}

\subsection{Customized Space Discretization}

\begin{figure}
\centering
\begin{tabular}{c}
\includegraphics[scale=0.6]{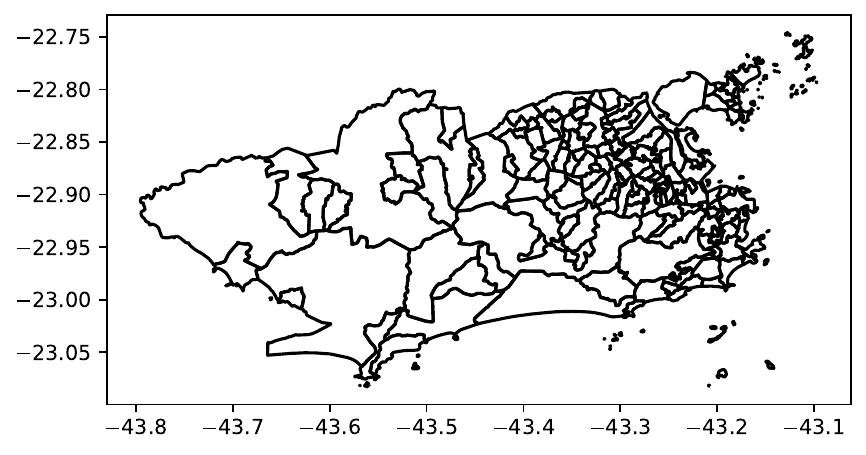}
\end{tabular}
\caption{\label{fig:figuredistrict}
Space discretization of a region containing the city of Rio de Janeiro into 160 administrative districts.\label{distfig}}
\end{figure}

LASPATED also facilitates space discretization with customized subregions.
For example, Figure~\ref{fig:figuredistrict} obtained with LASPATED displays a customized discretization of the city of Rio de Janeiro into $160$~administrative districts.

\subsection{Discretization using Voronoi diagrams}

\begin{figure}
\centering
\begin{tabular}{c}
\includegraphics[scale=0.18]{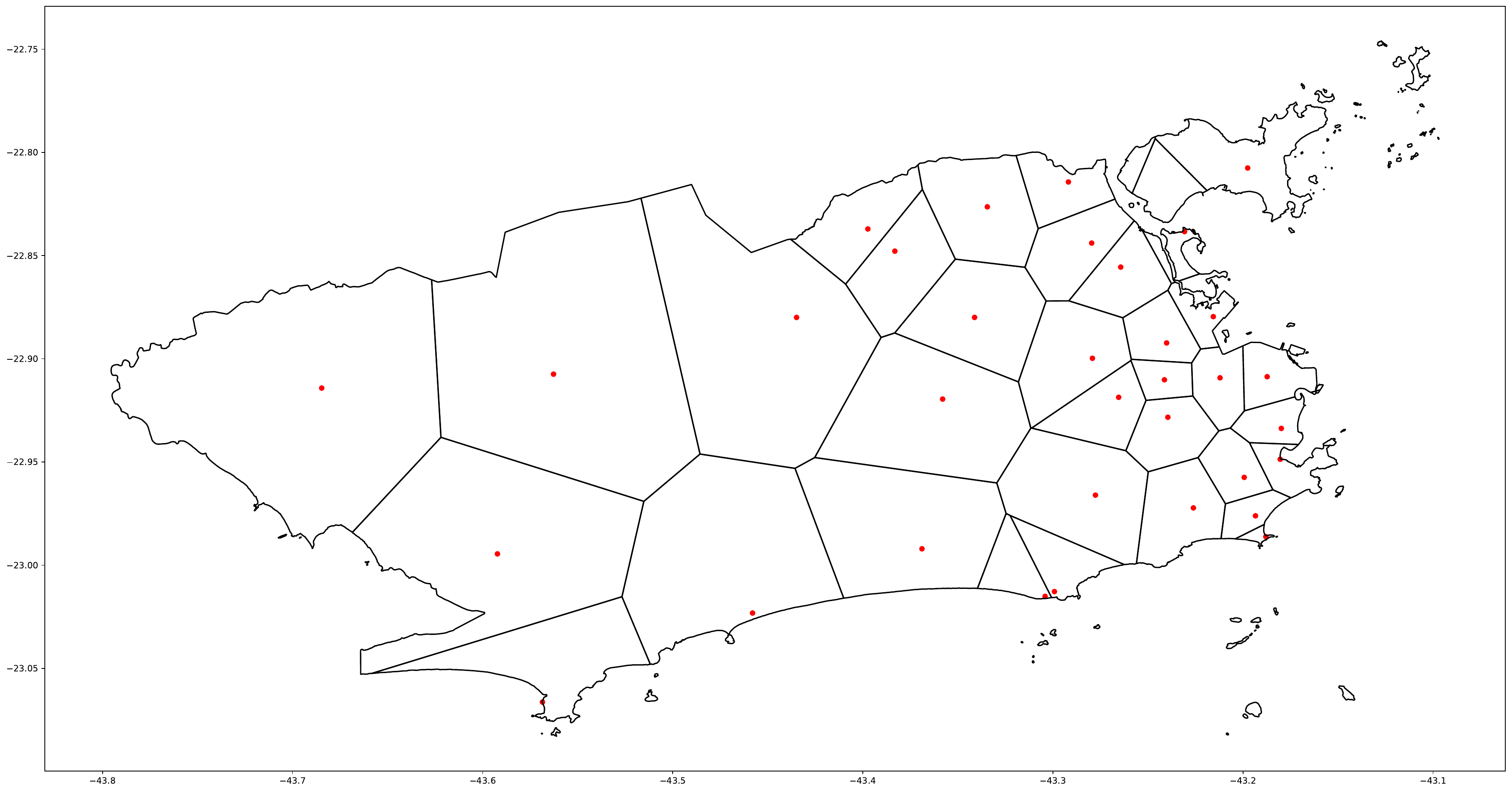}
\end{tabular}
\caption{\label{figurevoronoi} Discretization of a region containing the city of Rio de Janeiro into 34 subregions, given by the Voronoi diagram of ambulance stations in Rio de Janeiro.}
\end{figure}

LASPATED also provides space discretization with Voronoi diagrams.
Figure~\ref{figurevoronoi} obtained with LASPATED displays a discretization based on the Voronoi diagram given by the locations of ambulance stations in Rio de Janeiro.
Each subregion includes the set of points that are closest to a specific station.

\section{Additional Discretization Functionalities}
\label{sec:adddisc}

\begin{figure}
    \centering
    \includegraphics[width=\textwidth]{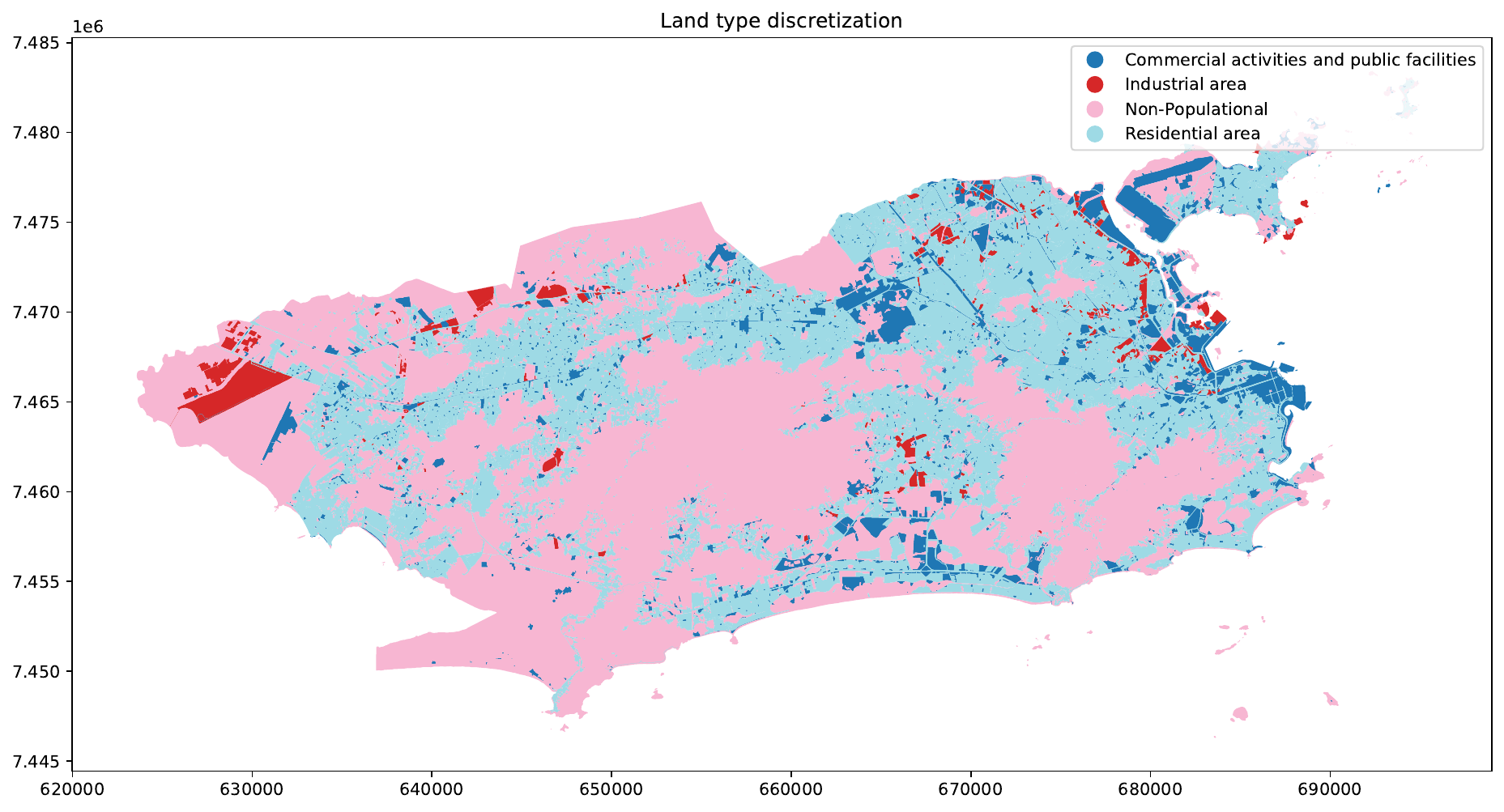}
    \caption{4 types of land use in Rio de Janeiro.}
    \label{fig:soil_use}
\end{figure}

Often different discretizations are used for different purposes.
For example, different spatial attribute data such as population count and land use type may be provided using different space discretizations.
Consider two discretizations $D_{1}$ and $D_{2}$.
Let $\mathcal{I}_{1}$ be the index set of the subregions of discretization $D_{1}$, and let $\mathcal{I}_{2}$ be the index set of the subregions of discretization $D_{2}$.
In this context, LASPATED provides the following functionalities:
\begin{itemize}
\item
Given subregion indices $i_{1} \in \mathcal{I}_{1}$ and $i_{2} \in \mathcal{I}_{2}$, LASPATED computes the area $\mathcal{A}(i_{1},i_{2})$ of the intersection of subregions $i_{1}$ and $i_{2}$.
As an example, Figure~\ref{fig:soil_use} displays a partition of Rio de Janeiro into 4 different types of land use.
LASPATED can be used to compute for every subregion of a given space discretization (such as rectangles or hexagons) the area of each land use type in the subregion.
\item
Consider a given attribute, such as population count, and assume that the attribute value $P_{i_{1}}$ in each subregion $i_{1} \in \mathcal{I}_{1}$ of discretization $D_{1}$ is uniformly distributed with density $d_{i_{1}} = P_{i_{1}} / \mathcal{A}(i_{1})$, where $\mathcal{A}(i_{1})$ is the area of the subregion~$i_{1}$.
We want to allocate this attribute value to the subregions of another discretization $D_{2}$.
LASPATED computes the value of the attribute allocated to subregion $i_{2} \in \mathcal{I}_{2}$ of discretization $D_{2}$ as $\displaystyle \sum_{i_{1} \in \mathcal{I}_{1}} d_{i_{1}} \mathcal{A}(i_{1},i_{2}) = \displaystyle \sum_{i_{1} \in \mathcal{I}_{1}} P_{i_{1}} \mathcal{A}(i_{1},i_{2}) / \mathcal{A}(i_{1})$.

\end{itemize}

\ignore{
\section{LASPATED Calibration Functions}
\label{sec:calib}

LASPATED provides \proglang{MATLAB} and \proglang{C++} functions for the calibration of models as in Section~\ref{sec:model1} for given values of weights $W_{G}$ and $w_{i,j}$, by solving
problem~\eqref{eqn:reformmodel1} using the projected gradient method with Armijo line search along a feasible direction or along the boundary.

Cross validation functions are also available for the model from section~\ref{sec:model1}. The cross validation allows the selection of weights $W_{G}$ and $w_{i,j}$ and the computation of the corresponding optimal intensities
$\lambda_{c.i.t}$ that solve problem~\eqref{eqn:reformmodel1}
for the selected weights.

LASPATED also provides \proglang{MATLAB} and \proglang{C++} functions for the
calibration of models with covariates, as in  Section~\ref{sec:modelcov}.
More precisely, it solves optimization problem~\eqref{optlikelihbeta} using projected gradient with Armijo line search along a feasible direction or along the boundary.

Our repository includes the examples for the calibration functions described in the next section.
}

\section{Numerical Examples}
\label{sec:examples}

We demonstrate the LASPATED calibration functions using several examples with simulated and real data.
In all examples, intensities of arrivals were calibrated using the projected gradient method with Armijo search along a feasible direction, as described in section~\ref{sec:opt_sol}. Experiments were performed using GCC version 11.4, and Python 3.10 in a AMD Ryzen 5 2600 processor with 20GB of RAM in an Ubuntu 22.04 OS.

\subsection{Numerical Examples with Artificial Data without Covariates}
\label{sec:simulated_example}

\begin{figure}
\centering
\begin{tabular}{c}
\includegraphics[scale=0.65]{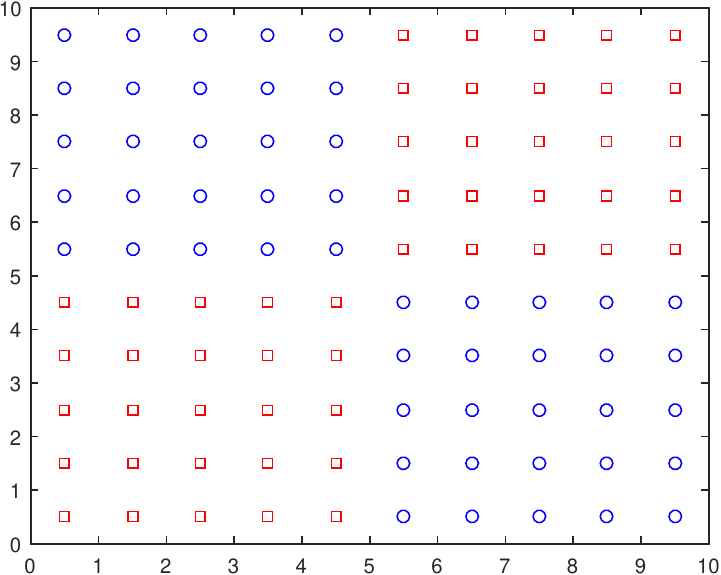}
\end{tabular}
\caption{\label{figureserr}
Example region $\mathcal{S} = [0,10]^2$.
The intensity function $\lambda$ is different in the blue and red subregions, but the user does not know about these subregions.
For estimation purposes, the user discretizes $\mathcal{S}$ into $10 \times 10 = 100$ square zones.
}
\end{figure}

In these examples, points arrive in $2$-dimensional space and over time according to a periodic non-homogeneous Poisson process.
There is only one type of point, thus $|\mathcal{C}| = 1$, and hence the type notation~$c$ is omitted.
The region under consideration is $\mathcal{S} = [0,10]^2$, as shown in Figure~\ref{figureserr}. 

The region $\mathcal{S}$ is partitioned into two subsets $\mathcal{S} = \mathcal{B} \cup \mathcal{R}$, with $\mathcal{B} = [0,5] \times (5,10] \cup (5,10] \times [0,5]$ and $\mathcal{R} = [0,5] \times [0,5] \cup (5,10] \times (5,10]$.

\subsubsection{Example~1}
\label{sec:example 1}

\begin{figure}
\centering
\resizebox{\textwidth}{!}{
\begin{tabular}{cc}
\includegraphics[scale=0.5]{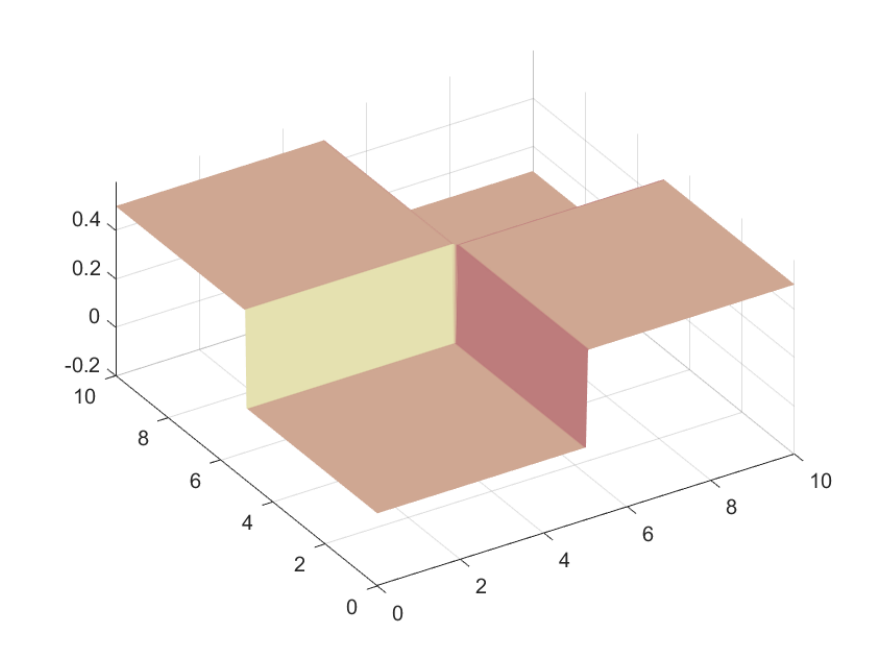} &
\includegraphics[scale=0.5]{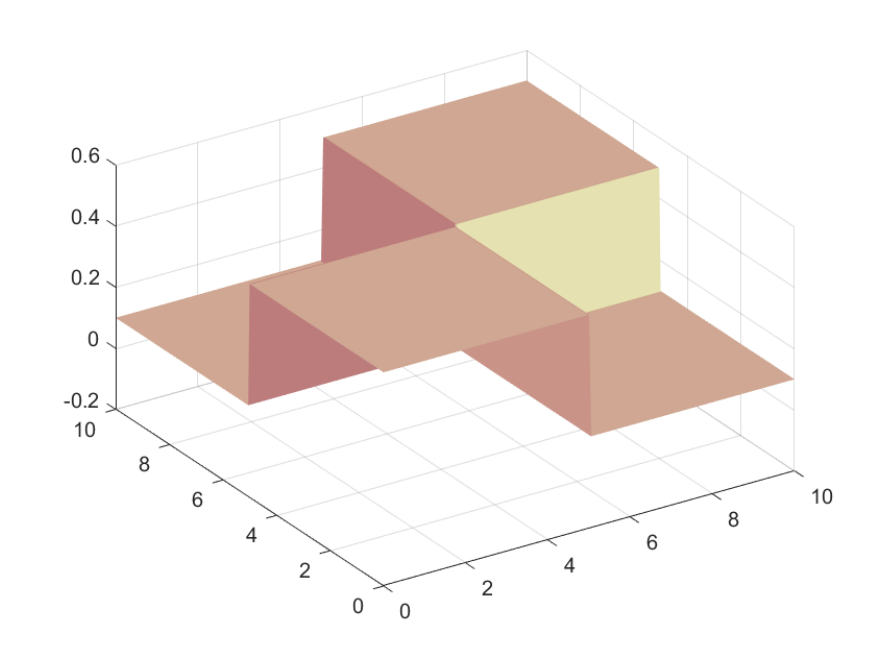} \\
$\hspace*{0.6cm}$(a) Time intervals that start at odd times & $\hspace*{0.6cm}$(b) Time intervals that start at even times
\end{tabular}}
\caption{\label{figuresex1}
Intensities as a function of location, for time intervals that start at odd times and even times.}
\end{figure}

In this example, the rate function $\lambda : \mathcal{S} \times [0,T] \mapsto \RR_{+}$ is different on $\mathcal{B}$ and on $\mathcal{R}$, and is periodic with period~$2$, as follows:
\[
\lambda(s,t) \ \ = \ \ \left\{\begin{array}{lcl}
0.1 & \mbox{if} & s \in \mathcal{B} \mbox{ and } t \in (2k,2k+1] \mbox{ for some } k \in \NN, \\
0.5 & \mbox{if} & s \in \mathcal{B} \mbox{ and } t \in (2k-1,2k] \mbox{ for some } k \in \NN, \\
0.5 & \mbox{if} & s \in \mathcal{R} \mbox{ and } t \in (2k,2k+1] \mbox{ for some } k \in \NN, \\
0.1 & \mbox{if} & s \in \mathcal{R} \mbox{ and } t \in (2k-1,2k] \mbox{ for some } k \in \NN.
\end{array}\right.
\]
This rate function is represented in Figure~\ref{figuresex1}(a) (for time intervals that start at odd times) and Figure~\ref{figuresex1}(b) (for time intervals that start at even times). 

The user knows the region $\mathcal{S}$ and that $|\mathcal{C}| = 1$, but does not know about the subregions $\mathcal{B}$ and $\mathcal{R}$ that affect the intensity function $\lambda$, and does not know that $\lambda$ is periodic with period~$2$.
For estimation purposes, the user discretizes $\mathcal{S}$ into $100$~square zones of unit area each, as shown in Figure~\ref{figureserr}.
Thus $\mathcal{I} = \mathcal{I}_{\mathcal{B}} \cup \mathcal{I}_{\mathcal{R}}$ indexes two types of zones, but this is not known by the user:
\begin{itemize}
\item
$\mathcal{I}_{\mathcal{B}}$ indexes $50$~blue zones (with blue circles in their centers in Figure~\ref{figureserr}); there are $25$~blue zones in the bottom right and $25$~blue zones in the upper left of the region;
\item
$\mathcal{I}_{\mathcal{R}}$ indexes $50$ red zones (with red squares in their centers in Figure~\ref{figureserr}); there are $25$~red zones on the bottom left and $25$~red zones in the upper right of the region.
\end{itemize}
Also, for estimation purposes, the user discretizes time into $28$~time intervals of length~$1$ each.
Thus, the estimates are denoted with $\lambda_{i,t}$ for $i \in \mathcal{I}$ and $t \in \mathcal{T} \defi \{1,\ldots,28\}$.
Given arrival data $M_{i,t,n}$ for $i \in \mathcal{I}$, $t \in \mathcal{T}$, and $n = 1,\ldots,N_{i,t}$, the regularized loss function in~\eqref{eqn:regularization1} is used to estimate $\lambda_{i,t}$.
The penalty coefficients are $w_{i,j} = w > 0$ when $i,j$ are neighboring zones, and $w_{i,j} = 0$ otherwise, where $w$ will be varied as described later.
Two zones are neighbors if their borders share an edge.

We consider two partitions $\mathcal{G}_{2}$ and $\mathcal{G}_{4}$ of time intervals $\mathcal{T}$ in~\eqref{eqn:regularization1}.
For partition $\mathcal{G}_{2}$, the time groups are $G_{0} = \{2k \, : \, k=1,\ldots,14\}$ and $G_{1} = \{2k+1 \, : \, k=0,1,\ldots,13\}$.
For partition $\mathcal{G}_{4}$, the time groups are $G_{0} = \{4k \, : \, k=1,\ldots,7\}$, $G_{1} = \{4k+1 \, : \, k=0,1,\ldots,6\}$, $G_{2} = \{4k+2 \, : \, k=0,1,\ldots,6\}$, and $G_{3} = \{4k+3 \, : \, k=0,1,\ldots,6\}$.
For both partitions, penalty coefficients $W_{G} = w$ for all groups~$G$ (note that all groups have the same cardinality so that it seems reasonable to choose the same weights for different groups).
Also, $\varepsilon = 0.001$ in~\eqref{eqn:Ceps}.

\begin{figure}
\centering
\resizebox{\textwidth}{!}{
\begin{tabular}{cc}
\includegraphics[scale=0.3]{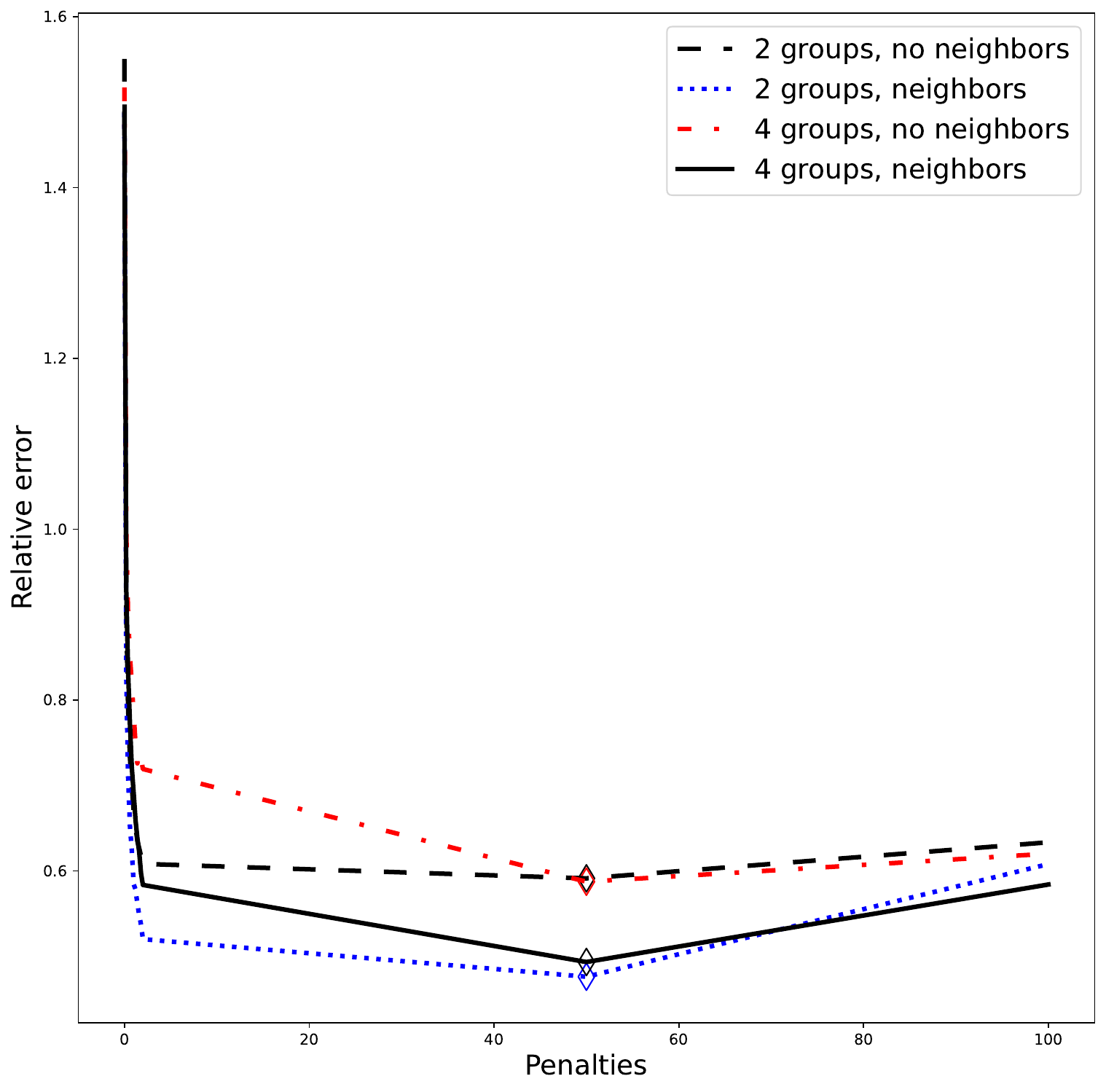} &
\includegraphics[scale=0.3]{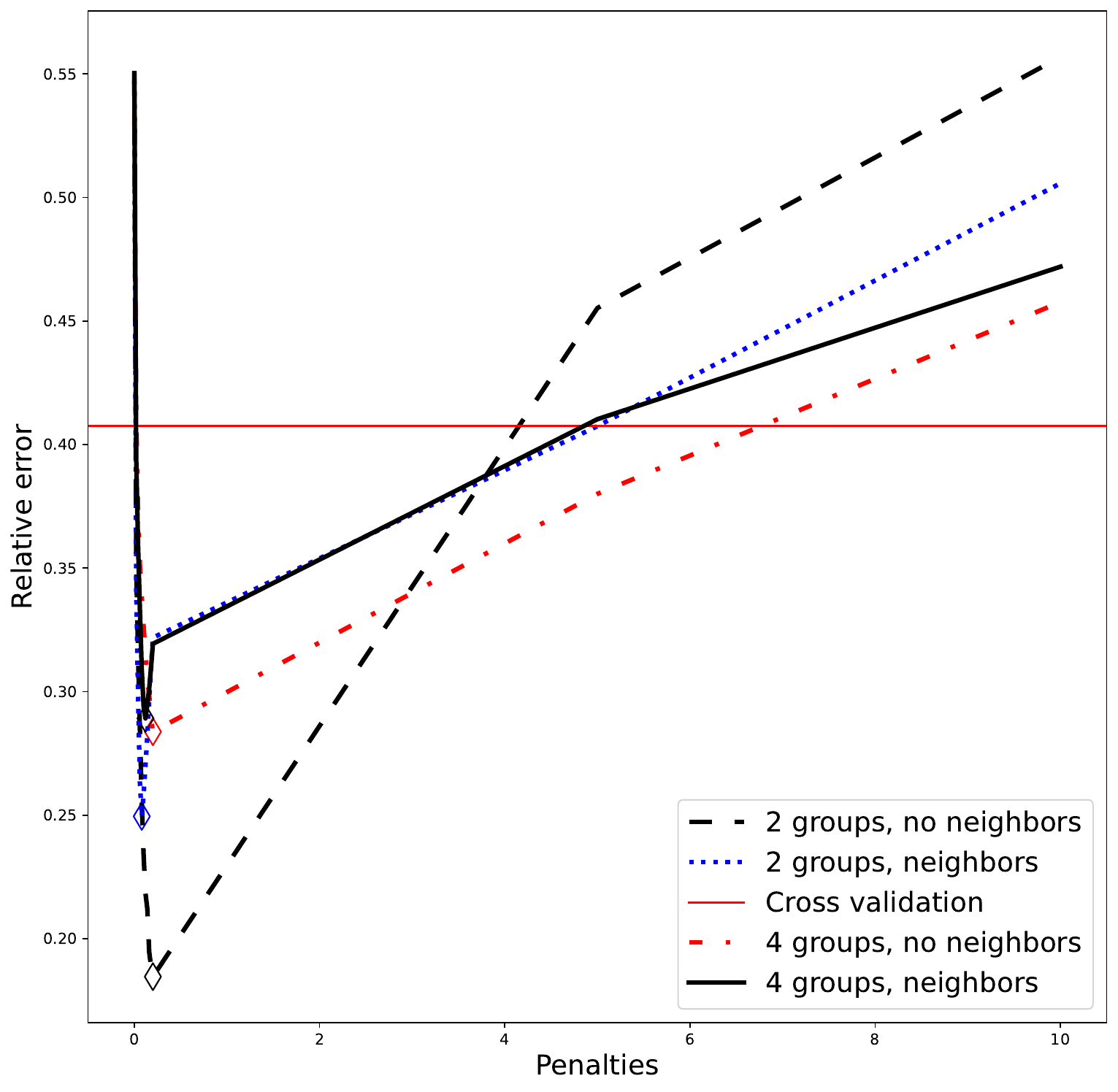} \\
$\hspace*{0.6cm}$ $N_{i,t} = 1$ & $\hspace*{0.6cm}$ $N_{i,t} = 10$
\vspace{5mm} \\
\includegraphics[scale=0.3]{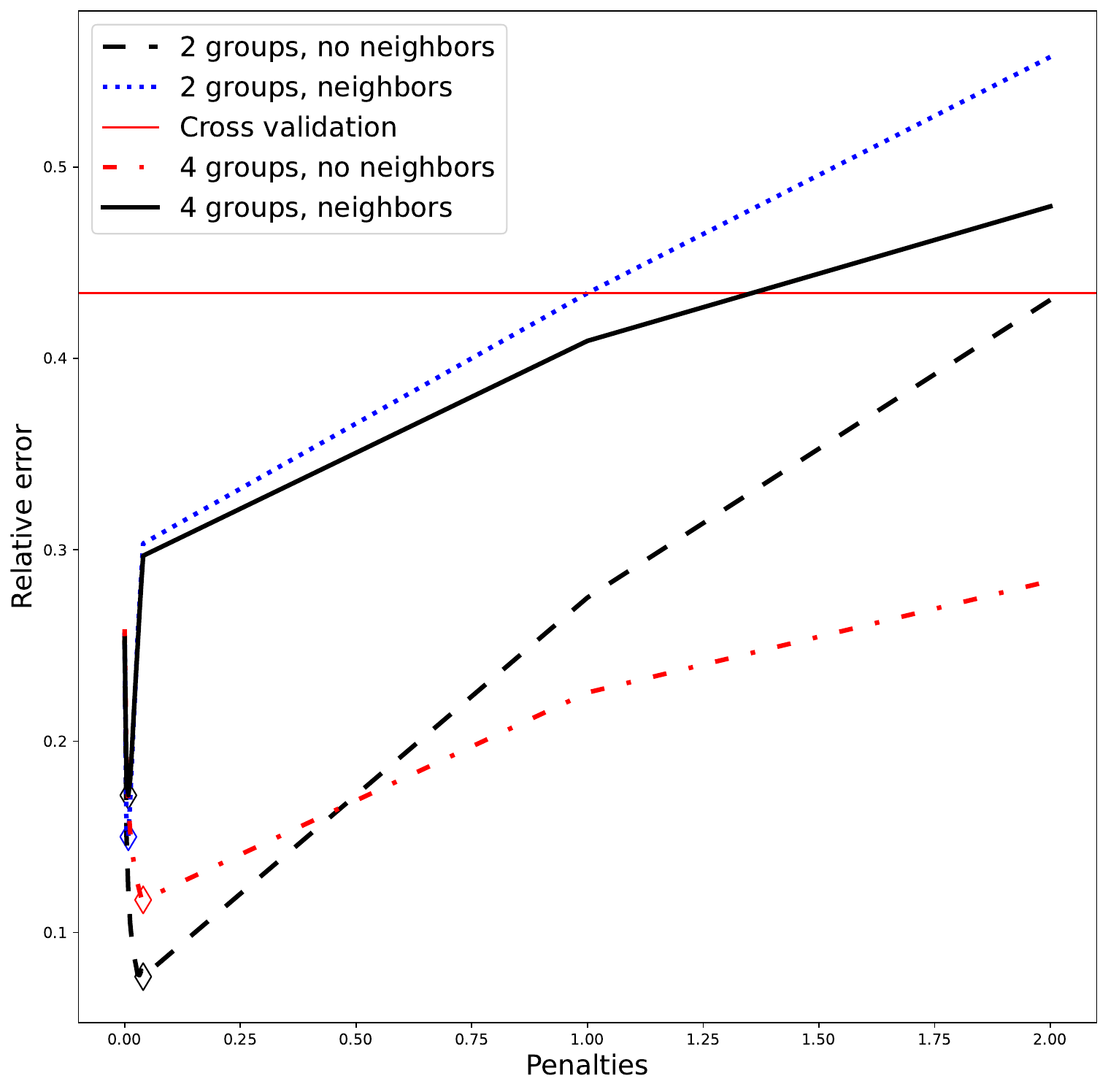} &
\includegraphics[scale=0.3]{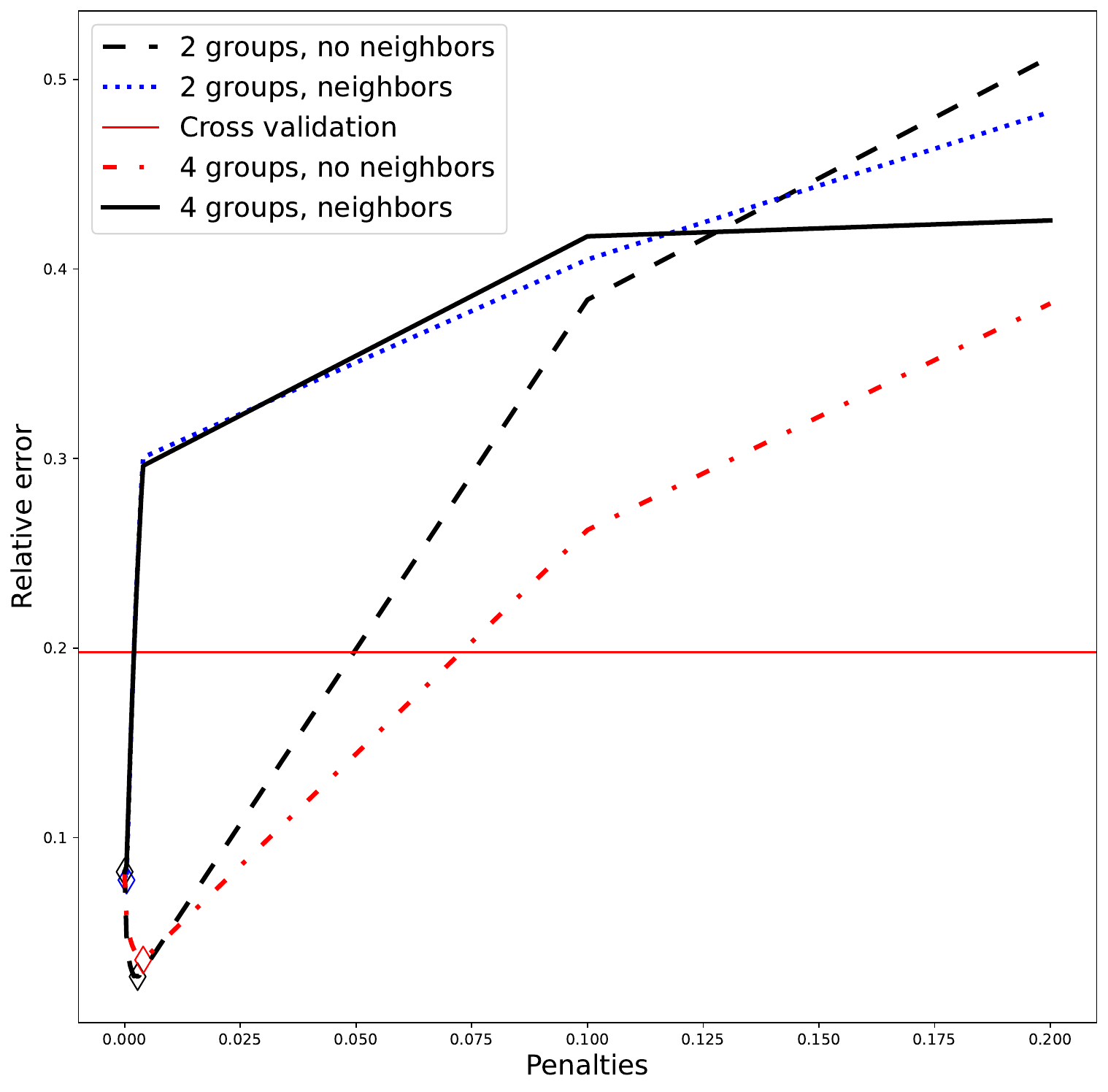}\\ 
$\hspace*{0.6cm}$ $N_{i,t} = 50$ \vspace{5mm} &
$\hspace*{0.6cm}$ $N_{i,t} = 500$ \vspace{5mm}
\end{tabular}}
\caption{\label{figureserr1}
Mean relative estimation errors $\frac{1}{|\mathcal{I}||\mathcal{T}|} \sum_{i \in \mathcal{I}} \sum_{t \in \mathcal{T}} \left|\frac{\lambda_{i,t} - \hat{\lambda}^{w}_{i,t}}{\lambda_{i,t}}\right|$ as a function of the penalty parameter~$w$, for four sample sizes and six estimators (including the empirical estimator at $w = 0$).
}
\end{figure}

For any value of the penalty parameter~$w$, let $\hat{\lambda}^{w}_{i,t}$ denote the estimator of $\lambda_{i,t}$ produced by~\eqref{eqn:model0}.
Note that when the penalty parameter $w = 0$, then $\hat{\lambda}^{0}_{i,t}$ reduces to the empirical estimator of the intensities, which is the mean rate of arrivals in zone~$i$ and time interval~$t$.
For a range of values of the penalty parameter~$w$, we computed the estimates $\hat{\lambda}^{w}_{i,t}$, and Figure~\ref{figureserr1} shows the mean (over all zones~$i$ and time intervals~$t$) relative error given by
\if{
$$\frac{100}{T|\mathcal{I}|}\sum_{t=1}^{T} \sum_{i \in \mathcal{I}} 
\left|\frac{ \lambda_{c,i,t} - {\hat{\lambda}}_{c,i,t}^{p}}{\lambda_{c,i,t}}\right|
$$
}\fi
\[
\frac{1}{|\mathcal{I}||\mathcal{T}|} \sum_{i \in \mathcal{I}} \sum_{t \in \mathcal{T}}
\left|\frac{\lambda_{i,t} - \hat{\lambda}^{w}_{i,t}}{\lambda_{i,t}}\right|
\]
as a function of penalty parameter~$w$, for different values of $N_{i,t}$.
Figure~\ref{figureserr1} shows the mean relative errors for the following four estimators: estimator with partition $\mathcal{G}_{4}$ of time intervals and with neighbor-based spatial regularization (legend ``4 groups, neighbors'' in the figure), estimator with partition $\mathcal{G}_{4}$ of time intervals and without spatial regularization (legend ``4 groups, no neighbors'' in the figure), estimator with partition $\mathcal{G}_{2}$ of time intervals and with neighbor-based spatial regularization (legend ``2 groups, neighbors'' in the figure), and estimator with partition $\mathcal{G}_{2}$ of time intervals and without spatial regularization (legend ``2 groups, no neighbors'' in the figure).
Figure~\ref{figureserr1} presents results for four values of the sample size: $N_{i,t} = 1$, $N_{i,t} = 10$, $N_{i,t} = 50$, and $N_{i,t} = 500$.
Figure~\ref{figureserr1} also shows, for the three larger sample sizes, the mean relative error obtained by choosing the penalty parameter~$w$ by cross validation as follows:
The data are partitioned into $5$ subsets.
Then, for each of $5$~replications we used one of the data subsets (a different subset for each replication, with 20\% of data used for training) to compute $\hat{\lambda}^{w}_{i,t}$ for a range of values of~$w$, and we used the remaining data to compute the out-of-sample likelihood value for each value of~$w$.
Then we determined the penalty parameter~$w^*$ that maximizes the average out-of-sample likelihood value over the $5$~replications.
Figure~\ref{figureserr1} shows the mean relative error for the resulting cross validation estimator with partition $\mathcal{G}_{2}$ of time intervals and with neighbor-based spatial regularization using the penalty parameter~$w^*$.
Note that the empirical estimator, with $w = 0$, does not depend on the partition of time intervals and the neighbor structure, and therefore is given by all four estimators described above at $w = 0$.
Table~\ref{tabnoreg} shows the minimum mean relative errors over different values of penalty parameter~$w$, as well as the values of~$w$ that attain the minimum, for each of the four estimators, plus the mean relative errors of the cross validation estimator (using the penalty parameter~$w^*$) and the empirical estimator ($w = 0$), for each of the four sample sizes considered in Figure~\ref{figureserr1}.

\begin{table}
\centering
\begin{tabular}{|c|c|c|c|c|c|c|}
\hline
Sample size $N_{i,t}$ & Reg 1 & Reg 2 & Reg 3 & Reg 4 & CV & Emp \\
\hline
1 & 0.45/50.0 & 0.62/50.0 & 0.42/50.0 & 0.58/2.0 & - & 1.54/0 \\ \hline
10 & 0.31/0.1 & 0.32/0.2 & 0.27/0.08 & 0.22/0.2 & 0.42/5.0 & 0.54/0 \\ \hline
50 & 0.17/0.008 & 0.12/0.04 & 0.15/0.008 & 0.08/0.032 & 0.44/1.0 & 0.25/0 \\ \hline
500 & 0.08/0.0 & 0.03/0.004 & 0.08/0.0004 & 0.02/0.004 & 0.20/0.002 & 0.08/0 \\ \hline
\end{tabular}
\caption{\label{tabnoreg}
The column headings denote the estimators as follows: Reg 1: estimator with partition $\mathcal{G}_{4}$ of time intervals and with neighbor-based spatial regularization; Reg 2: estimator with partition $\mathcal{G}_{4}$ of time intervals and without spatial regularization; Reg 3: estimator with partition $\mathcal{G}_{2}$ of time intervals and with neighbor-based spatial regularization; Reg 4: estimator with partition $\mathcal{G}_{2}$ of time intervals and without spatial regularization; CV: cross validation; Emp: empirical estimator.
For each sample size and for each estimator Reg 1, Reg 2, Reg 3, Reg 4, each cell of the table gives two numbers separated by a slash (/): the first number is the minimum mean relative error over different values of penalty parameter~$w$, and the second number is the value of~$w$ that attains the minimum.
For the cross validation estimator, the first number is the mean relative error using the penalty parameter~$w^*$, and the second number is the value of~$w^*$.
For the empirical estimator, the first number is the mean relative error using the penalty parameter~$w = 0$.
}
\end{table}

\if{

\begin{table}
\centering
\begin{tabular}{|c|c|c|c|c|c|c|}
\hline
Sample size $N_{i,t}$ & Reg 1 & Reg 2 & Reg 3 & Reg 4 & CV & Emp \\
\hline
14 & 0.30/10 & 0.46/10 & 0.35/5 & 0.26/0.01 & - & 1.19/0  \\  
\hline
140  & 0.21/1  & 0.22/3   &  0.22/1 &  0.26/1  &  0.21/1  & 0.39/0 \\
\hline
700 &  0.05/0.02  & 0.07/0.03  & 0.04/0.02 & 0.05/0.03  &  0.12/0.1  & 0.08/0.01 \\
\hline
7000 &  0.05/0 & 0.04/0.01 & 0.05/0 & 0.05/0.01 & 0.15/0.01 & 0.05/0 \\ 
\hline
\end{tabular}
\caption{\label{tabnoreg2} For each sample size and for each estimator, each cell of the table gives two numbers separated by a slash (/): the first number is the minimum mean relative error over different values of penalty parameter~$w$, and the second number is the value of~$w$ that attain the minimum.
The column headings denote the estimators as follows: Reg 1: estimator with partition $\mathcal{G}_{4}$ of time intervals and with neighbor-based spatial regularization; Reg 2: estimator with partition $\mathcal{G}_{4}$ of time intervals and without spatial regularization; Reg 3: estimator with partition $\mathcal{G}_{2}$ of time intervals and with neighbor-based spatial regularization; Reg 4: estimator with partition $\mathcal{G}_{2}$ of time intervals and without spatial regularization; CV: cross validation; Emp: empirical estimator.}
\end{table}

}\fi


In Figure~\ref{figureserr1}, note that the vertical and horizontal scales of the sub-figures are different.
As expected, the estimation error is decreasing in the sample size $N_{i,t}$.
Note that the mean relative errors shown in Figure~\ref{figureserr1}, and the values of the penalty parameter~$w$ with the smallest mean relative errors (shown with diamonds), were based on using the correct values $\lambda_{i,t}$ in the calculations, which are not known by the user.
Even when regularization is not based on the correct subregions and partition of time intervals, it can result in better estimates than the empirical estimates (corresponding to $w = 0$), as long as reasonable values are chosen for~$w$.
\if{
As explained before, we also used cross validation to determine good values of the penalty parameter~$w$ (without knowing the correct values $\lambda_{i,t}$), resulting in mean relative errors for most experiments much smaller than the mean relative error of the empirical estimator.
}\fi
The results indicate that the improvement of the estimates using regularization is more pronounced when the number of observations is small: for $N_{i,t} = 1$, the estimation error decreases from about 149\% (for the empirical estimator) to about 48\% for the best regularized estimator, and for $N_{i,t} = 10$, the estimation error decreases from about 55\% to about 18\%.
Therefore, when a small amount of data are available, the regularized estimator results in a smaller estimation error, even if the regularization is not based on the correct model structure, as long as relevant information and reasonable weights are used for the regularization.

\begin{figure}
\centering
\resizebox{\textwidth}{!}{
\begin{tabular}{cc}
\includegraphics[scale=0.3]{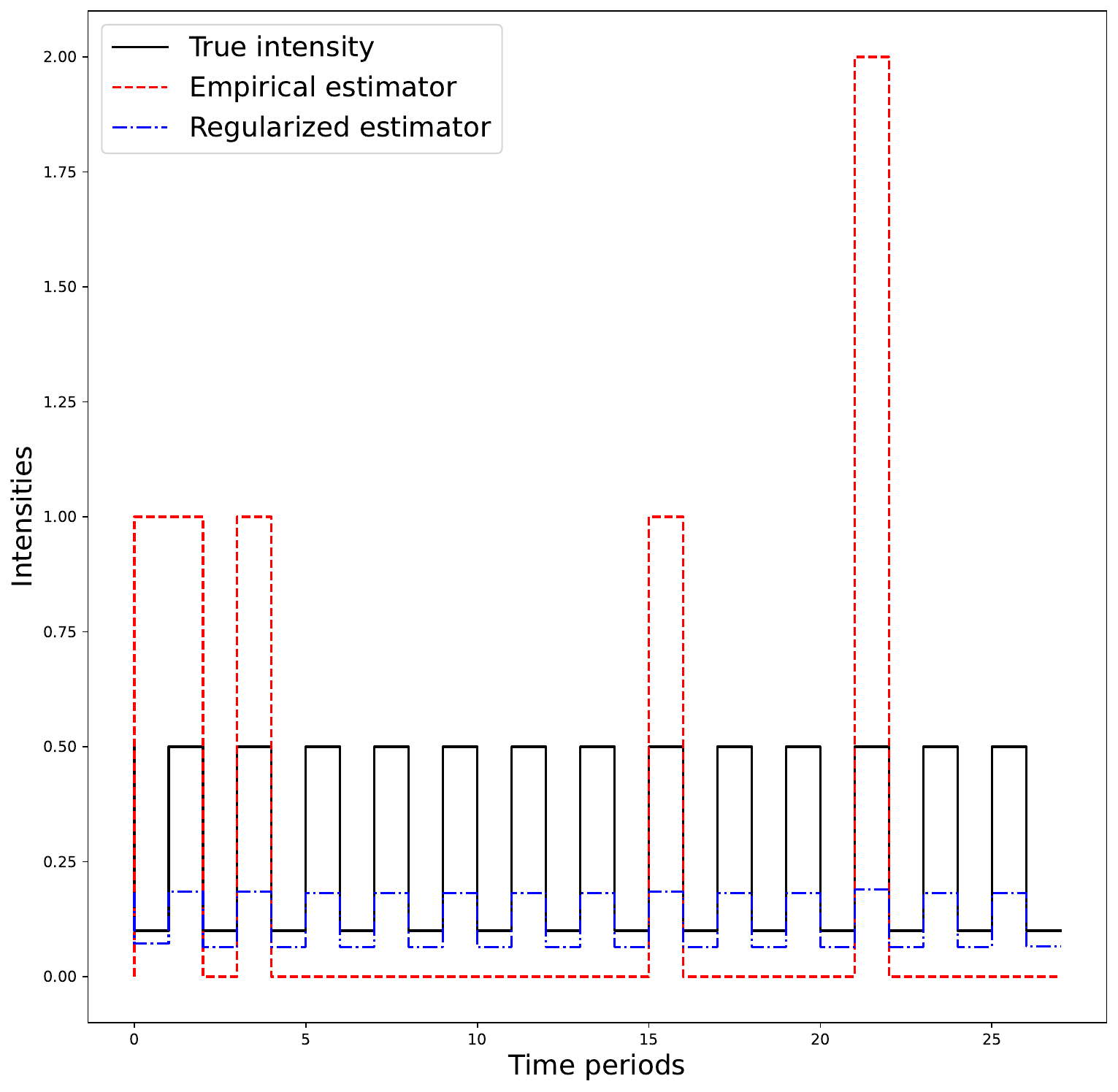} &
\includegraphics[scale=0.3]{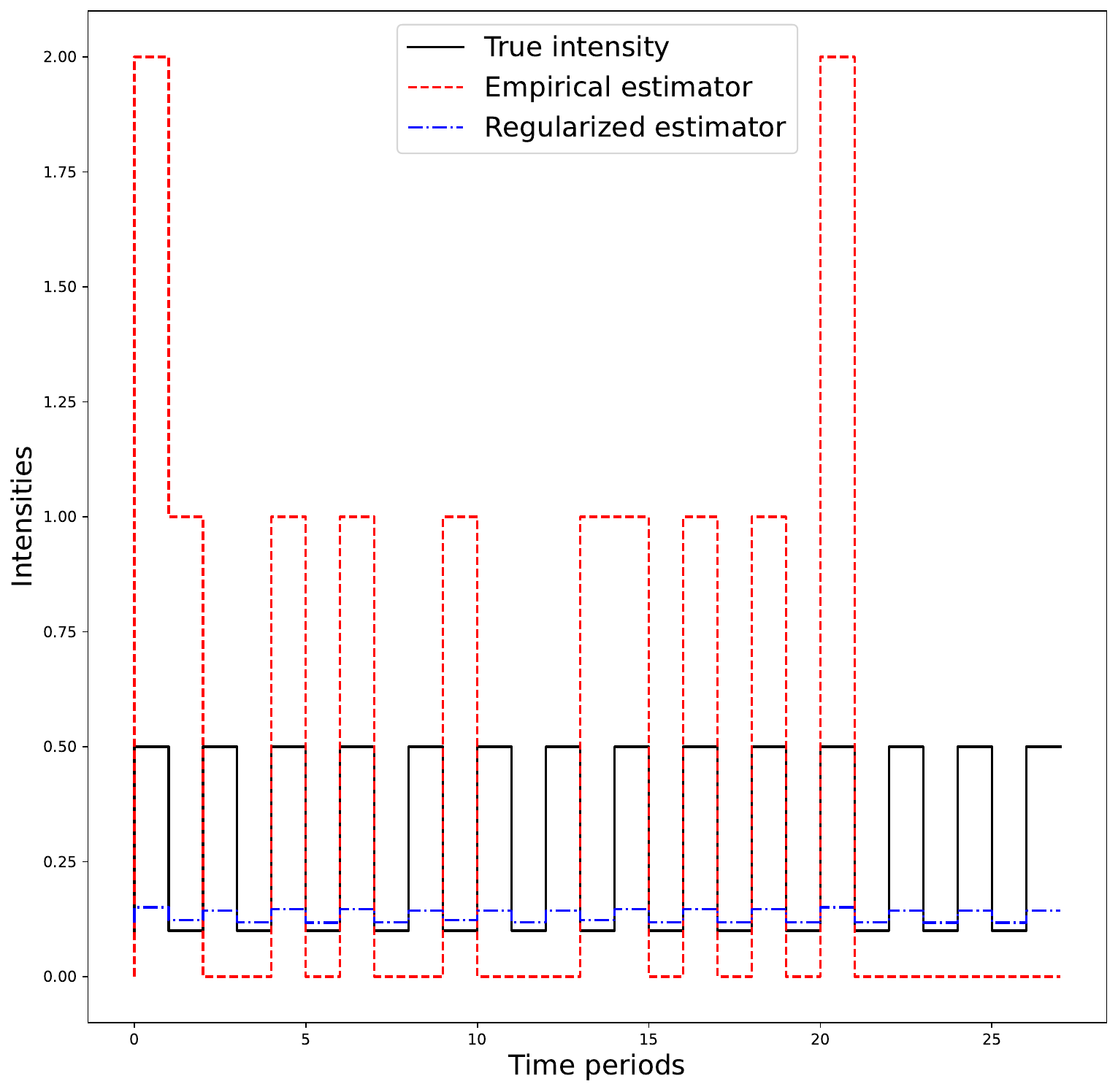} \\
$\hspace*{0.6cm}$ $N_{i,t} = 1$, zone $1 \in \mathcal{I}_{\mathcal{R}}$ & $\hspace*{0.6cm}$ $N_{i,t} = 1
$, zone $6 \in \mathcal{I}_{\mathcal{B}}$
\vspace{5mm} \\
\includegraphics[scale=0.3]{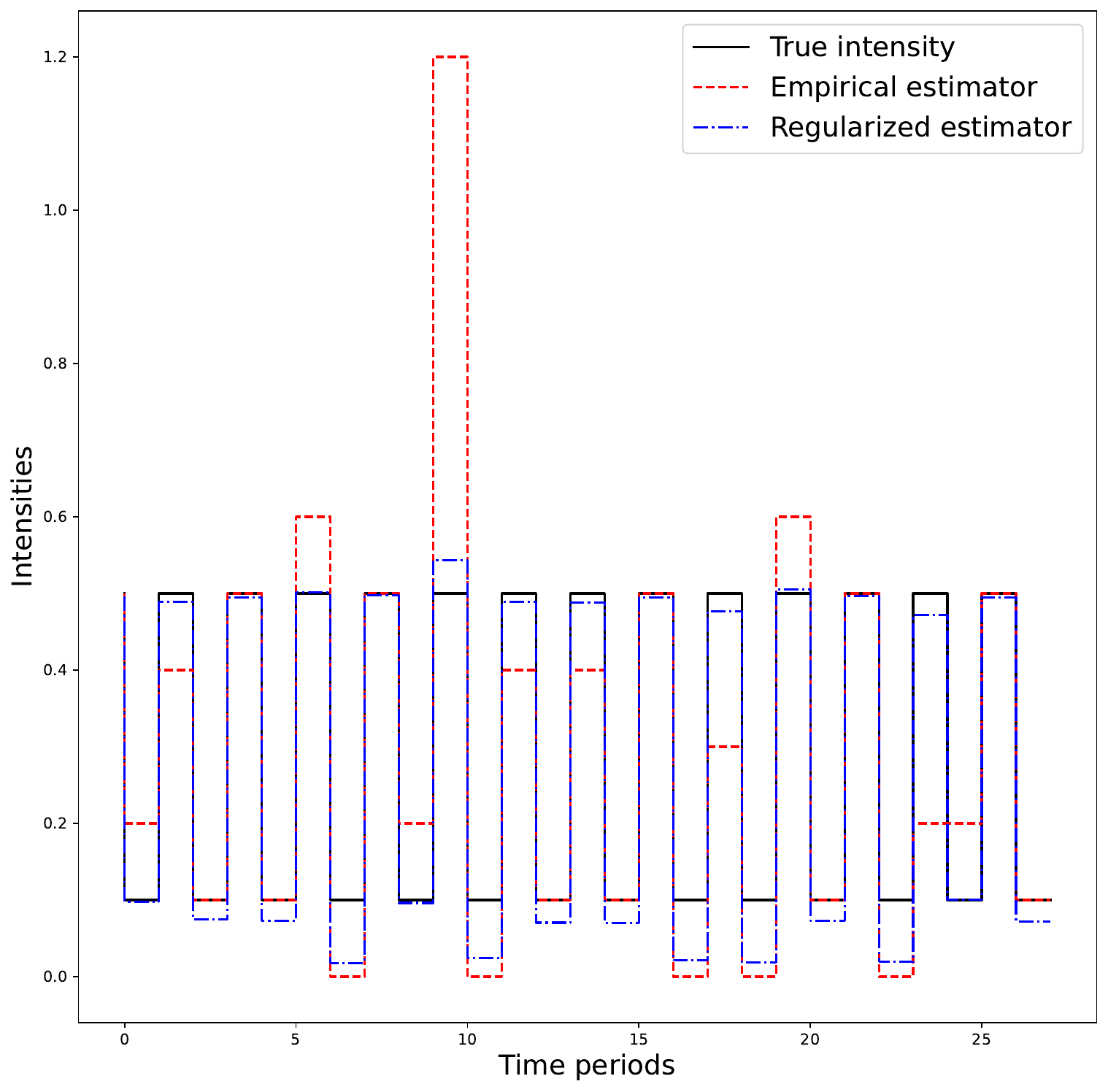}&
\includegraphics[scale=0.3]{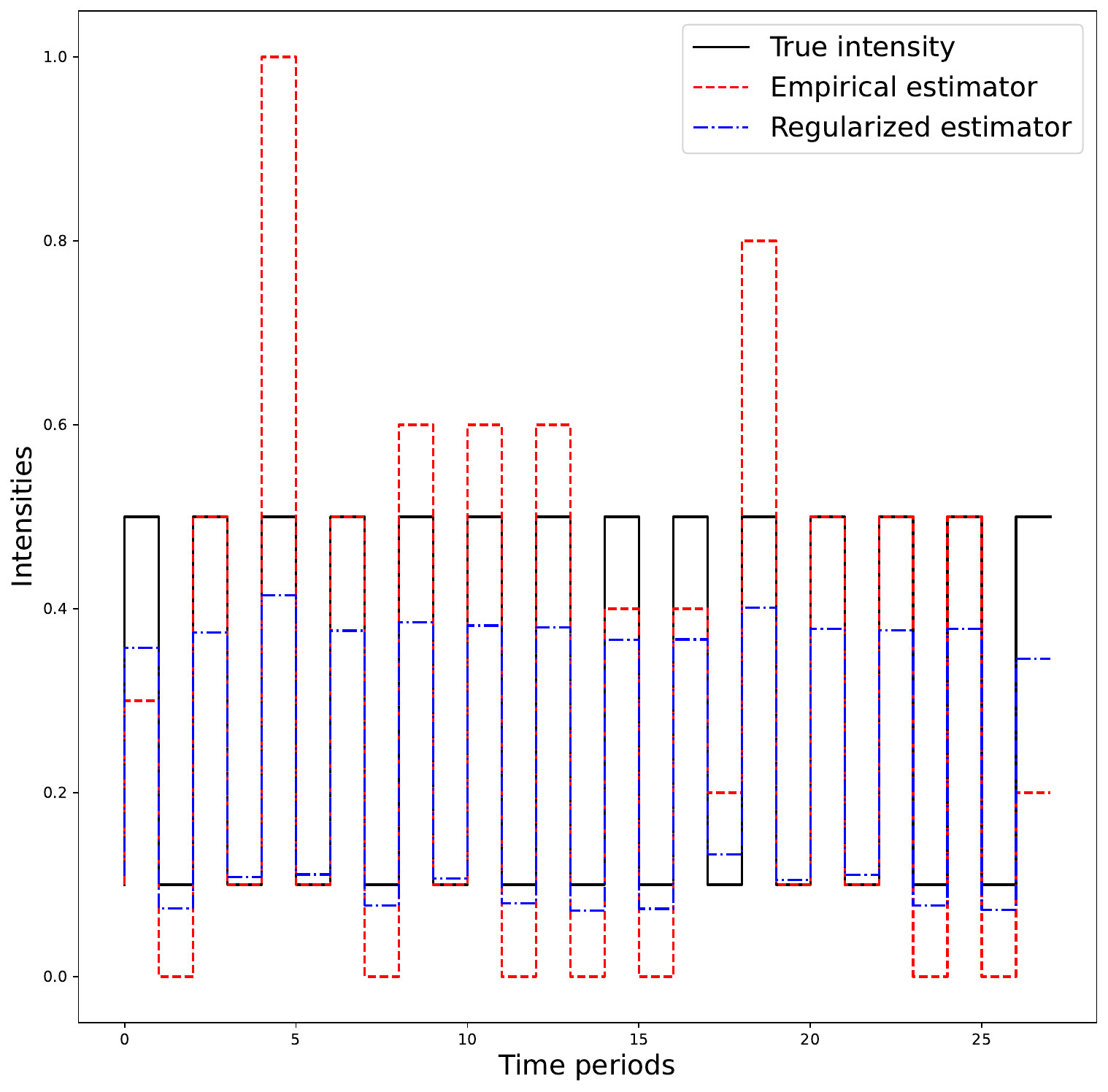}\\
$\hspace*{0.6cm}$ $N_{i,t} = 10$, zone $1 \in \mathcal{I}_{\mathcal{R}}$ & $\hspace*{0.6cm}$ $N_{i,t} = 10$, zone $6 \in \mathcal{I}_{\mathcal{B}}$
\vspace{5mm}
\end{tabular}}
\caption{\label{figureserrth}
Comparison of true, empirical, and regularized estimates of the Poisson process intensities.
True intensities as shown in Figure~\ref{figuresex1}.}
\end{figure}

Figure~\ref{figureserrth} shows the true intensities $\lambda(s,t)$ and the estimates $\hat{\lambda}^{w^*}_{i,t}$ obtained with partition $\mathcal{G}_{2}$ of time intervals and with neighbor-based spatial regularization using the cross validation penalty parameters~$w^*$, for $s \in \mathcal{R}$, zone $1 \in \mathcal{I}_{\mathcal{R}}$, $N_{i,t} = 1$ (top left plot), for $s \in \mathcal{R}$, zone $1 \in \mathcal{I}_{\mathcal{R}}$, $N_{i,t} = 10$ (bottom left plot), for $s \in \mathcal{B}$, zone $6 \in \mathcal{I}_{\mathcal{B}}$, $N_{i,t} = 1$ (top right plot), and for $s \in \mathcal{B}$, zone $6 \in \mathcal{I}_{\mathcal{B}}$, $N_{i,t} = 10$ (bottom right plot).
It can be seen that if little data are available, then the regularized estimates are closer to the true intensities than the empirical estimates.

\subsubsection{Example~2}
\label{sec:example 2}

\begin{figure}
\centering
\resizebox{\textwidth}{!}{
\begin{tabular}{cc}
\includegraphics[scale=0.6]{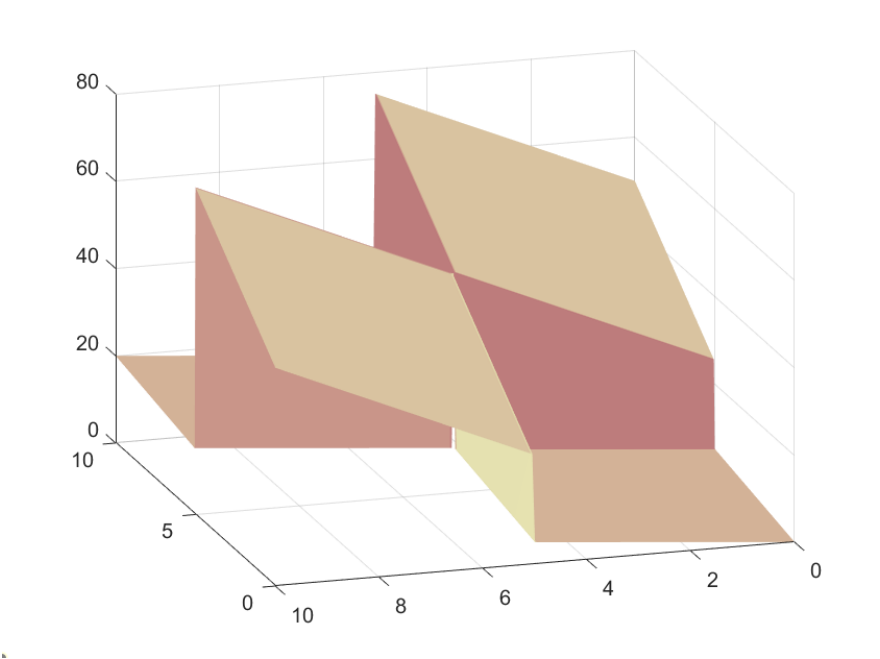}&
\includegraphics[scale=0.6]{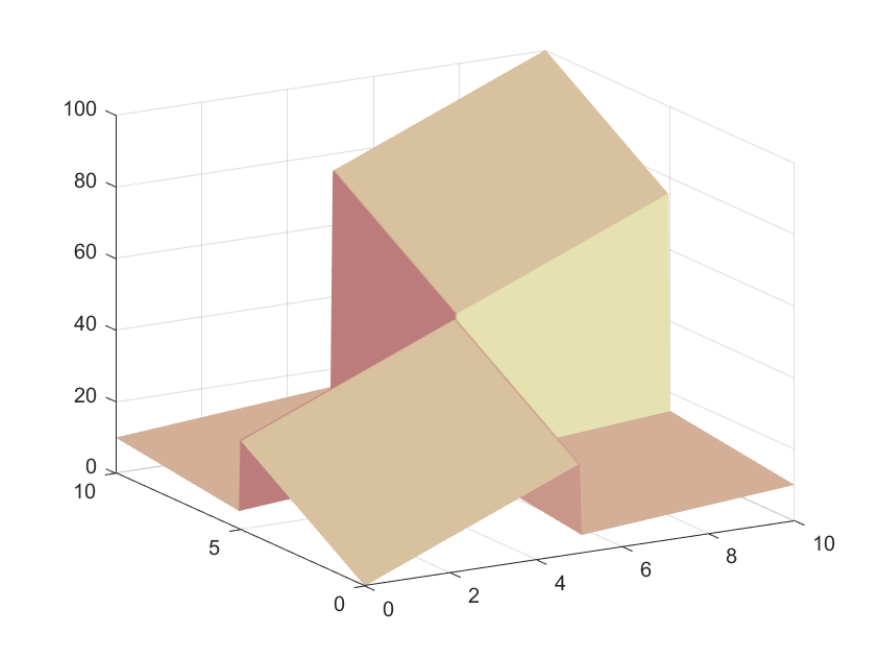}\\
$\hspace*{0.6cm}$  (c) Time intervals that start at odd times& $\hspace*{0.6cm}$(d) Time intervals that start at even times
\end{tabular}}
\caption{\label{figuresex2}
Intensities as a function of location, for time intervals that start at odd times and even times.}
\end{figure}

In this example, we consider the same region $\mathcal{S}$ and the same subregions $\mathcal{B}$ and $\mathcal{R}$ as in Example~1.
The intensity function $\lambda$ is also periodic, but not piecewise constant, as follows:
\[
\lambda(x,y,t) \ \ = \ \ \left\{
\begin{array}{lcl}
x + y & \mbox{if} & (x,y) \in \mathcal{B} \mbox{ and } t \in (2k,2k+1] \mbox{ for some } k \in \mathbb{N}, \\
5 (x + y) & \mbox{if} & (x,y) \in \mathcal{B} \mbox{ and } t \in (2k-1,2k] \mbox{ for some } k \in \mathbb{N}, \\
5 (x + y) & \mbox{if} & (x,y) \in \mathcal{R} \mbox{ and } t \in (2k,2k+1] \mbox{ for some } k \in \mathbb{N}, \\
x + y & \mbox{if} & (x,y) \in \mathcal{R} \mbox{ and } t \in (2k-1,2k] \mbox{ for some } k \in \mathbb{N}.
\end{array}\right.
\]
This rate function is represented in Figure~\ref{figuresex2}(c) (for time intervals that start at odd times) and Figure~\ref{figuresex2}(d) (for time intervals that start at even times). 

As in Example~1, the user knows the region $\mathcal{S}$ and that $|\mathcal{C}| = 1$, but does not know about the subregions $\mathcal{B}$ and $\mathcal{R}$ that affect the intensity function $\lambda$, and does not know that $\lambda$ is periodic with period~$2$.
For estimation purposes, the user discretizes $\mathcal{S}$ into $100$~square zones of unit area each.
\if{

\noindent\rule[0.5ex]{1\columnwidth}{1pt}
{\textbf{Simulation of red and blue distributions for $t \in (2k-1,2k]$}}\\
    \noindent\rule[0.5ex]{1\columnwidth}{1pt}
$S_{\mathcal{B}}=\{\}, S_{\mathcal{R}}=\{\}$.\\
Generate $N_{t, \mathcal{B}}$ from the Poisson$(\Lambda_{t, \mathcal{B}})$ distribution. \\
Generate $N_{t, \mathcal{R}}$ from the Poisson$(\Lambda_{t, \mathcal{R}})$ distribution. \\
{\textbf{For }} $i = 1,\ldots,N_{t, \mathcal{B}}$ \\
\hspace*{0.8cm}ContBlue=1;\\
\hspace*{0.8cm}{\textbf{While }}ContBlue\\
\hspace*{2cm}Generate $(x,y)$ uniformly in $\mathcal{B}$. \\
\hspace*{2cm}Generate $U$ uniformly in $(0,\overline{\lambda}_{t, \mathcal{B}})$.\\
\hspace*{2cm}{\textbf{If }}$U \le \lambda(x,y,t)$, then\\ 
\hspace*{2.2cm}$S_{\mathcal{B}}=S_{\mathcal{B}} \cup \{(x,y)\}$, contBlue=0.\\
\hspace*{2cm}{\textbf{End If}}\\
\hspace*{0.8cm}{\textbf{End While}}\\
{\textbf{End For}}\\
{\textbf{For }} $i = 1,\ldots,N_{t, \mathcal{R}}$ \\
\hspace*{0.8cm}ContRed=1;\\
\hspace*{0.8cm}{\textbf{While }}ContRed\\
\hspace*{2cm}Generate $(x,y)$ uniformly in $\mathcal{R}$. \\
\hspace*{2cm}Generate $U$ uniformly in $(0,\overline{\lambda}_{t, \mathcal{R}})$.\\
\hspace*{2cm}{\textbf{If }}$U \le \lambda(x,y,t)$, then\\ 
\hspace*{2.2cm}$S_{\mathcal{R}}=S_{\mathcal{R}} \cup \{(x,y)\}$, contRed=0.\\
\hspace*{2cm}{\textbf{End If}}\\
\hspace*{0.8cm}{\textbf{End While}}\\
{\textbf{End For}}\\
\noindent\rule[0.5ex]{1\columnwidth}{1pt}

}\fi
The user computes the estimators $\hat{\lambda}^{w}_{i,t}$ for each $i \in \mathcal{I}$ and each $t \in \mathcal{T}$, by solving optimization problem~\eqref{eqn:reformmodel1} using regularization with penalty parameter~$w$ and $\varepsilon = 0.001$.
Each estimator $\hat{\lambda}^{w}_{i,t}$ now approximates the rate $\int_{R_{i}} \lambda(x,y,t) \diff(x,y)$, where $R_{i}$ denotes the square of unit area for the zone~$i$.
For example, for a zone~$i$ with $R_{i} = (a_{i} - 1, a_{i}] \times (b_{j} - 1, b_{j}]$ and $\lambda(x,y,t) = x+y$ for $(x,y) \in R_{i}$, it holds that
\[
\int_{R_{i}} \lambda(x,y,t) \diff(x,y)
\ \ = \ \ \int_{a_{i} - 1}^{a_{i}} \int_{b_{j} - 1}^{b_{j}} (x+y) \diff y \diff x
\ \ = \ \ a_{i} + b_{j} - 1
\ \ = \ \ \lambda(x_{i},y_{i},t)
\ \ = \ \ x_{i} + y_{i}
\]
where $(x_{i},y_{i}) = (a_{i} - 0.5, b_{j} - 0.5)$ is the centroid of the zone~$i$.

The mean relative error of estimator $\hat{\lambda}^{w}_{i,t}$ is now given by
\[
\frac{1}{|\mathcal{I}| |\mathcal{T}|} \sum_{i \in \mathcal{I}} \sum_{t \in \mathcal{T}} \frac{\left|\displaystyle \int_{R_{i}} \lambda(x,y,t) \diff(x,y) - \hat{\lambda}^{w}_{i,t}\right|}{\displaystyle \int_{R_{i}} \lambda(x,y,t) \diff(x,y)}
\ \ = \ \ \frac{1}{|\mathcal{I}| |\mathcal{T}|} \sum_{i \in \mathcal{I}} \sum_{t \in \mathcal{T}} \frac{\left|\lambda(x_{i},y_{i},t) - \hat{\lambda}^{w}_{i,t}\right|}{\lambda(x_{i},y_{i},t)}.
\]
Parameter estimates $\hat{\lambda}^{w}_{i,t}$ were computed for a range of values of the penalty parameter~$w$, for the same four estimators as in Example~1: estimator with partition $\mathcal{G}_{4}$ of time intervals and with neighbor-based spatial regularization (legend ``4 groups, neighbors'' in Figure~\ref{figureserr2pl}), estimator with partition $\mathcal{G}_{4}$ of time intervals and without spatial regularization (legend ``4 groups, no neighbors'' in Figure~\ref{figureserr2pl}), estimator with partition $\mathcal{G}_{2}$ of time intervals and with neighbor-based spatial regularization (legend ``2 groups, neighbors'' in Figure~\ref{figureserr2pl}), and estimator with partition $\mathcal{G}_{2}$ of time intervals and without spatial regularization (legend ``2 groups, no neighbors'' in Figure \ref{figureserr2pl}).
As before, four values were used for the sample size: $N_{i,t} = 1$, $N_{i,t} = 10$, $N_{i,t} = 50$, and $N_{i,t} = 500$.
Figure~\ref{figureserr2pl} shows the mean relative error for these four estimators, as well as the estimator with partition $\mathcal{G}_{2}$ of time intervals and with neighbor-based spatial regularization using the penalty parameter~$w^*$ chosen with cross validation.
As before, the empirical estimator corresponds to~$w = 0$.
In this example intensities are very different between neighboring zones, and as a result space regularization is not helpful, but if little data are available then time regularization still provides better estimates than the empirical estimator.



\begin{figure}
\centering
\resizebox{\textwidth}{!}{
\begin{tabular}{cc}
\includegraphics[scale=0.3]{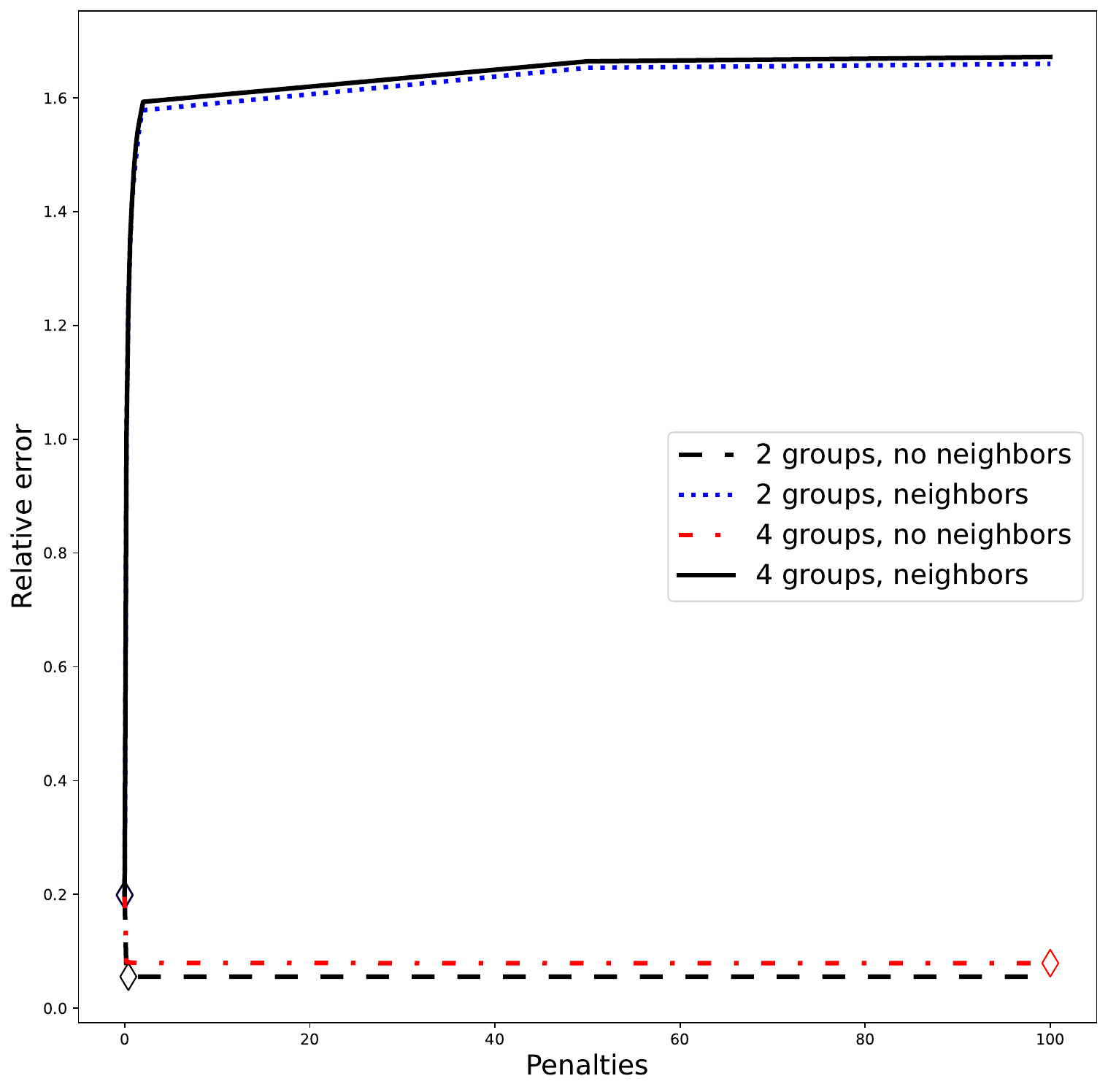} &
\includegraphics[scale=0.3]{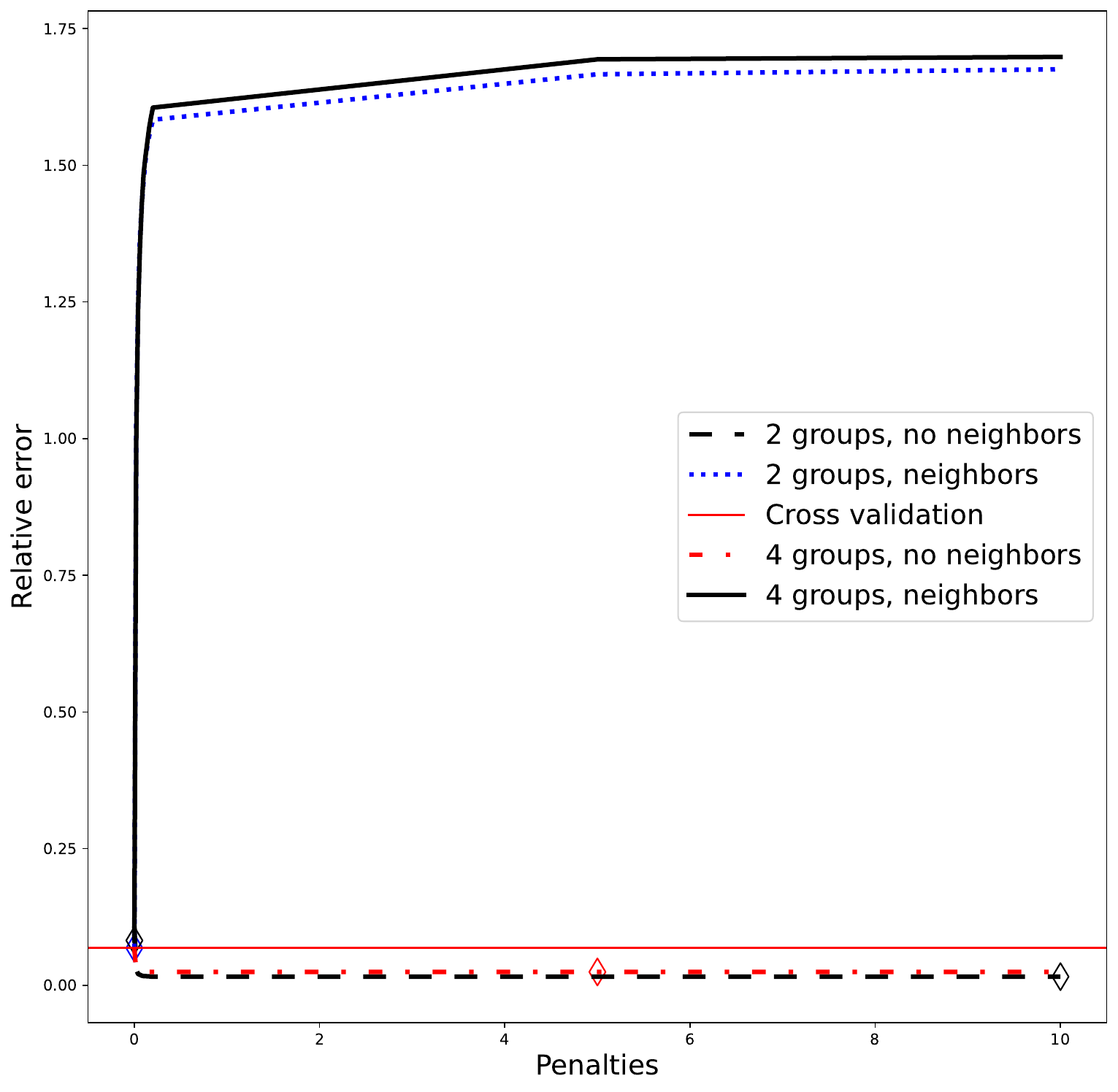} \\
$\hspace*{0.6cm}$ $N_{i,t} = 1$ & $\hspace*{0.6cm}$ $N_{i,t} = 10$
\vspace{5mm} \\
\includegraphics[scale=0.3]{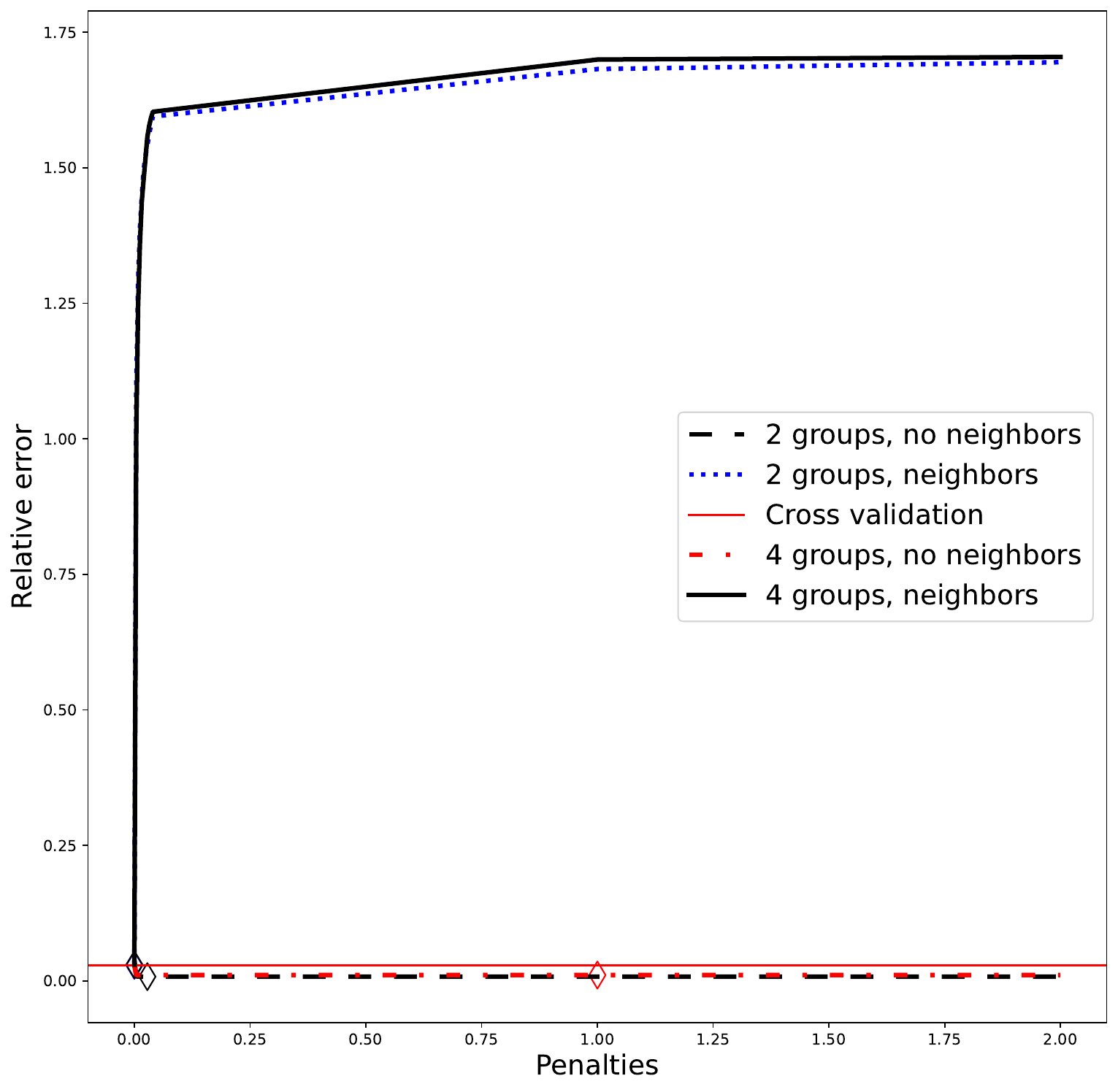} &
\includegraphics[scale=0.3]{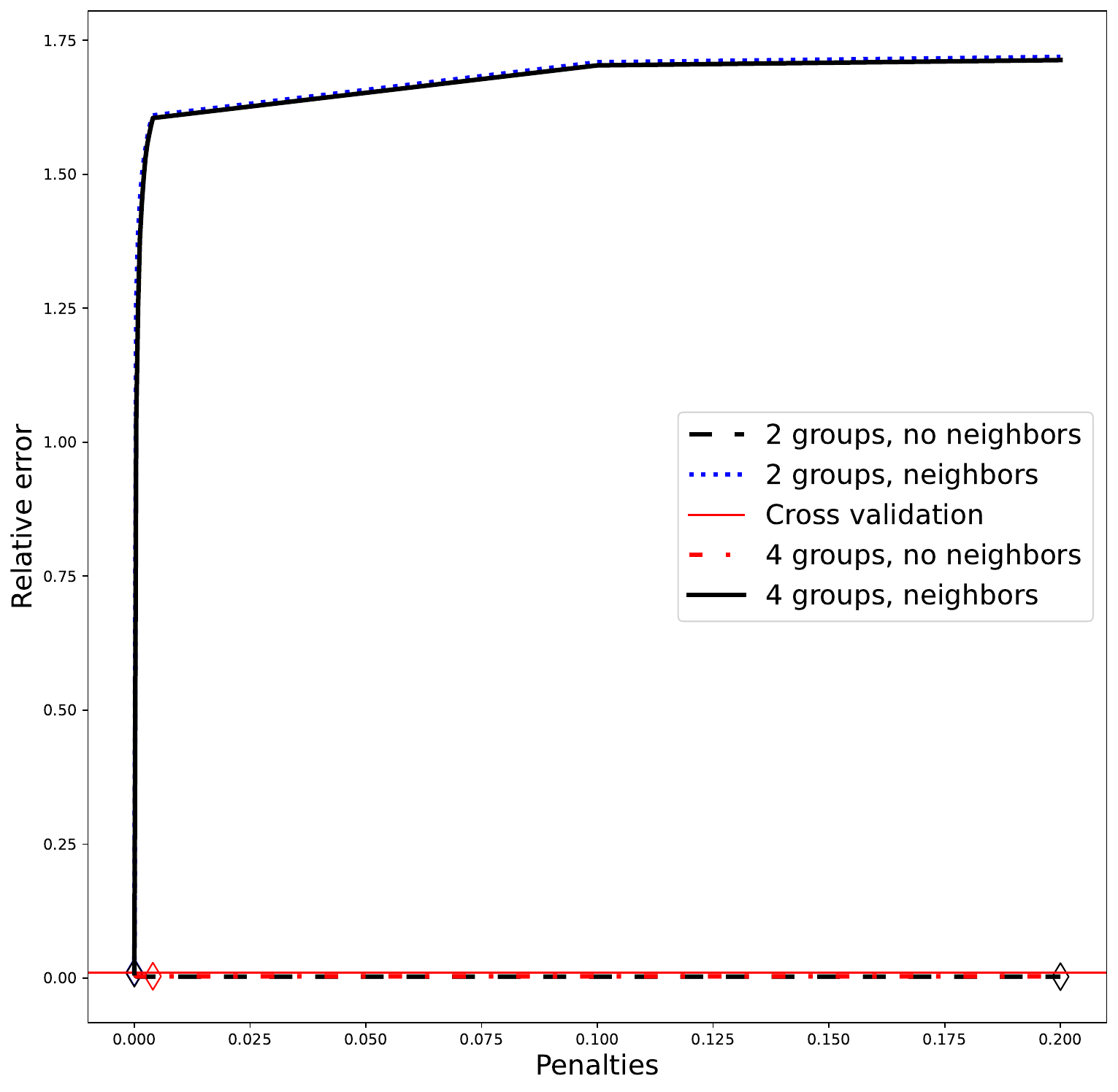} \\
$\hspace*{0.6cm}$ $N_{i,t} = 50$ & $\hspace*{0.6cm}$ $N_{i,t} = 500$
\end{tabular}}
\caption{\label{figureserr2pl}
Mean relative errors $\displaystyle \frac{1}{|\mathcal{I}| |\mathcal{T}|} \sum_{t \in \mathcal{T}} \sum_{i \in \mathcal{I}} \frac{\left|\int_{R_i} \lambda(x,y,t) \diff(x,y) - \hat{\lambda}_{t,i}^{w}\right|}{\int_{R_i} \lambda(x,y,t) \diff(x,y)}$ as a function of the penalty parameter~$w$. Intensities $\lambda(x,y,t)$ are as shown in Figure~\ref{figuresex2}.
}
\end{figure}


\if{
\begin{figure}
\centering
\begin{tabular}{cc}
\includegraphics[scale=0.3]{results_cov_n2_crop.pdf} &
\includegraphics[scale=0.3]{results_cov_n20_crop.pdf} \\
$\hspace*{0.6cm}$ $N_{i,t} = 14$ periods observed & $\hspace*{0.6cm}$ $N_{i,t} = 140$ periods observed
\vspace{5mm} \\
\includegraphics[scale=0.3]{results_cov_n100_crop.pdf} &
\includegraphics[scale=0.3]{results_cov_n1000_crop.pdf} \\
$\hspace*{0.6cm}$ $N_{i,t} = 700$ periods observed & $\hspace*{0.6cm}$ $N_{i,t} = 7000$ periods observed
\end{tabular}
\caption{\label{figureserr2pl}
Relative estimation  errors for covariates without holidays. The horizontal line denotes the relative error for the model with regressors. \(N1\) considers each zone has eight neighbors and only apply penalties if neighboring zones are of the same color. \(N2\) considers each zone only has four neighbors and apply penalties regardless of color.
}
\end{figure}

\begin{figure}
\centering
\begin{tabular}{cc}
\includegraphics[scale=0.3]{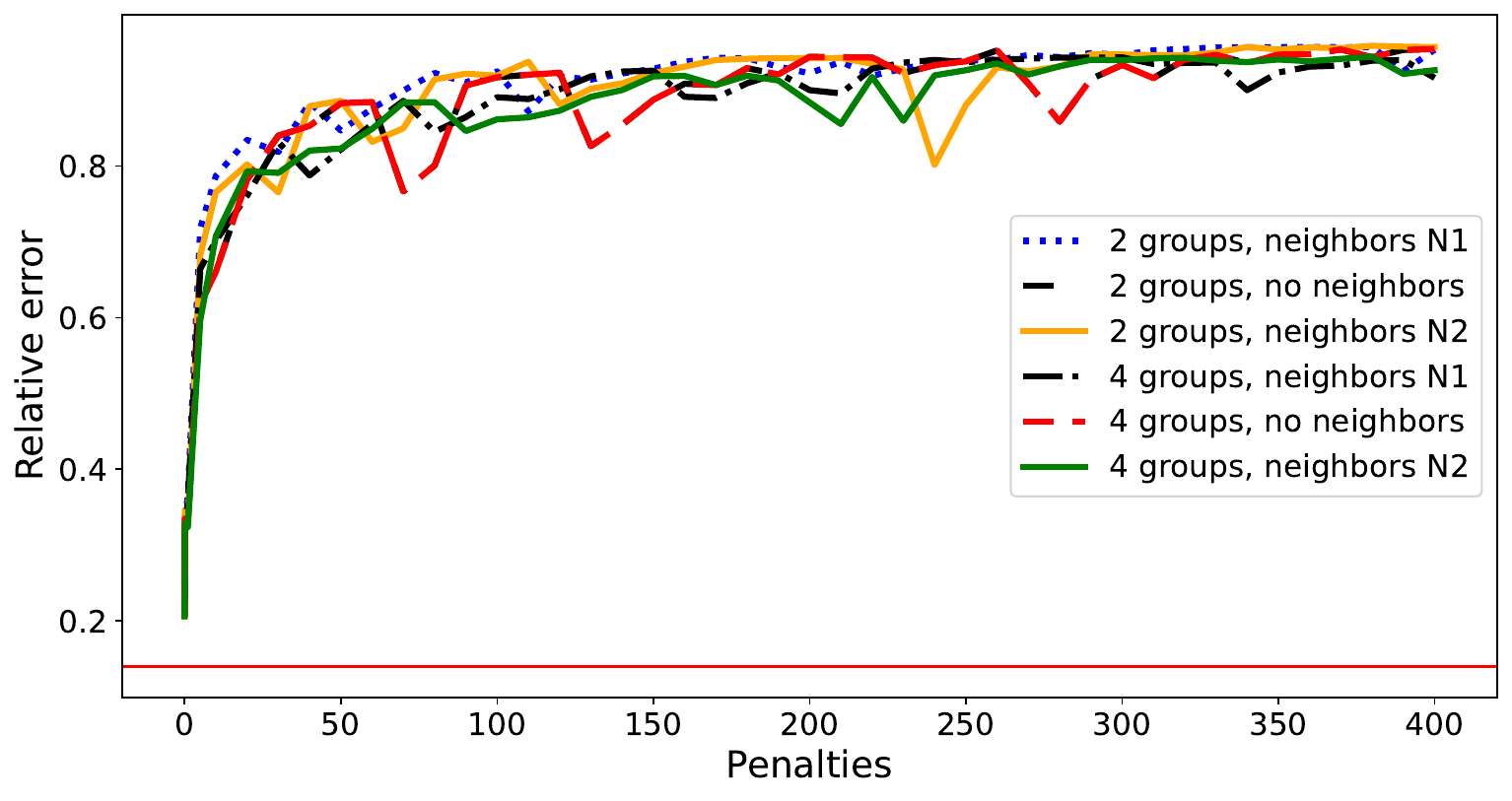} &
\includegraphics[scale=0.3]{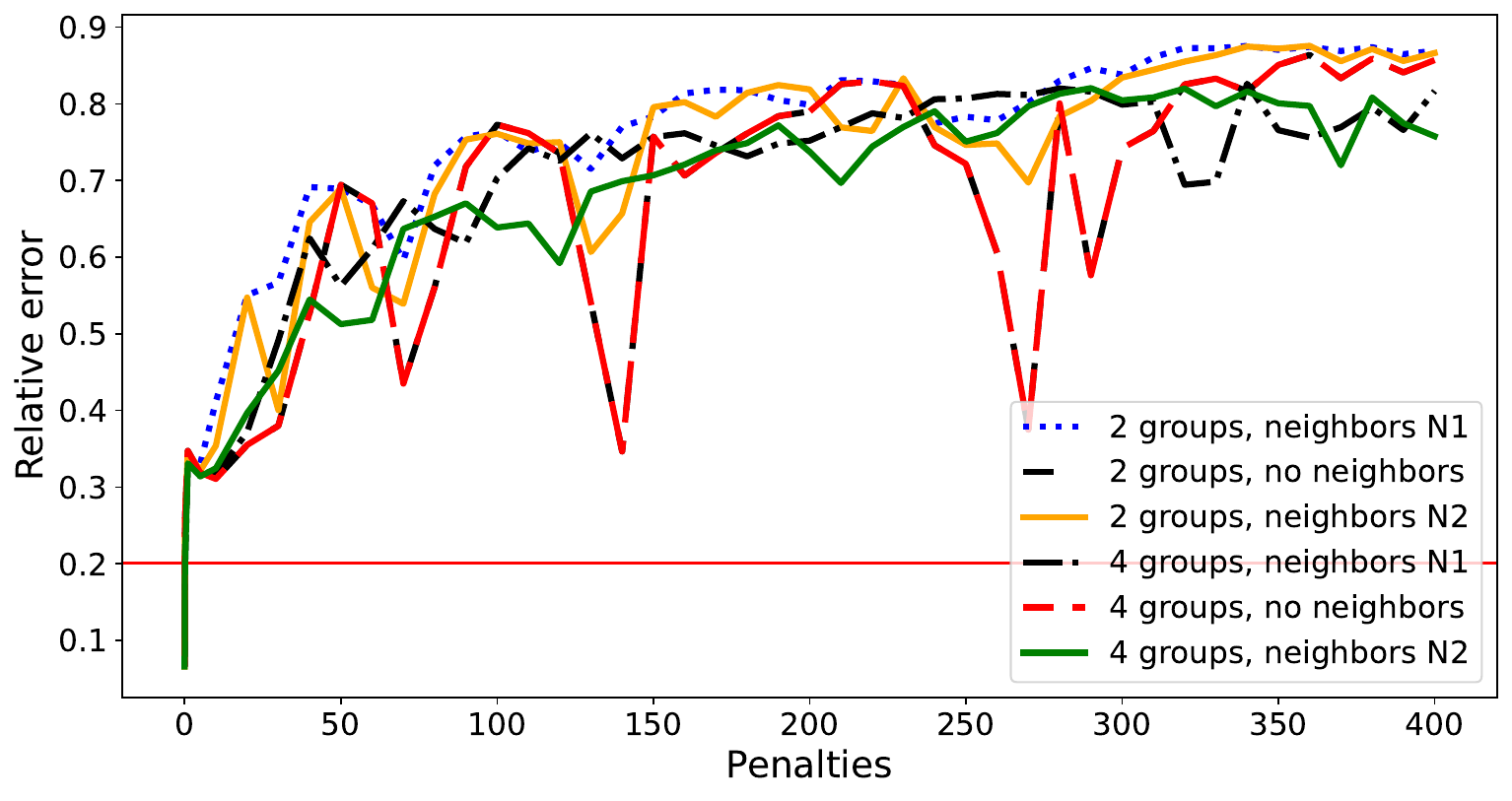} \\
$\hspace*{0.6cm}$ $N_{i,t} = 14$ periods observed & $\hspace*{0.6cm}$ $N_{i,t} = 140$ periods observed
\vspace{5mm} \\
\includegraphics[scale=0.3]{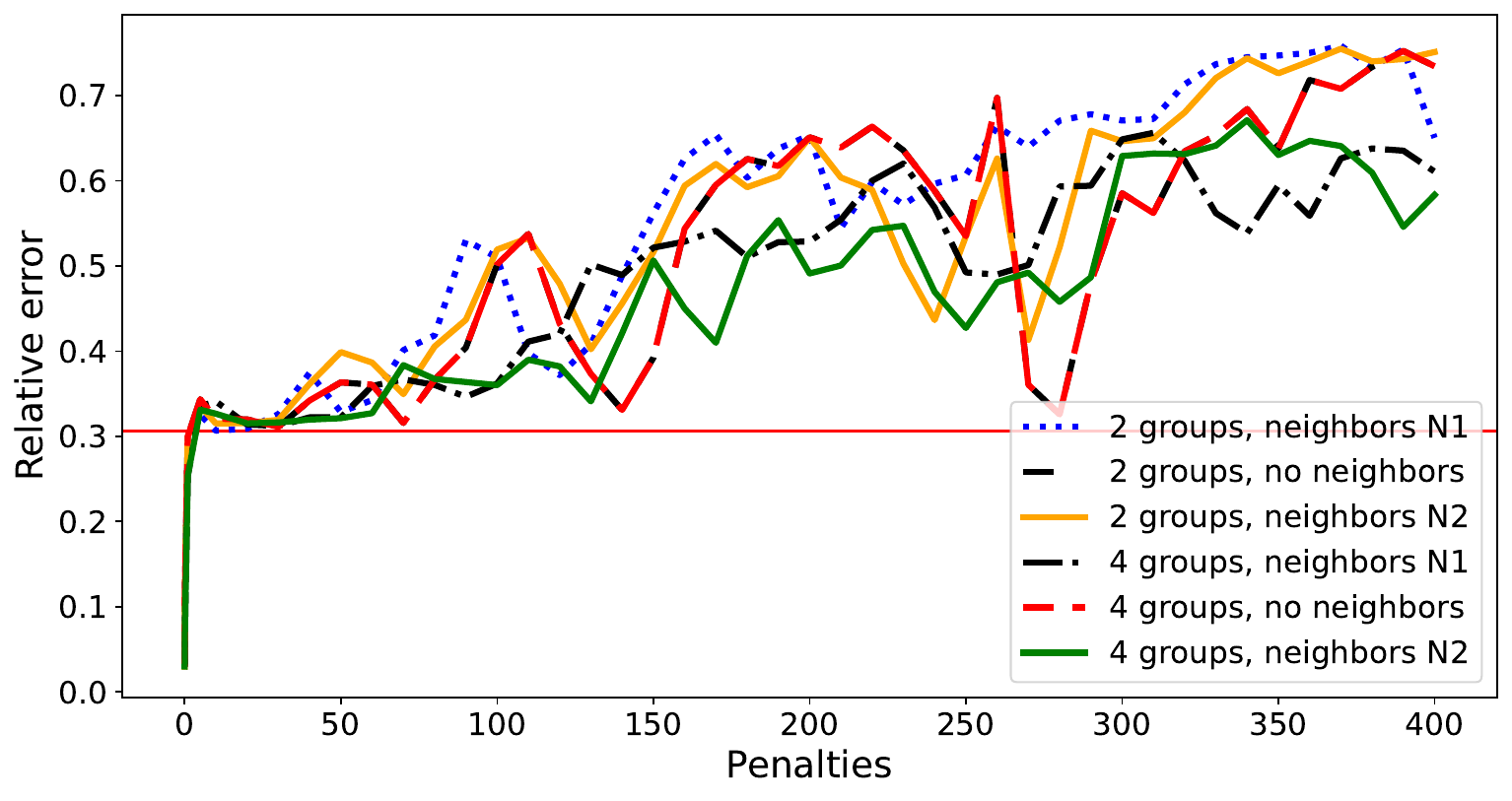} &
\includegraphics[scale=0.3]{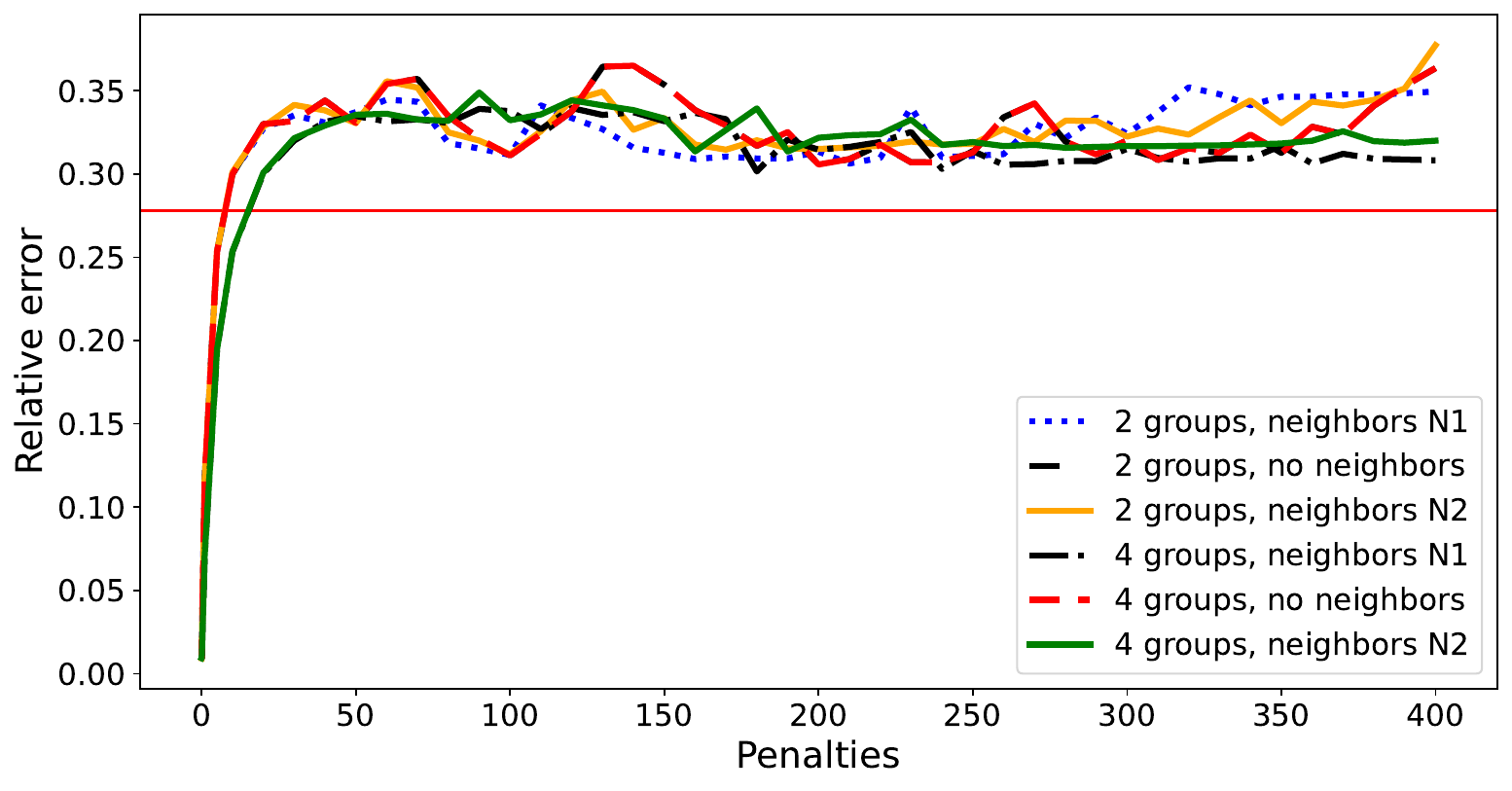} \\
$\hspace*{0.6cm}$ $N_{i,t} = 700$ periods observed & $\hspace*{0.6cm}$ $N_{i,t} = 7000$ periods observed
\end{tabular}
\caption{\label{figureserr2pl}
Relative estimation errors for covariates with holidays. The horizontal line denotes the relative error for the model with regressors. \(N1\) considers each zone has eight neighbors and only apply penalties if neighboring zones are of the same color. \(N2\) considers each zone only has four neighbors and apply penalties regardless of color.
}
\end{figure}

\begin{figure}
\centering
\begin{tabular}{cc}
\includegraphics[scale=0.3]{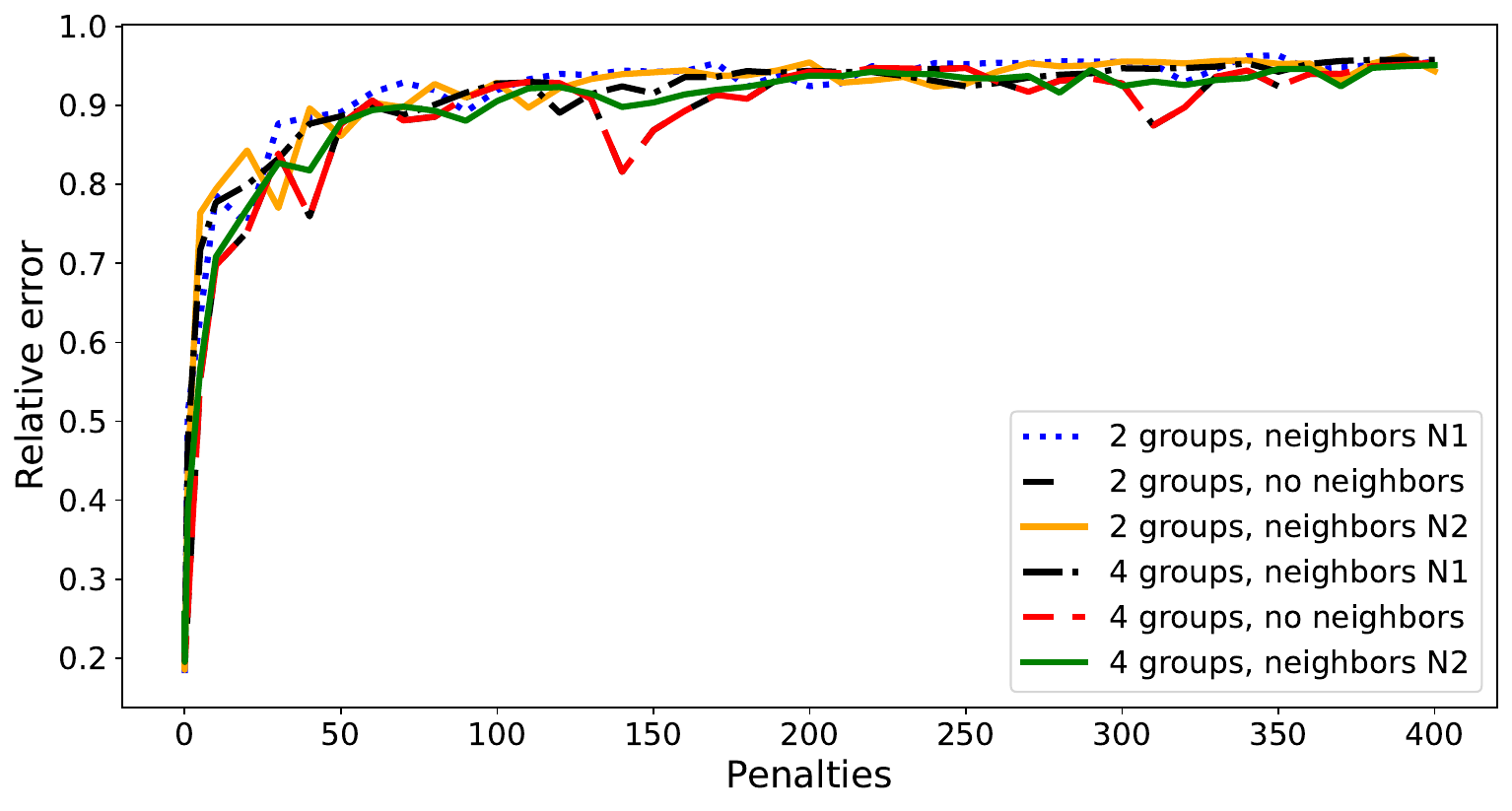} &
\includegraphics[scale=0.3]{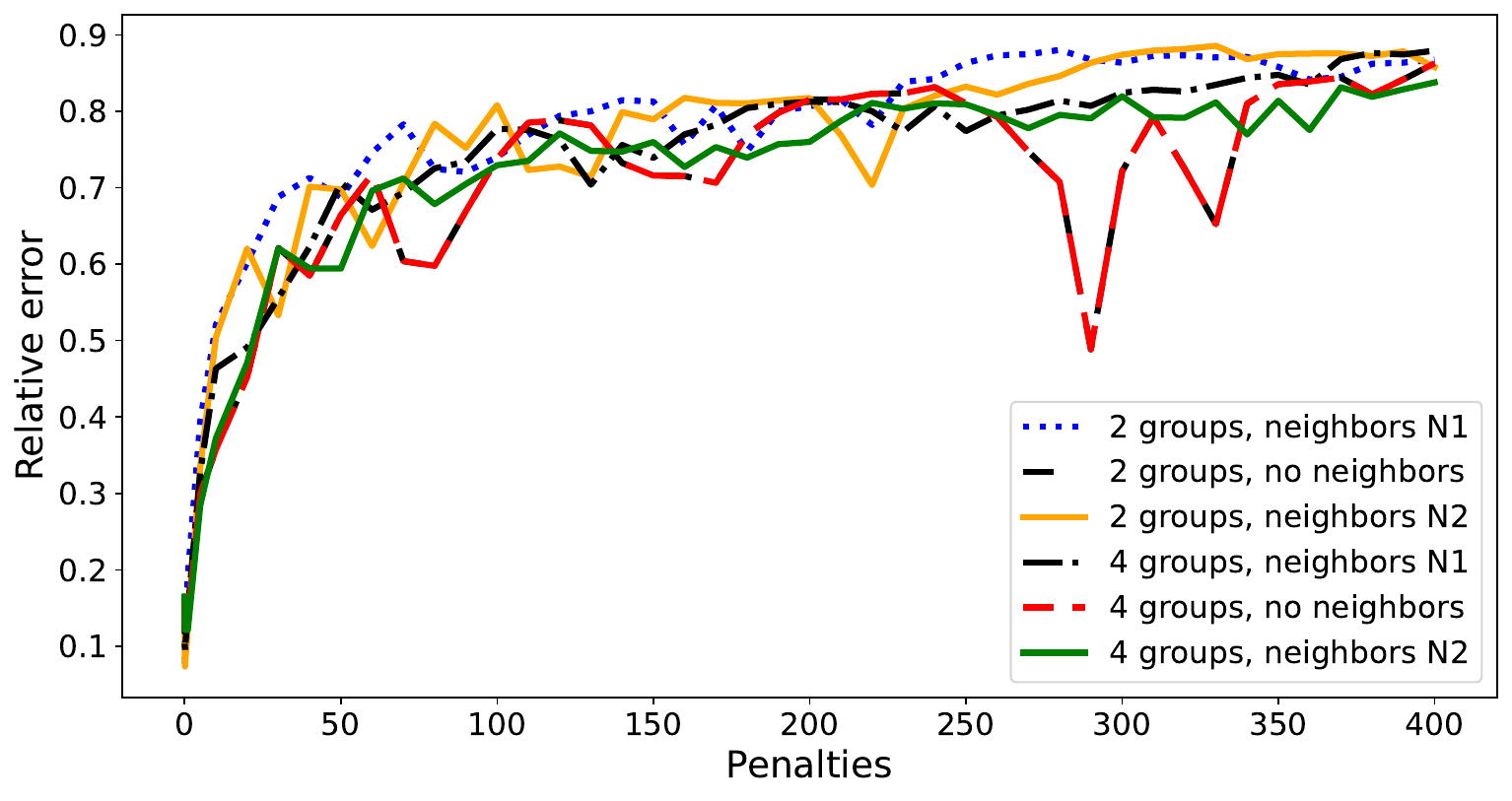} \\
$\hspace*{0.6cm}$ $N_{i,t} = 14$ periods observed & $\hspace*{0.6cm}$ $N_{i,t} = 140$ periods observed
\vspace{5mm} \\
\includegraphics[scale=0.3]{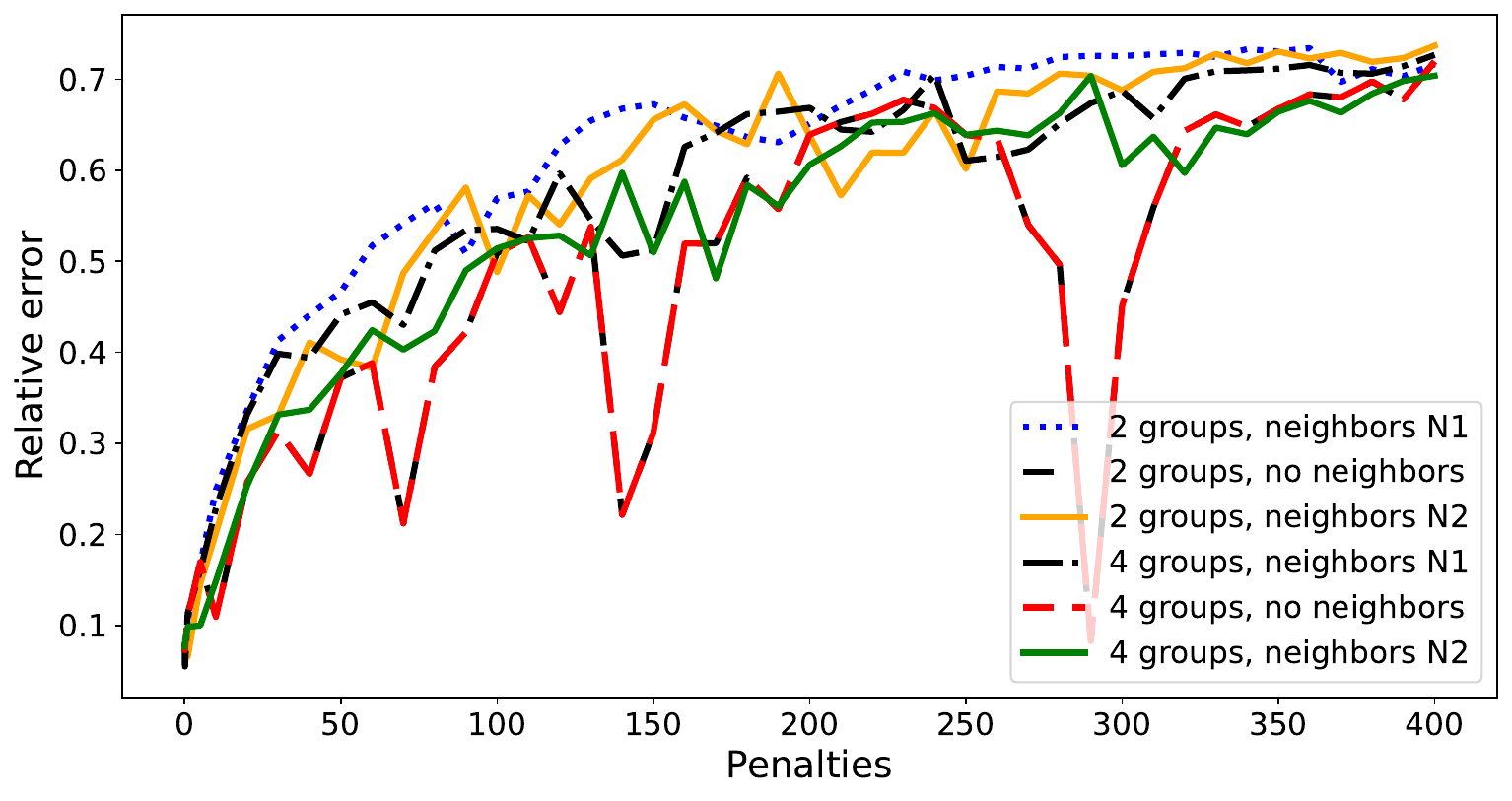} &
\includegraphics[scale=0.3]{results_clam_n1000_crop.pdf} \\
$\hspace*{0.6cm}$ $N_{i,t} = 700$ periods observed & $\hspace*{0.6cm}$ $N_{i,t} = 7000$ periods observed
\end{tabular}
\caption{\label{figureserr2pl}
Relative estimation  errors for parameters according to figure~\ref{figuresex2}. \(N1\) considers each zone has eight neighbors and only apply penalties if neighboring zones are of the same color. \(N2\) considers each zone only has four neighbors and apply penalties regardless of color.
}
\end{figure}

}\fi

\subsection{Numerical Examples with Artificial Data with Covariates}

As in the previous examples, points arrive in $\mathcal{S} = [0,10]^2$ and over time according to a periodic non-homogeneous Poisson process.
The region $\mathcal{S}$ is partitioned into two subregions $\mathcal{S} = \mathcal{B} \cup \mathcal{R}$, with $\mathcal{B} = [0,5] \times (5,10] \cup (5,10] \times [0,5]$ and $\mathcal{R} = [0,5] \times [0,5] \cup (5,10] \times (5,10]$.
There is only one type of point, thus $|\mathcal{C}| = 1$, and hence the type notation~$c$ is omitted.
Each point $s \in \mathcal{S}$ has three attributes, denoted $x(s) \defi (x_{1}(s),x_{2}(s),x_{3}(s))$.
For the examples, $x(s)$ was generated as follows.
Let $U^{b}_{j}$, $U^{r}_{j}$, $j = 1,\ldots, 20$, be independent random variables uniformly distributed on $(0,1)$.
For $s \in \mathcal{B}$, let
\[
x_{1}(s) \ \ = \ \ \sum_{j=1}^{10} \left(\frac{1}{2}\right)^{j} \left\{U^{b}_{2j-1} |\sin(2 \pi j s_{1} / 10)| + U^{b}_{2j} |\sin(2 \pi j s_{2} / 10)|\right\},
\]
and for $s \in \mathcal{R}$, let
\[
x_{1}(s) = \sum_{j=1}^{10} \left(\frac{1}{2}\right)^{j} \left\{(U^{r}_{2j-1}+1) |\sin(2 \pi j s_{1} / 10)| + (U^{r}_{2j}+1) |\sin(2 \pi j s_{2} / 10)|\right\}.
\]
Let $x_{2}(s) = 0.25$ and $x_{3}(s) = 0.5$ for $s \in \mathcal{B}$, and $x_{2}(s) = 0.5$ and $x_{3}(s) = 0.25$ for $s \in \mathcal{R}$.

\begin{figure}
\centering
\includegraphics[scale=0.7]{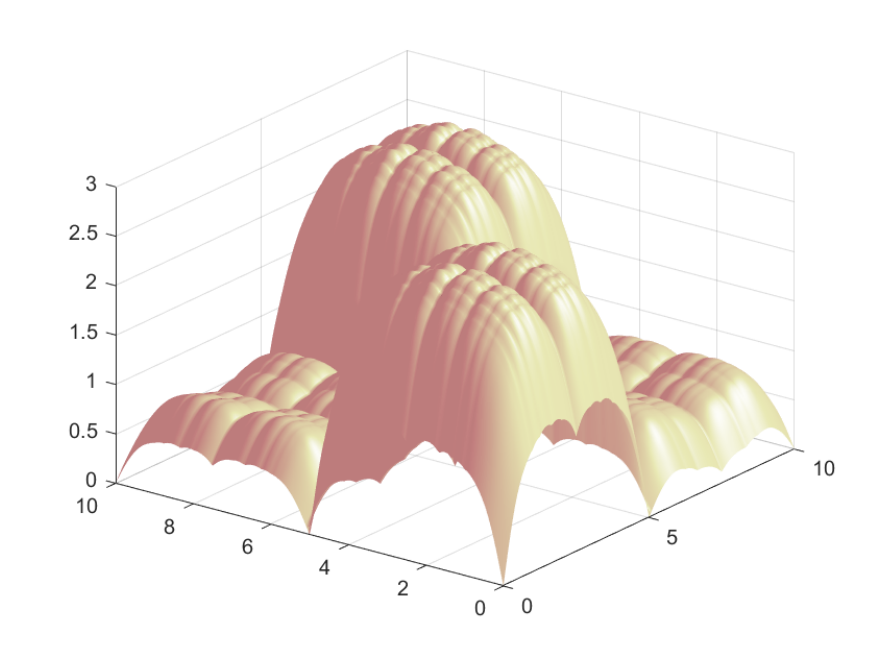}
\caption{Graph of covariate function $x_{1}$.\label{fig:x1plot}}
\end{figure}

We consider $2$~settings, one without a holiday effect, and one with a holiday effect.
In both settings, the rate function $\lambda$ is periodic with period~$2$.

\subsubsection{Example~3}

In the setting without a holiday effect, the rate function $\lambda : \mathcal{S} \times [0,T] \mapsto \RR_{+}$ is as follows:
\[
\lambda(s,t) \ \ = \ \ \left\{\begin{array}{lcl}
\beta(0)^{\top} x(s) & \mbox{if} & t \in (2k,2k+1] \mbox{ for some } k \in \NN, \\
\beta(1)^{\top} x(s) & \mbox{if} & t \in (2k-1,2k] \mbox{ for some } k \in \NN
\end{array}\right.
\]
where
\[
\begin{array}{ll}
\beta_{1}(0) = 0 & \beta_{1}(1) = 0.05, \\
\beta_{2}(0) = 6 & \beta_{2}(1) = 18, \\
\beta_{3}(0) = 3 & \beta_{3}(1) = 6.
\end{array}
\]
Thus, if $\mathcal{E} \defi \cup_{k \in \NN} (2k,2k+1]$ denotes the even-indexed time intervals and $\mathcal{O} \defi \cup_{k \in \NN} (2k-1,2k]$ denotes the odd-indexed time intervals, and $\mathcal{A} \subset \mathcal{S}$ and $\mathcal{D} \subset [0,\infty)$, then the expected number of arrivals in $\mathcal{A} \times \mathcal{D}$ is given by $\int_{\mathcal{D}} \int_{\mathcal{A}} \lambda(s,t) \diff s \diff t = \left|\mathcal{D} \cap \mathcal{E}\right| \beta(0)^{\top} \int_{\mathcal{A}} x(s) \diff s + \left|\mathcal{D} \cap \mathcal{O}\right| \beta(1)^{\top} \int_{\mathcal{A}} x(s) \diff s$, where $\left|\mathcal{D} \cap \mathcal{E}\right| \defi \int_{\mathcal{D} \cap \mathcal{E}} \diff t$ and $\left|\mathcal{D} \cap \mathcal{O}\right| \defi \int_{\mathcal{D} \cap \mathcal{O}} \diff t$.

\subsubsection{Example~4}

In the setting with a holiday effect, the rate function $\lambda : \{0,1\} \times \mathcal{S} \times [0,T] \mapsto \RR_{+}$ is as follows:
\[
\lambda(h,s,t) \ \ = \ \ \left\{\begin{array}{lcl}
\beta(h,0)^{\top} x(s) & \mbox{if} & t \in (2k,2k+1] \mbox{ for some } k \in \NN, \\
\beta(h,1)^{\top} x(s) & \mbox{if} & t \in (2k-1,2k] \mbox{ for some } k \in \NN
\end{array}\right.
\]
where $h = 1$ if $t$ falls in a holiday, and $h = 0$ otherwise, and
\[
\begin{array}{llll}
\beta_{1}(0,0) = 0, & \beta_{1}(0,1) = 0.05, & \beta_{1}(1,0) = 0, & \beta_{1}(1,1) = 0.1, \\
\beta_{2}(0,0) = 6, & \beta_{2}(0,1) = 18, & \beta_{2}(1,0) = 12, & \beta_{2}(1,1) = 36, \\
\beta_{3}(0,0) = 3, & \beta_{3}(0,1) = 6, & \beta_{3}(1,0) = 6, & \beta_{3}(1,1) = 12.
\end{array}
\]

We consider $2$~estimators for each of the $2$~settings: one estimator knows and uses aggregated covariate data, and the other estimator does not use covariate data.
Below, we describe the $2$~estimators in more detail.

\subsubsection{Estimators that use the covariate data}

The user knows the region $\mathcal{S}$ and that $|\mathcal{C}| = 1$, but does not know that $\lambda$ is periodic with period~$2$.
In the setting with a holiday effect, the user knows that there is a holiday effect, but the user distinguishes $8$~different holidays and allows them to have different parameters.
For estimation purposes, $\mathcal{S}$ is discretized into $100$~square zones of unit area each, as shown in Figure~\ref{figureserr}.
Thus $\mathcal{I} = \mathcal{I}_{\mathcal{B}} \cup \mathcal{I}_{\mathcal{R}}$ indexes two types of zones, but this is not known by the estimator.
For each zone $i \in \mathcal{I}$, the user observes only the aggregate value $y_{i,l} \defi \int_{R_{i}} x_{l}(s) \diff s$ of the covariate $x_{l}$ for the zone.
For example, if a zone in $\mathcal{I}_{\mathcal{B}}$ is $[i-1,i) \times (j-1,j]$, then the aggregate value of the covariate $x_{1}$ for the zone is
\begin{eqnarray*}
& & \int_{i-1}^{i} \int_{j-1}^{j} x_{1}(s_{1},s_{2}) \diff s_{2} \diff s_{1} \\
& = & \sum_{k=1}^{10} \left(\frac{1}{2}\right)^{k} \left\{U^{b}_{2k-1} \int_{i-1}^{i} |\sin(2 \pi k s_{1} / 10)| \diff s_{1} + U^{b}_{2k} \int_{j-1}^{j} |\sin(2 \pi k s_{2} / 10)| \diff s_{2}\right\}.
\end{eqnarray*}
Therefore, we need to compute integrals of form $\displaystyle \int_{i-1}^{i} |\sin(2 \pi k s_{1} / 10)| \diff s_{1}$.
\if{
For $0 \leq s_{1}, s_{1} \leq 10$,
we have
$\sin(2 \pi k s_{1} / 10)\geq 0$
for
$$
\frac{10 \ell }{k} \leq s_{1} \leq 
\frac{10 \ell }{k}  + \frac{5}{k},\;
\ell=0,1,\ldots,k-1,
$$
while 
$\sin(2 \pi k s_{1} / 10)) \leq 0$
for
$$
\frac{10 \ell }{k}-\frac{5}{k} \leq s_{1} \leq 
\frac{10 \ell }{k},\;\ell=1,2,\ldots,k.
$$
Therefore 
$
\displaystyle \int_{s_{1}=i-1}^{s_{1}=i}|\sin(2 \pi k s_{1} / 10)|ds_{1}
$ is given by
$$
\begin{array}{l}
\displaystyle \int_{s_{1}=i-1}^{s_{1}=i}|\sin(2 \pi k s_{1} / 10)|ds_{1} \\
=\displaystyle \int_{i-1}^{\frac{5}{k}\left \lceil \frac{k}{5}(i-1) \right \rceil}|\sin(2 \pi k s_{1} / 10)|ds_{1} 
+ 
\displaystyle
\sum_{\ell=\left \lceil \frac{k}{5}(i-1) \right \rceil}^{\left \lceil \frac{i k}{5} \right \rceil-2}
\int_{\frac{5}{k}\ell}^{\frac{5}{k}(\ell+1)}|\sin(2 \pi k s_{1} / 10)|ds_{1} +
\int_{\frac{5}{k}(\left \lceil \frac{i k}{5} \right \rceil-1)}^{i}|\sin(2 \pi k s_{1} / 10)|ds_{1} \\
=-\frac{5}{\pi k}\left( 
\cos\left(\frac{\pi}{k} 
\lceil 
\frac{k}{5}(i-1)\rceil \right)
-\cos\left(\frac{\pi k(i-1)}{5}\right)
\right)
+ \frac{10}{\pi k}|\ell \mbox{ even}:
\left \lceil \frac{k}{5}(i-1) \right \rceil 
\leq \ell \leq \left \lceil \frac{i k}{5} \right \rceil-1 |\\
\;\;\;+
\frac{5}{\pi k} (-1)^{1 + \left\lceil \frac{ik}{5} \right\rceil}
\left(  -\cos(\frac{\pi i k}{5})
+\cos(\pi(-1+\lceil  \frac{i k}{5} \rceil ))
\right).
\end{array}
$$

Therefore, we need to compute integrals of form 
$\displaystyle \int_{i-1}^{i} |\sin(2 \pi k s_{1} / 10)| \diff s_{1}$. 
}\fi

Note that, for $0 \leq s_{1} \leq 10$, it holds that $\sin(2 \pi k s_{1} / 10) \geq 0$ if
\[
\frac{5}{k} \ell \ \ \leq \ \ s_{1} \ \ \leq \ \ \frac{5}{k} (\ell + 1)
\quad \mbox{and} \quad \ell \in \{0,2,\ldots,2k-2\} \mbox{ is even,}
\]
and $\sin(2 \pi k s_{1} / 10) \leq 0$ if
\[
\frac{5}{k} \ell \ \ \leq \ \ s_{1} \ \ \leq \ \ \frac{5}{k} (\ell + 1)
\quad \mbox{and} \quad \ell \in \{1,\ldots,2k-1\} \mbox{ is odd.}
\]
Also note that
\[
\frac{5}{k} \left\lceil \frac{k}{5} (i-1) \right\rceil \ \ \le \ \ i
\qquad \Longleftrightarrow \qquad \left\lceil \frac{k}{5} (i-1) \right\rceil \ \ \le \ \ \frac{k}{5} i
\qquad \Longleftrightarrow \qquad \left\lceil \frac{k}{5} (i-1) \right\rceil \ \ \le \ \ \left\lfloor \frac{k}{5} i \right\rfloor.
\]
Therefore, if $\frac{5}{k} \left\lceil \frac{k}{5} (i-1) \right\rceil \le i$, then
\begin{eqnarray*}
& & \int_{i-1}^{i} |\sin(2 \pi k s_{1} / 10)| \diff s_{1} \\
& = & \int_{i-1}^{\frac{5}{k} \left\lceil \frac{k}{5} (i-1) \right\rceil} |\sin(2 \pi k s_{1} / 10)| \diff s_{1} + \sum_{\ell = \left\lceil \frac{k}{5} (i-1) \right\rceil}^{\left\lfloor \frac{k}{5} i \right\rfloor - 1} \int_{\frac{5}{k} \ell}^{\frac{5}{k} (\ell+1)} |\sin(2 \pi k s_{1} / 10)| \diff s_{1} +
\int_{\frac{5}{k} \left\lfloor \frac{k}{5} i \right\rfloor}^{i} |\sin(2 \pi k s_{1} / 10)| \diff s_{1} \\
& = & \frac{5}{\pi k} \left(-1\right)^{\left\lceil \frac{k}{5} (i-1) \right\rceil} \left[\cos\left(\pi \left\lceil \frac{k}{5} (i-1) \right\rceil\right) - \cos\left(\pi \frac{k}{5} (i-1)\right)\right] \\
& & {} + \frac{10}{\pi k} \left(\left\lfloor \frac{k}{5} i \right\rfloor - \left\lceil \frac{k}{5} (i-1) \right\rceil\right) - \frac{5}{\pi k} \left(-1\right)^{\left\lfloor \frac{k}{5} i \right\rfloor} \left[\cos\left(\pi \frac{k}{5} i\right) - \cos\left(\pi \left\lfloor \frac{k}{5} i \right\rfloor\right)\right]
\end{eqnarray*}
and if $\frac{5}{k} \left\lceil \frac{k}{5} (i-1) \right\rceil > i$, then
\[
\int_{i-1}^{i} |\sin(2 \pi k s_{1} / 10)| \diff s_{1}
\ \ = \ \ \frac{5}{\pi k} \left(-1\right)^{\left\lfloor \frac{k}{5} i \right\rfloor} \left[\cos\left(\pi \frac{k}{5} (i-1)\right) - \cos\left(\pi \frac{k}{5} i\right)\right].
\]
Also, $y_{i,2} \defi \int_{R_{i}} x_{2}(s) \diff s = 0.25$, and $y_{i,3} \defi \int_{R_{i}} x_{3}(s) \diff s = 0.5$ for $i \in \mathcal{I}_{\mathcal{B}}$, and $y_{i,2} = 0.5$, and $y_{i,3} = 0.25$ for $i \in \mathcal{I}_{\mathcal{R}}$.

As in the previous examples, the user discretizes time into $28$~time intervals of length~$1$ each.
The user's model is the same as the model specified in Example~\ref{excov2}, with $|\mathcal{C}| = 1$, $|\mathcal{T}| = 28$, $K_{1} = 28$, and $|\mathcal{O}| = 3$.
Thus, in the setting without a holiday effect, the user's intensity function $\hat{\lambda} : \mathcal{I} \times \mathcal{T} \mapsto \RR_{+}$ given by
\[
\hat{\lambda}(i,t) \ \ = \ \ \hat{\beta}(t)^{\top} y_{i}
\]
estimates $\int_{R_{i}} \int_{(t-1,t]} \lambda(s,\tau) \diff \tau \diff s$, where $\hat{\beta}(t) = \big(\hat{\beta}_{1}(t),\hat{\beta}_{2}(t),\hat{\beta}_{3}(t)\big)$.
Thus, this estimated model has $28 \times 3$ parameters.
In the setting with a holiday effect, the user's intensity function $\hat{\lambda} : \{0,1,\ldots,8\} \times \mathcal{I} \times \mathcal{T} \mapsto \RR_{+}$ given by
\[
\hat{\lambda}(h,i,t) \ \ = \ \ \hat{\beta}(h,t)^{\top} y_{i}
\]
estimates $\int_{R_{i}} \int_{(t-1,t]} \lambda(h',s,\tau) \diff \tau \diff s$, where $h' = 0$ if $h = 0$ and $h' = 1$ if $h \in \{1,\ldots,8\}$, and $\hat{\beta}(h,t) = \big(\hat{\beta}_{1}(h,t),\hat{\beta}_{2}(h,t),\hat{\beta}_{3}(h,t)\big)$.
Thus, this estimated model has $9 \times 28 \times 3$ parameters.

\subsubsection{Estimators that do not use the covariate data}

The user knows the region $\mathcal{S}$ and that $|\mathcal{C}| = 1$, but does not know the covariate values $y_{i}$, and does not know that $\lambda$ is periodic with period~$2$.
In the setting with a holiday effect, the user knows that there is a holiday effect, but the user distinguishes $8$~different holidays and allows them to have different parameters.

The estimators are similar to those in Section~\ref{sec:simulated_example}.
In the setting without a holiday effect, the user's intensity function $\hat{\lambda}^{w}_{i,t}$, computed by solving~\eqref{eqn:model0} with the penalty parameter~$w$, estimates $\int_{R_{i}} \int_{(t-1,t]} \lambda(s,\tau) \diff \tau \diff s$.
In the setting with a holiday effect, the user's intensity function $\hat{\lambda}^{w}_{h,i,t}$, computed by solving~\eqref{eqn:model0} with the penalty parameter~$w$, estimates $\int_{R_{i}} \int_{(t-1,t]} \lambda(h',s,\tau) \diff \tau \diff s$.

\subsubsection{Numerical results}

Table~\ref{tablecompwomodels1} shows the mean relative errors of regularized estimators with best penalty parameter~$w^*$ without covariates from Section~\ref{sec:model1}, and the estimator with covariates from Section~\ref{sec:modelcov} (without regularization).
The column ``Holiday'' indicates whether the intensities used to generate the data are different on holidays or not.
For the examples with holidays, in the ``Sample size $N_{i,t}$'' column the first number is the number of observations for days which are not holidays and the second number is the number of observations for days which are holidays (for instance, for the first experiment with holidays, $N_{i,t} = 51$ for periods~$t$ which are not on holidays and $N_{i,t} = 1$ for periods~$t$ which are on holidays).
The column ``Cov'' shows results for the estimator with covariates from Section~\ref{sec:modelcov}.
The table also shows results for three regularized estimators without covariates:
an estimator that does not use space regularization (column ``No Cov 1''),
an estimator that uses space regularization with penalizing weights for neighbors of the same color (column ``No Cov 2''), and
an estimator that uses space regularization with penalizing weights for all neighbors regardless of their color (column ``No Cov 3'').
For each estimator, the cells in the table show the values of the mean relative error for the estimator and the instance (for the regularized estimators without covariates, the best mean relative error among all considered penalty weights is shown).

\begin{table}[]
\centering
\begin{tabular}{|c|c|c|c|c|c|c|}
\hline
Sample size $N_{i,t}$ & Holiday & Cov  & No Cov 1 & No Cov 2 & No Cov 3 & Emp  \\ \hline
52     &   No      &   0.006   &    0.013  &   0.001     &    0.048    &   0.048   \\ \hline
520    &   No      &   0.002   &    0.003  &   0.0006    &    0.015    &   0.015   \\ \hline
780    &   No      &   0.001   &    0.001  &   0.0001    &    0.006    &   0.006   \\ \hline
51/1     &   Yes     &   0.073   &    0.039  &   0.042     &    0.104    &  0.156    \\ \hline
510/10    &   Yes     &   0.049   &    0.015  &   0.011     &    0.048    &  0.048    \\ \hline
765/15    &   Yes     &   0.032   &    0.005  &   0.003     &    0.021    &  0.021    \\ \hline
\end{tabular}
\caption{Mean relative error $e_{*}$.
In column~1, $m$/$n$ denotes $N_{i,t} = m$ for non-holiday time periods~$t$, and $N_{i,t} = n$ for holiday time periods~$t$.}
\label{tablecompwomodels1}
\end{table}

\subsection{Numerical Example with Real Data}
\label{sec:case-study}

LASPATED was used with real data to calibrate arrival intensity functions for the process of medical emergencies reported to the Rio de Janeiro emergency medical service.
The data included the date and time of the phone call, with the date ranging from 2016/01/01 to 2018/01/08, the location of the emergency, and the type of the emergency.
The emergency type data included only a ``priority level'': high, intermediate, and low-priority emergencies.

Different models of arrival intensity $\lambda_{c,i,t}$ as a function of emergency type~$c$, location~$i$, and time~$t$, were calibrated.
The calibrated models were periodic with a period of one week.
The time of the week was discretized into time intervals of length $30$~minutes (thus the model had $T = 7 \times 48$ time intervals).
Three different space discretizations are demonstrated:
\begin{itemize}
\item
A $10 \times 10$ space discretization of a rectangle containing the city of Rio de Janeiro, shown in Figure~\ref{figurerect1010} (76 of these rectangles have nonempty intersection with the city and are shown in the figure);
\item
A hexagonal discretization of the city of Rio de Janeiro (obtained with scale parameter 7, with 297 hexagons intersecting Rio de Janeiro) shown in Figure~\ref{fig:uber7};
\item
A discretization by administrative districts of the city of Rio de Janeiro (with 160 subregions), shown in Figure~\ref{fig:figuredistrict}.
\end{itemize}

For each space discretization, the model without covariates described in Section~\ref{sec:model1} (called Model M1 in what follows), and the model with covariates described in Section~\ref{sec:modelcov} (called Model M2 in what follows), were calibrated with LASPATED.
For Model M2, we used the following four covariates: (a)~the population of the zone, (b)~the land area of commercial activities and public facilities in the zone, (c)~the land area of industrial activities in the zone, and (d)~the non-populated land area (such as forests, beaches, and water) in the zone.
For the calibration of Model M1, we set $w_{i,j} = w$ if zones~$i$ and~$j$ are neighbors, that is, if zones~$i$ and~$j$ share an edge or a vertex, and $w_{i,j} = 0$ otherwise.
The best value for $w$ was selected from $\{0, 0.01, 0.02, 0.03, 0.04, 0.05 \}$ using cross validation.
We did not regularize by time groups. 
The optimization problems for calibration were solved using the projected gradient method with Armijo line search along a feasible direction.

\begin{figure}
\centering
\resizebox{\textwidth}{!}{
\begin{tabular}{cc}
\includegraphics[scale=0.3]{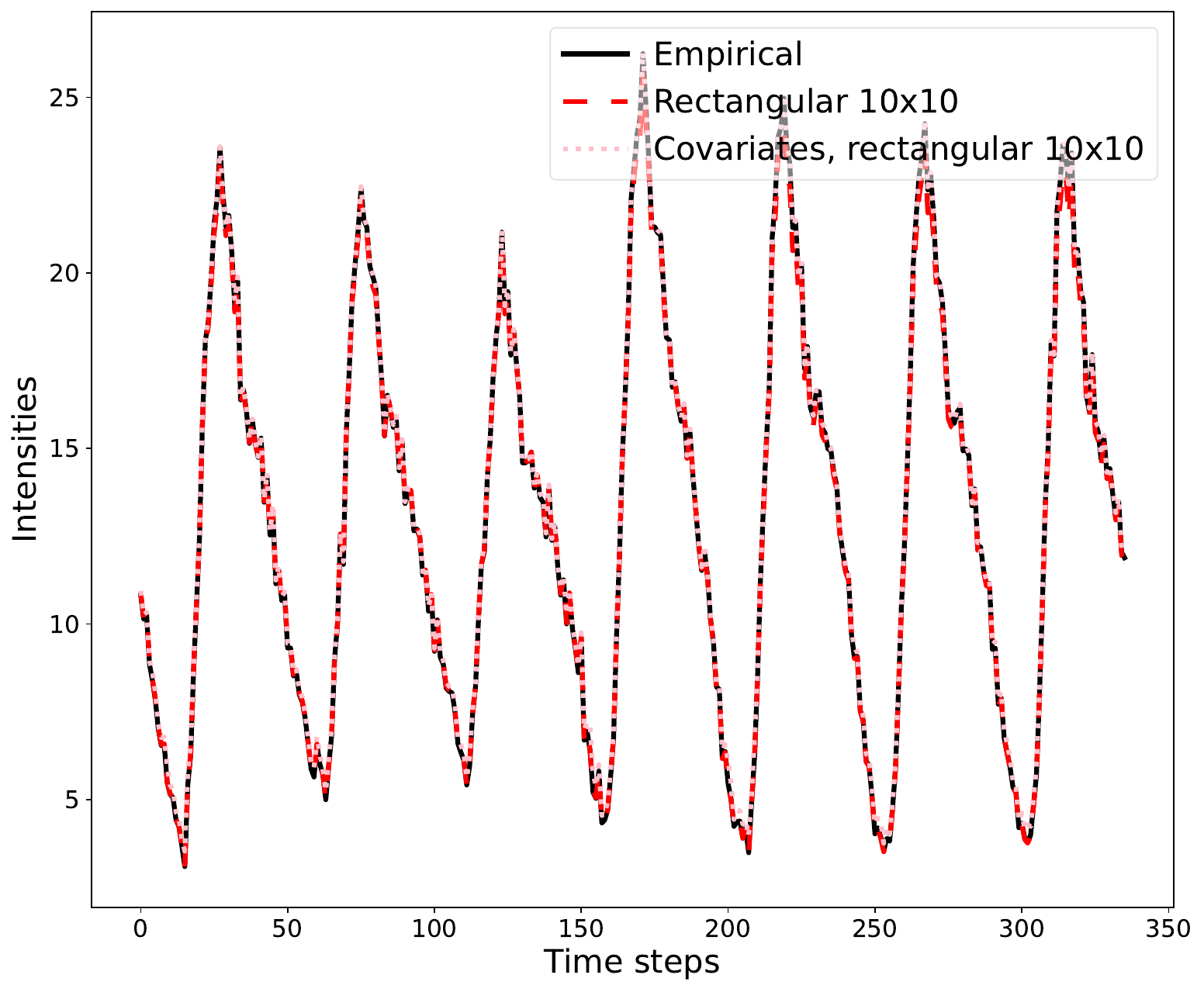}&
\includegraphics[scale=0.3]{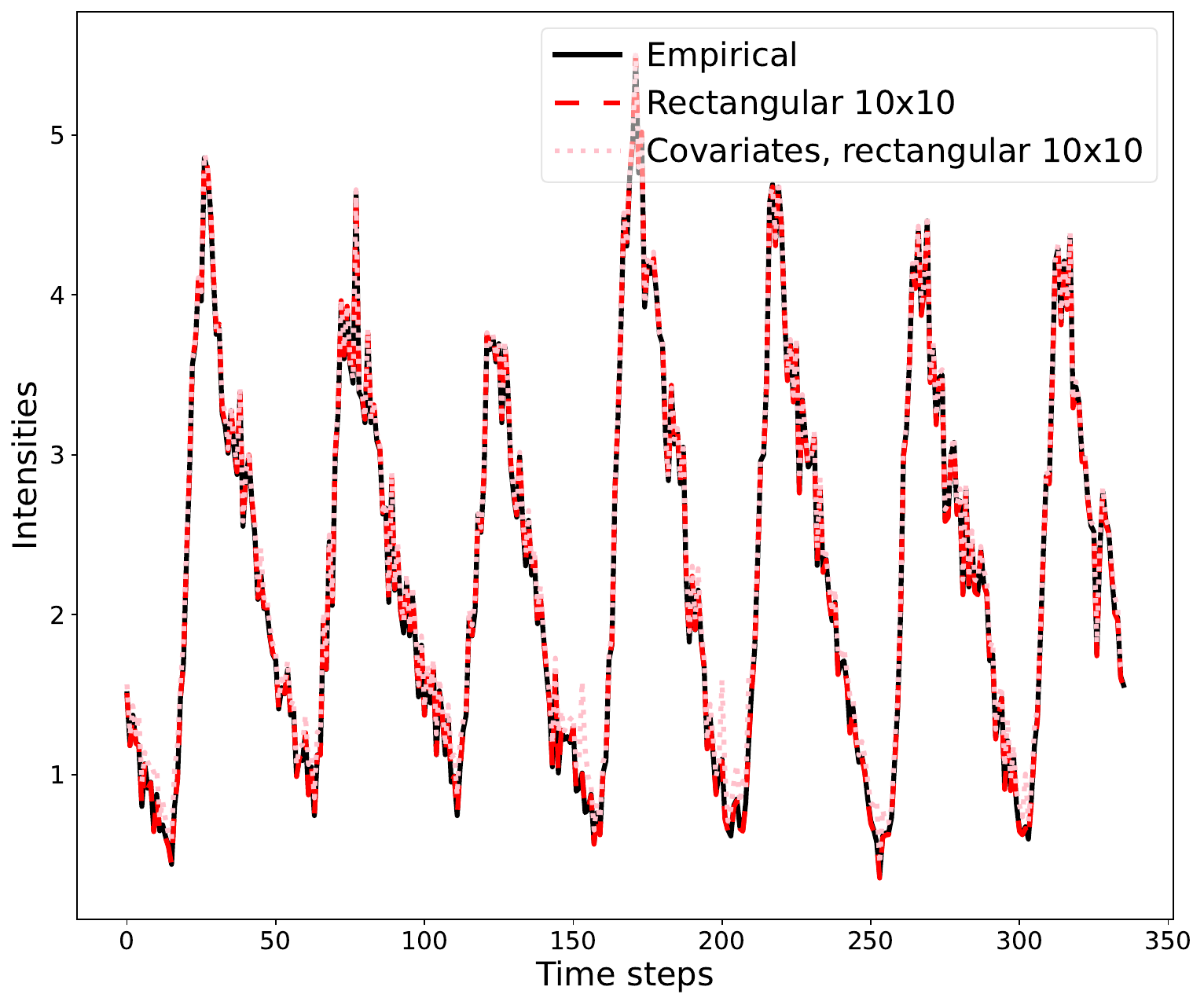}\\
\includegraphics[scale=0.3]{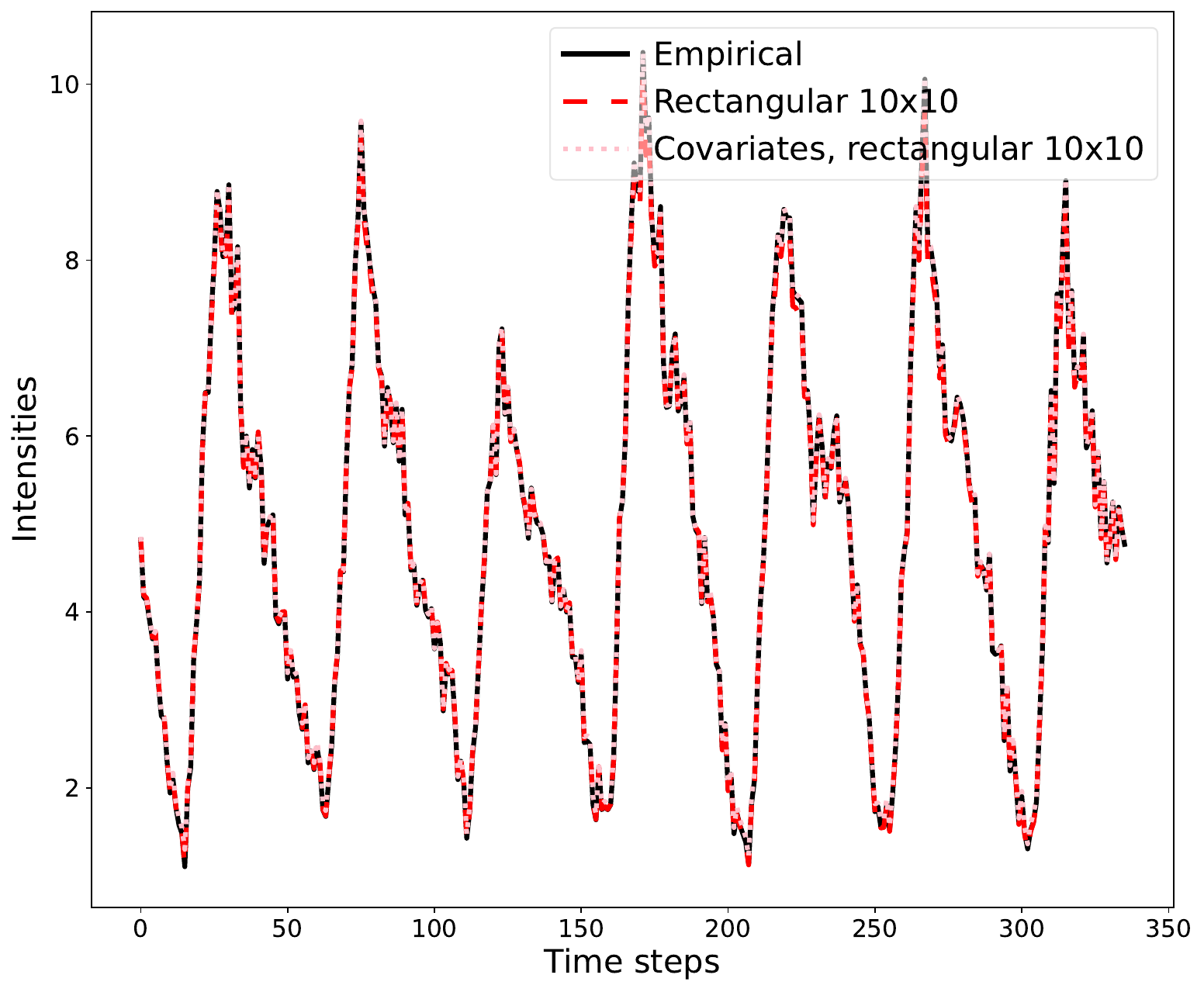}&
\includegraphics[scale=0.3]{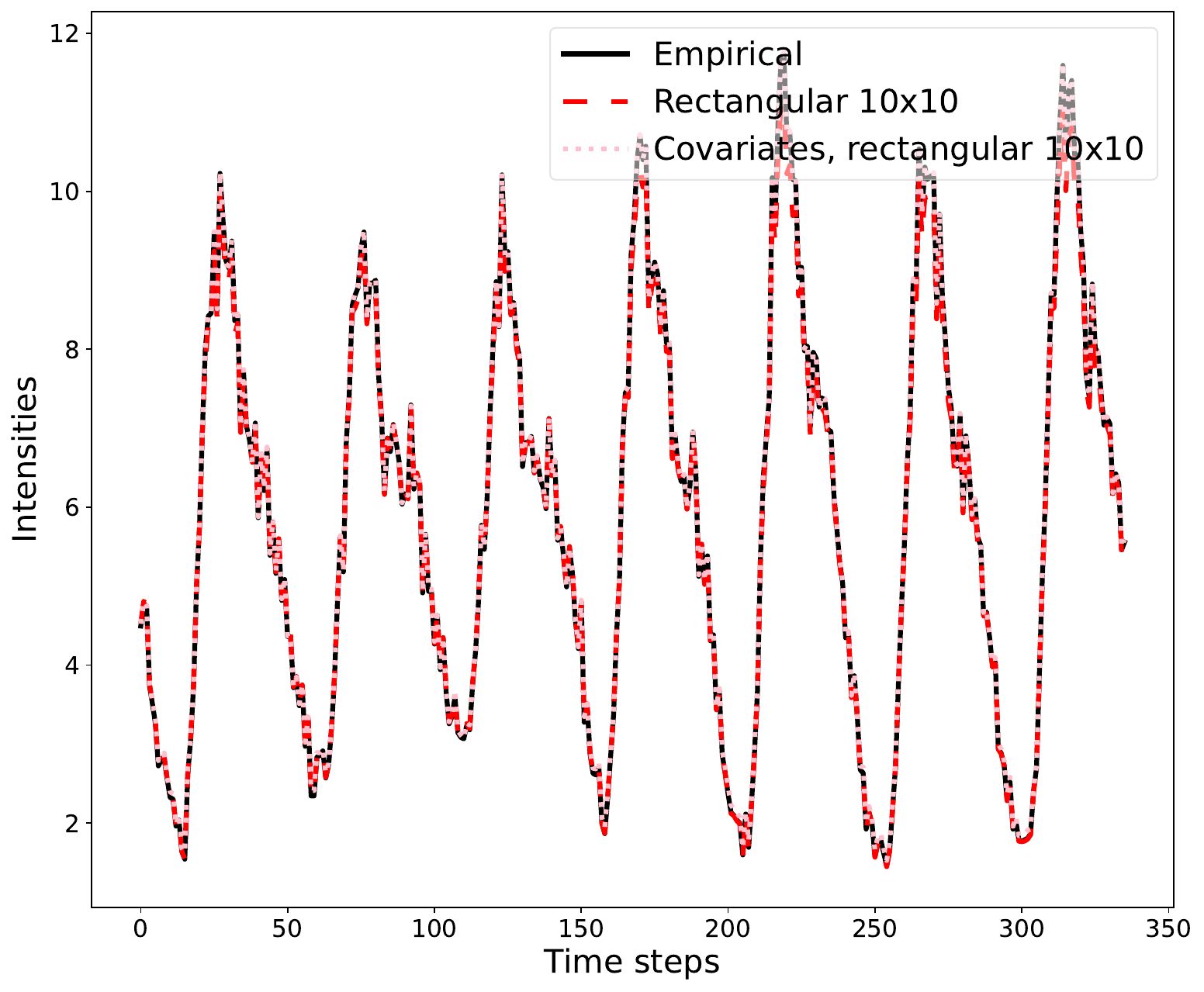}
\end{tabular}}
\caption{Aggregated estimates of the intensities for 3 estimators: (i)~Empirical: the empirical intensities, (ii)~Rectangular $10 \times 10$: Model M1 with the $10 \times 10$ rectangular space discretization, and (iii)~Covariates, rectangular $10 \times 10$: Model M2 with the $10 \times 10$ rectangular space discretization.
Top left: emergencies of all priorities, top right: high-priority emergencies, bottom left: intermediate priority emergencies, bottom right: low-priority emergencies.
\label{fig:allstepsrect}}
\end{figure}

Figure~\ref{fig:allstepsrect} shows the aggregated estimated intensities using the $10 \times 10$ rectangular discretization.
More specifically, the top left plot of Figure~\ref{fig:allstepsrect} shows $\sum_{c \in \mathcal{C}} \sum_{i \in \mathcal{I}} \hat{\lambda}_{c,i,t}$ as a function of~$t$, the top right plot of Figure~\ref{fig:allstepsrect} shows $\sum_{i \in \mathcal{I}} \hat{\lambda}_{c_{2},i,t}$ where $c_{2}$ denotes the high-priority emergencies, the bottom left plot of Figure~\ref{fig:allstepsrect} shows $\sum_{i \in \mathcal{I}} \hat{\lambda}_{c_{1},i,t}$ where $c_{1}$ denotes the intermediate priority emergencies, and the bottom right plot of Figure~\ref{fig:allstepsrect} shows $\sum_{i \in \mathcal{I}} \hat{\lambda}_{c_{0},i,t}$ where $c_{0}$ denotes the low-priority emergencies.
The intensity estimates shown in this figure are obtained with (i)~the empirical mean number of calls for each call type, zone, and time interval, (ii)~Model M1 with the $10 \times 10$ rectangular space discretization, and (iv)~Model M2 with the $10 \times 10$ rectangular space discretization.
In  this example, all estimators give similar estimates of the aggregated intensities.

\begin{figure}
\centering
\resizebox{\textwidth}{!}{
\begin{tabular}{cc}
\includegraphics[scale=0.3]{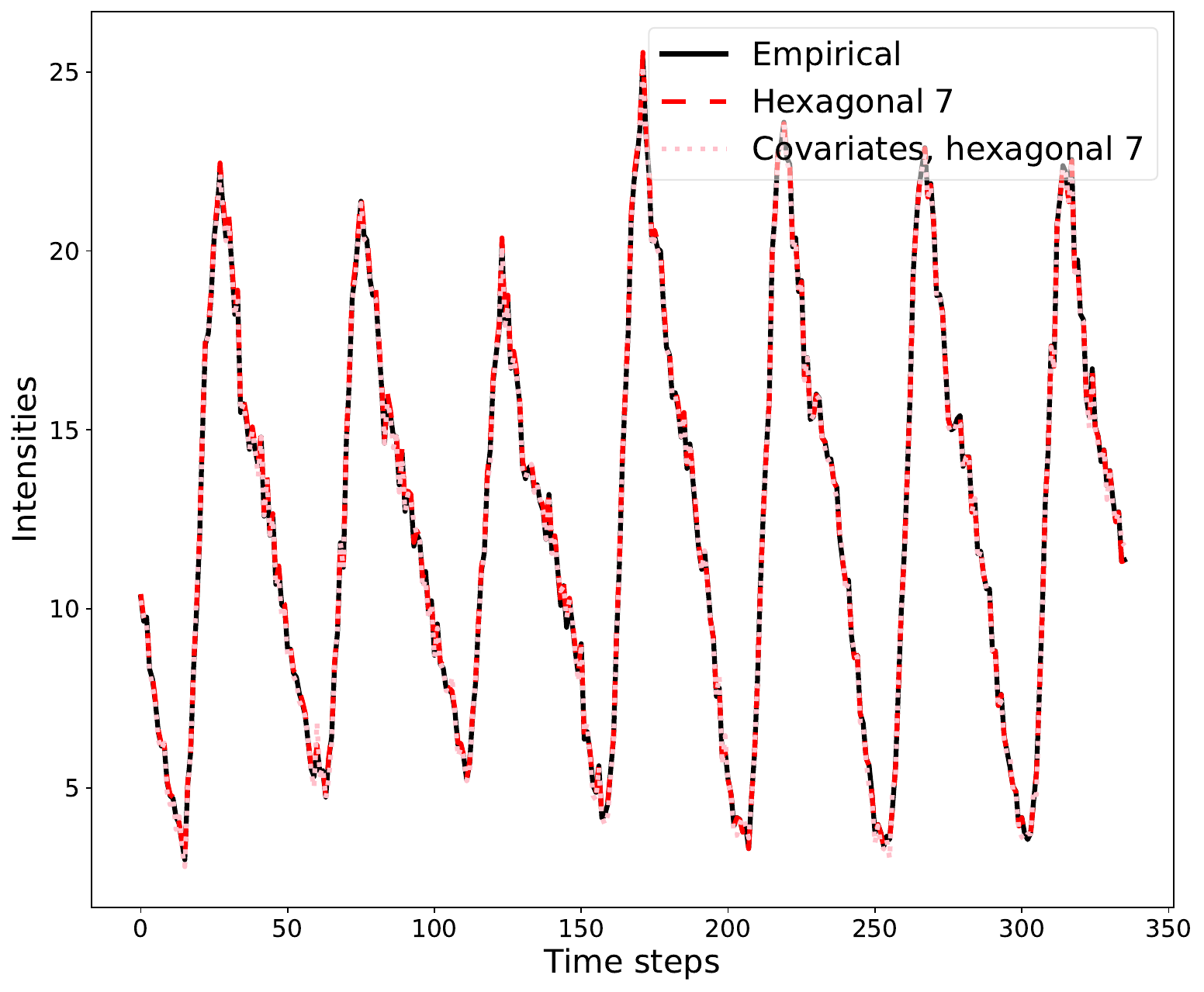}&
\includegraphics[scale=0.3]{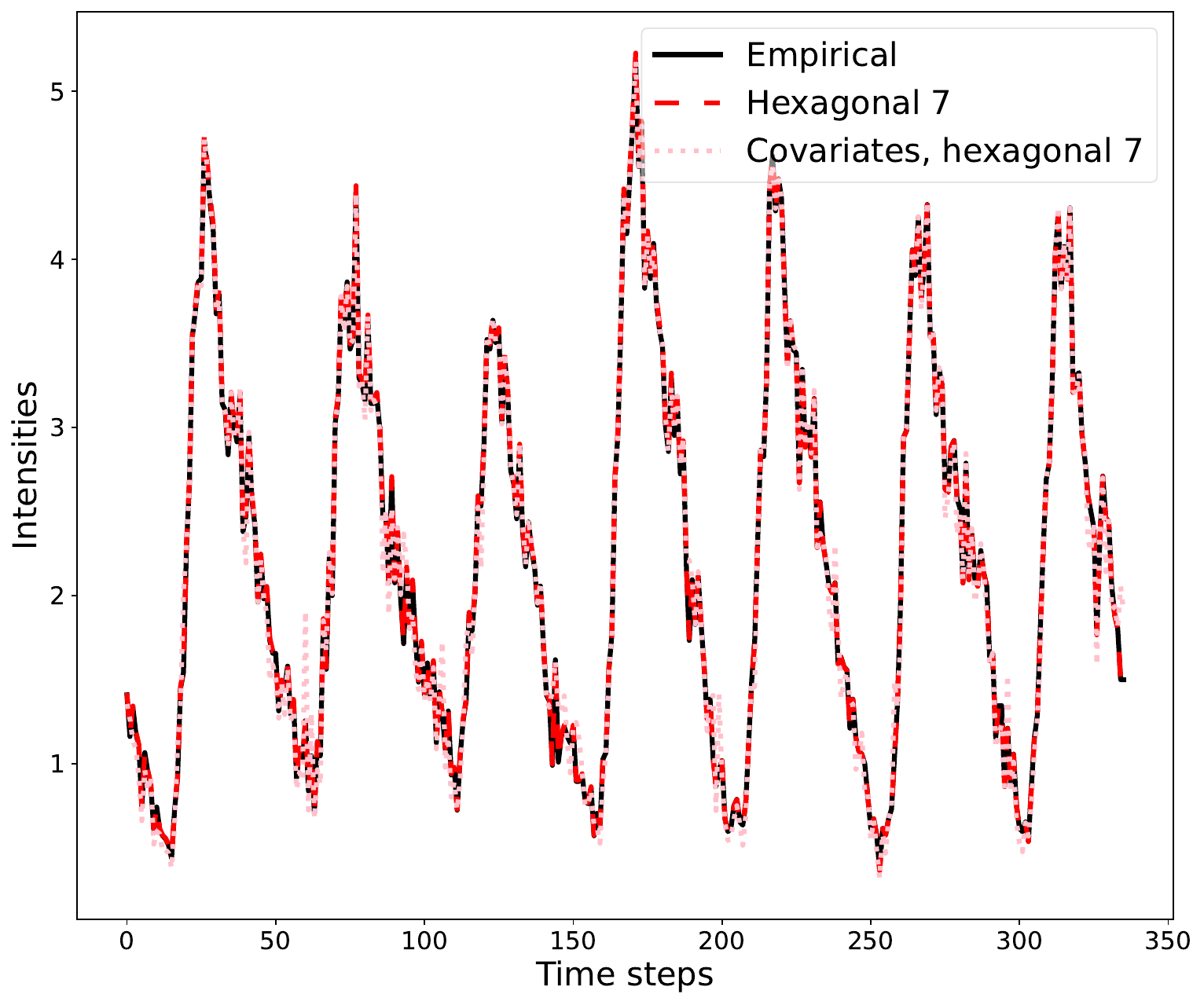}\\
\includegraphics[scale=0.3]{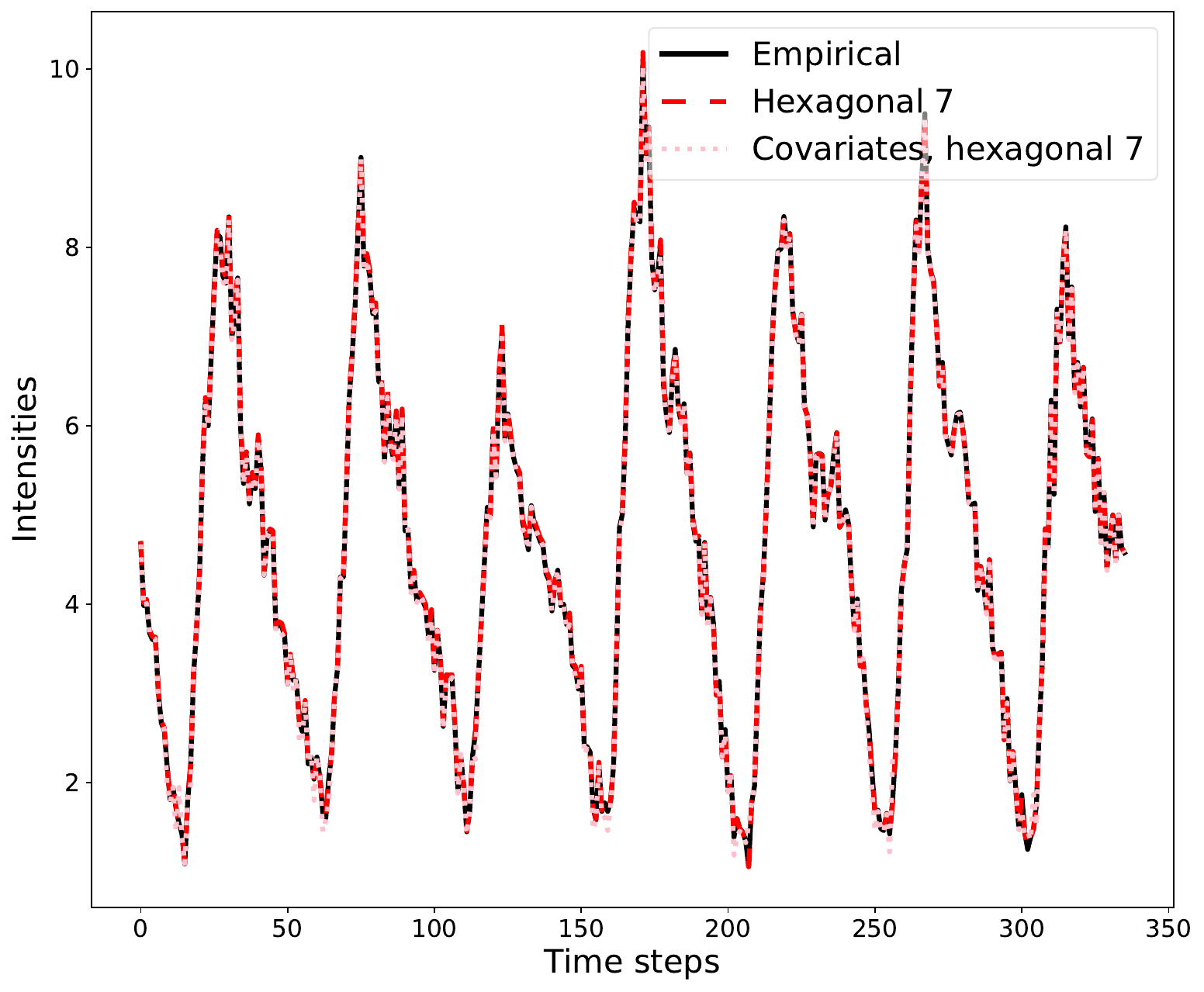}&
\includegraphics[scale=0.3]{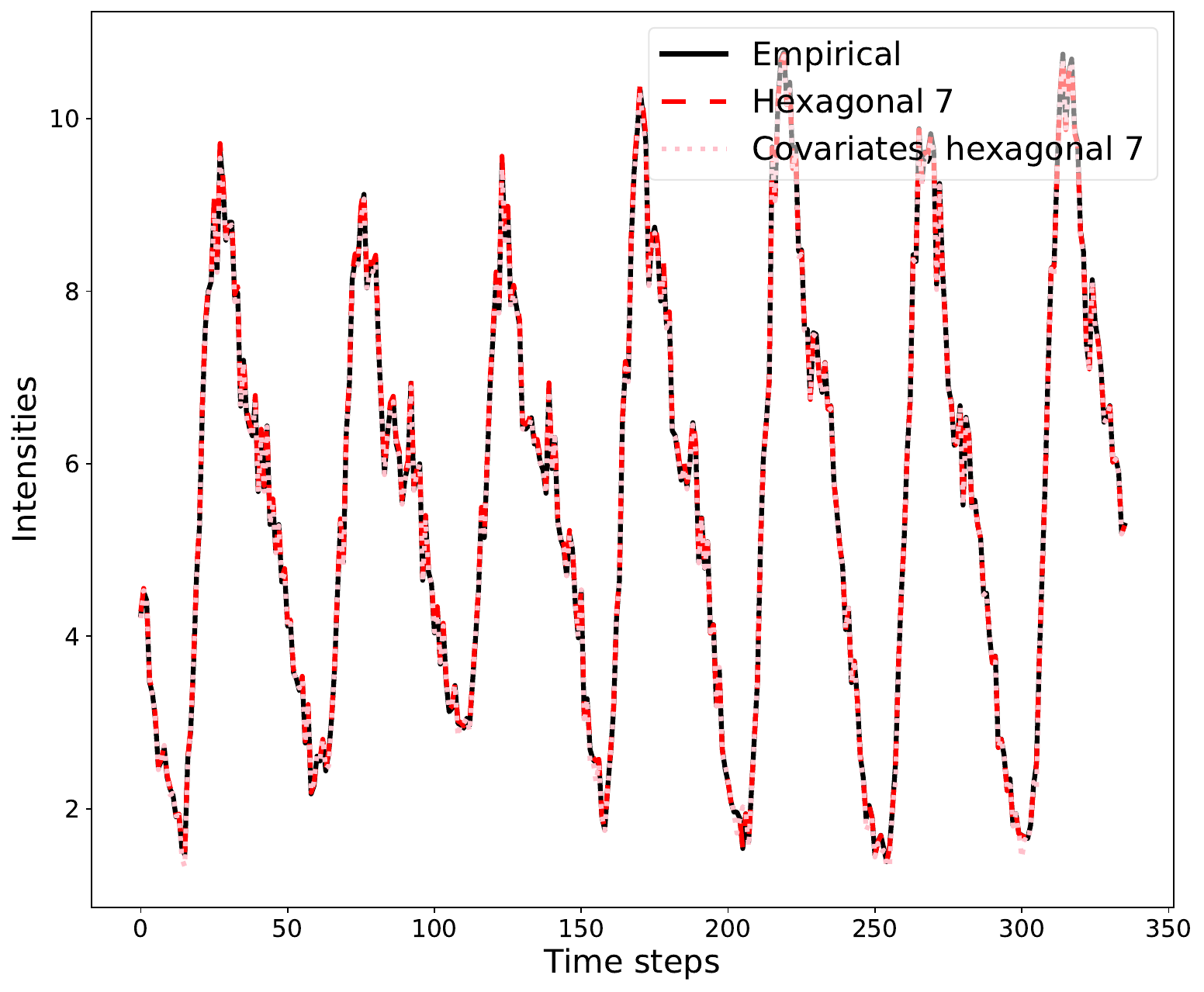}
\end{tabular}}
\caption{Aggregated estimates of the intensities for 3 estimators: (i)~Empirical: the empirical intensities, (ii)~Hexagonal 7: Model M1 with hexagonal discretization with scale parameter 7, and (iii)~Covariates, hexagonal 7: Model M2 with hexagonal discretization with scale parameter 7.
Top left: emergencies of all priorities, top right: high-priority emergencies, bottom left: intermediate priority emergencies, bottom right: low-priority emergencies.
\label{fig:allstepshexa}}
\end{figure}

Figure~\ref{fig:allstepshexa} shows the aggregated estimated intensities using hexagonal discretization with scale parameter 7.
The top left plot of Figure~\ref{fig:allstepshexa} shows $\sum_{c \in \mathcal{C}} \sum_{i \in \mathcal{I}} \hat{\lambda}_{c,i,t}$, the top right plot of Figure~\ref{fig:allstepshexa} shows $\sum_{i \in \mathcal{I}} \hat{\lambda}_{c_{2},i,t}$, the bottom left plot of Figure~\ref{fig:allstepshexa} shows $\sum_{i \in \mathcal{I}} \hat{\lambda}_{c_{1},i,t}$, and the bottom right plot of Figure~\ref{fig:allstepshexa} shows $\sum_{i \in \mathcal{I}} \hat{\lambda}_{c_{0},i,t}$.
The intensity estimates shown in this figure are obtained with (i)~the empirical mean number of calls for each call type, zone, and time interval, (ii)~Model M1 with hexagonal discretization with scale parameter 7, and (iii)~Model M2 with hexagonal discretization with scale parameter 7.

\begin{figure}
\centering
\resizebox{\textwidth}{!}{
\begin{tabular}{cc}
\includegraphics[scale=0.3]{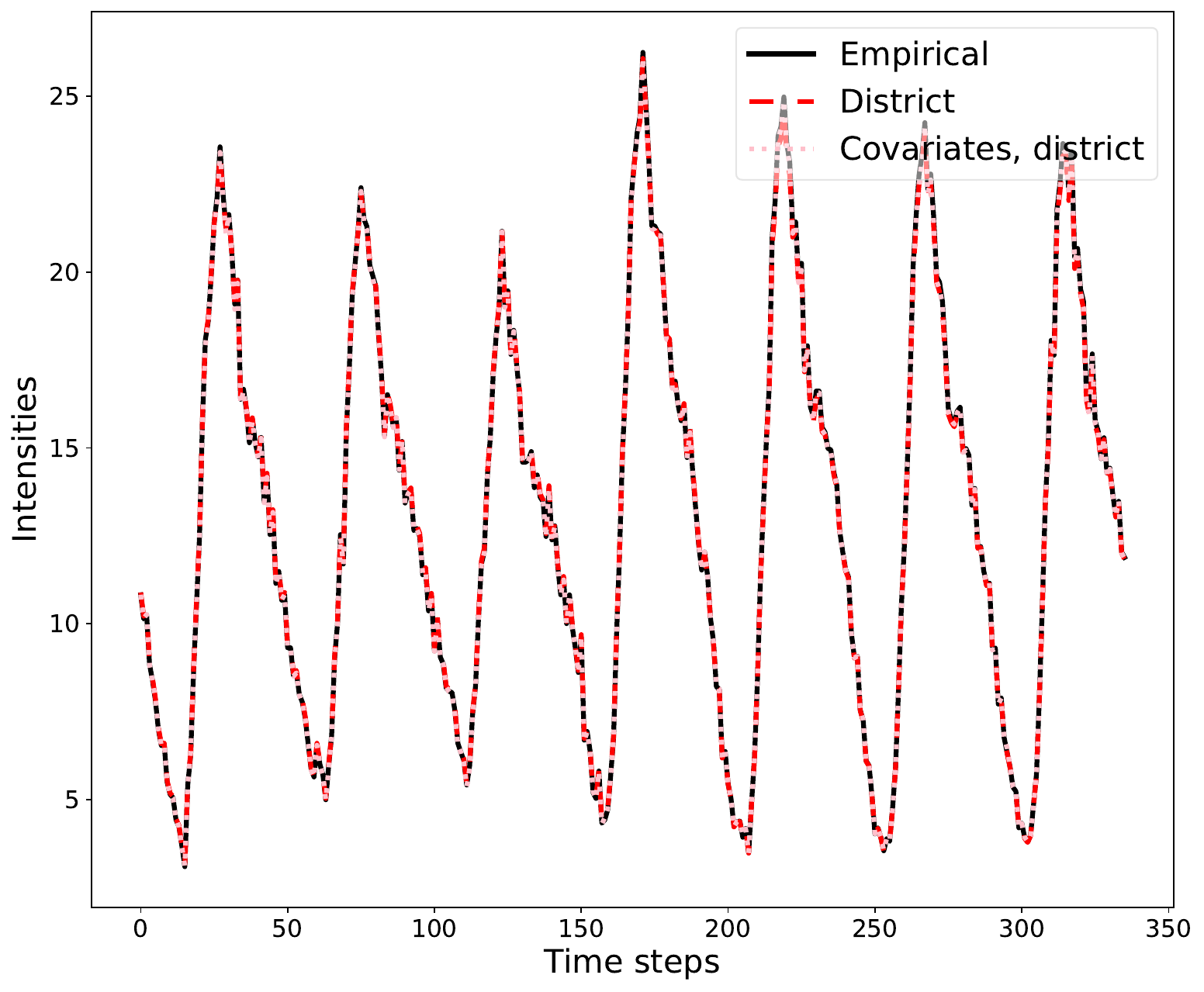}&
\includegraphics[scale=0.3]{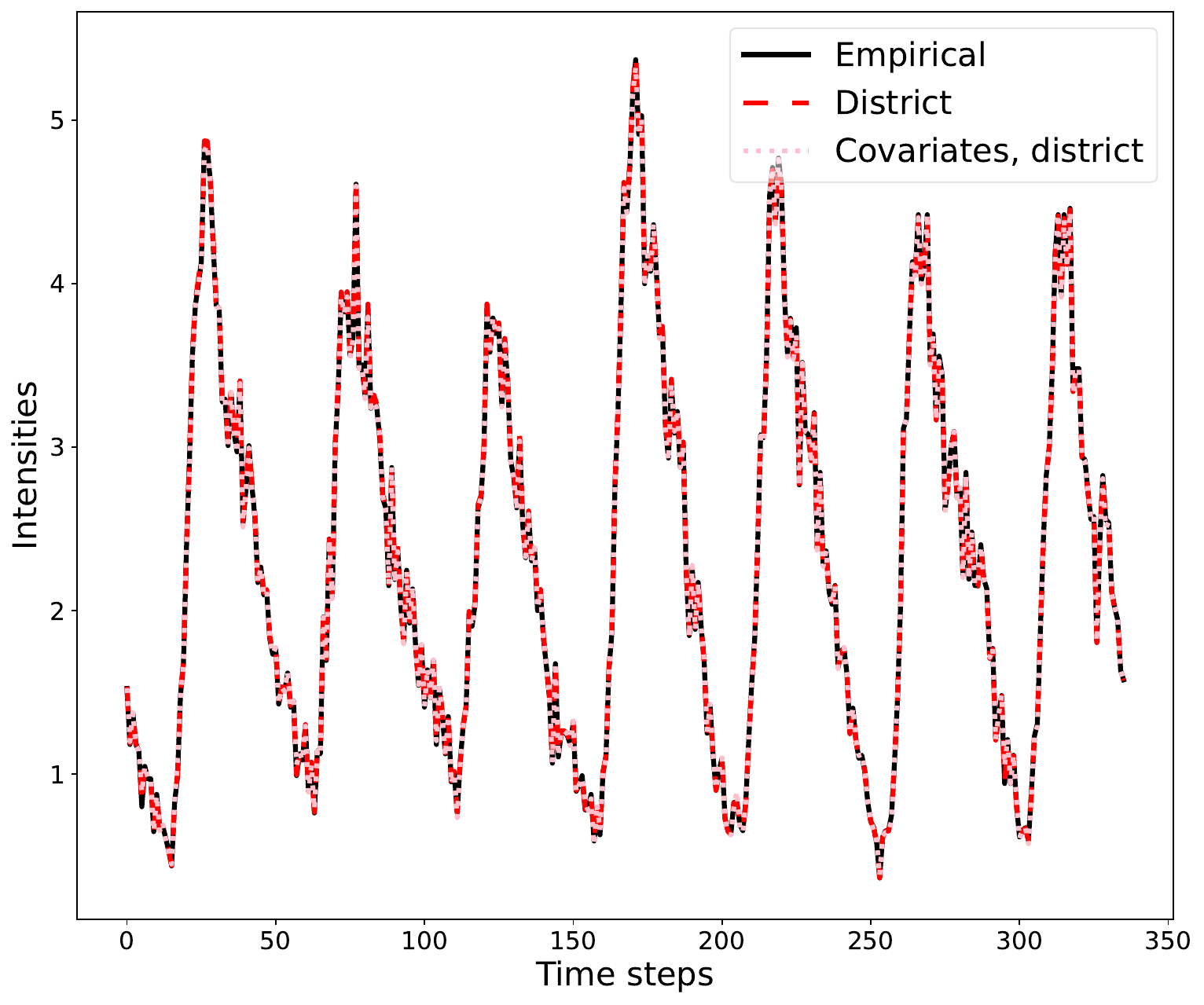}\\
\includegraphics[scale=0.3]{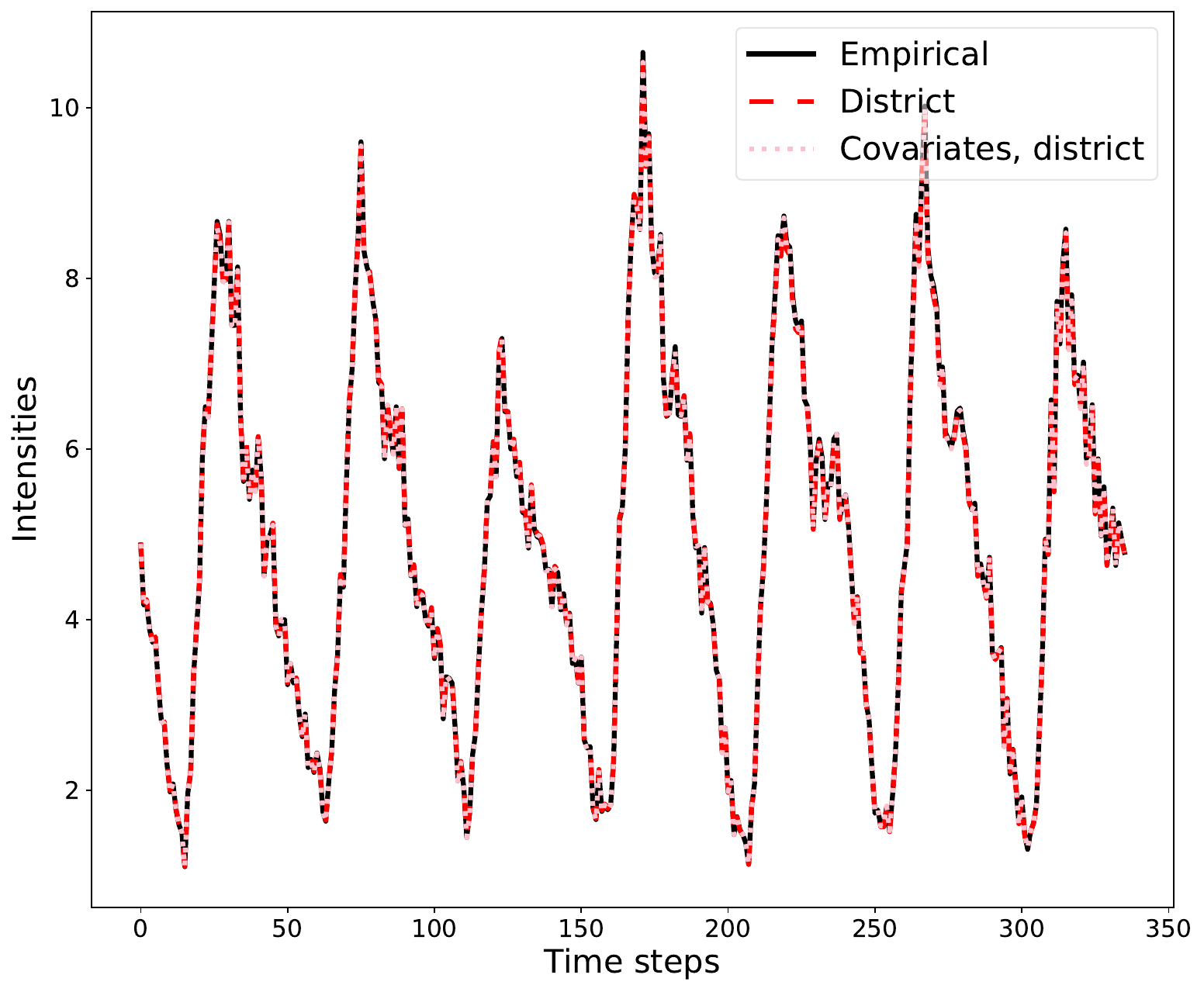}&
\includegraphics[scale=0.3]{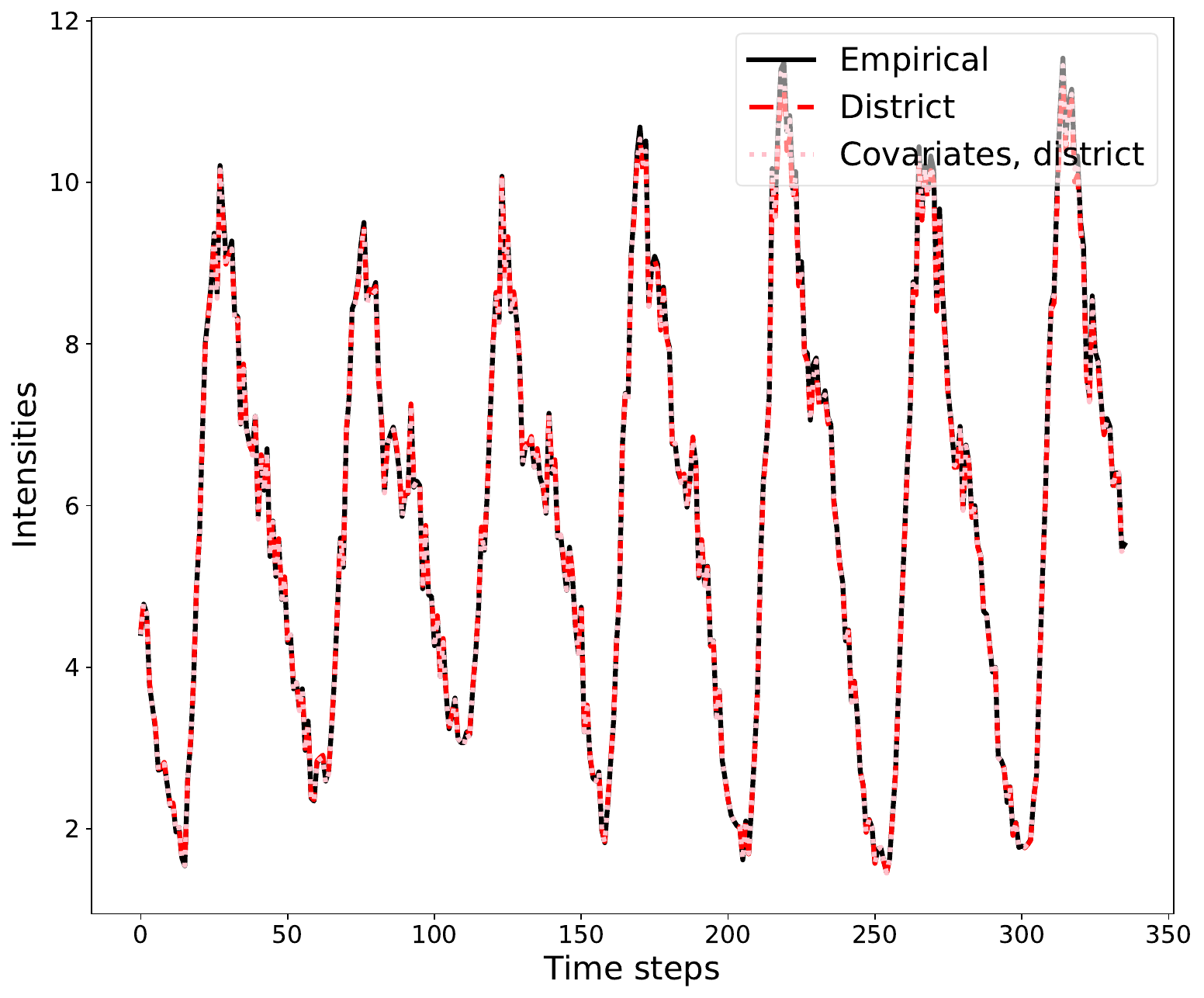}
\end{tabular}}
\caption{Aggregated estimates of the intensities for 3 estimators: (i)~Empirical: the empirical intensities, (ii)~District: Model M1 with the space discretization by districts, (iii)~Covariates, district: Model M2 with the space discretization by districts.
Top left: emergencies of all priorities, top right: high-priority emergencies, bottom left: intermediate priority emergencies, bottom right: low-priority emergencies.
\label{fig:allstepsdist}}
\end{figure}

Figure~\ref{fig:allstepsdist} shows the aggregated estimated intensities using discretization by district.
The top left plot of Figure~\ref{fig:allstepsdist} shows $\sum_{c \in \mathcal{C}} \sum_{i \in \mathcal{I}} \hat{\lambda}_{c,i,t}$, the top right plot of Figure~\ref{fig:allstepsdist} shows $\sum_{i \in \mathcal{I}} \hat{\lambda}_{c_{2},i,t}$, the bottom left plot of Figure~\ref{fig:allstepsdist} shows $\sum_{i \in \mathcal{I}} \hat{\lambda}_{c_{1},i,t}$, and the bottom right plot of Figure~\ref{fig:allstepsdist} shows $\sum_{i \in \mathcal{I}} \hat{\lambda}_{c_{0},i,t}$.
The intensity estimates shown in this figure are obtained with (i)~the empirical number of calls for each call type, zone, and time interval, (ii)~Model M1 with the space discretization by districts, and (iii)~Model M2 with the space discretization by districts.

\section{Replication script}\label{sec:rep}

Along with the LASPATED source code, we also provide a Python script that performs all experiments from section~\ref{sec:examples}. Before running the script, the first step is to compile the C++ code. If you have a GCC compiler installed,  this can be done by running the command:
\begin{verbatim}
    $ cd LASPATED/Replication 
    $ make -C cpp_tests
\end{verbatim}

The cpp\_tests directory contains a Makefile that compiles the C++ code. The Makefile accesses the environment variable \$GUROBI\_HOME, and if it is set, then the code for the model with Covariates inside laspated.h is accessible. Otherwise, the script will not run the experiments that use covariates. The Makefile can be edited to match the installed Gurobi version.

After the C++ code is successfully compiled, you can run the replication script with:

\begin{verbatim}
    $ python replication_script.py
\end{verbatim}

This command will generate the rectangular, hexagonal, and district discretizations using the laspated Python module, run the experiments with the C++ code and process the results, generating the figures and tables presented in Section \ref{sec:examples}.

All results are saved in directory Replication/replication\_results, with subdirectories plots and tables 
containing .pdf and .txt files given in
Table \ref{table_script}. 

\begin{table}[h]
    \centering
    \begin{tabular}{|c|c|c}
    \hline
     File Name                                                  &   Figure/Table\\ \hline
     plots/plot\_err\_by\_weight\_m1\_obs1.pdf                  &   Upper-left plot of Figure~\ref{figureserr1}\\ \hline
     plots/plot\_err\_by\_weight\_m1\_obs10.pdf                 &   Upper-right plot of Figure~\ref{figureserr1}\\ \hline
     plots/plot\_err\_by\_weight\_m1\_obs50.pdf                 &   Lower-left plot of Figure~\ref{figureserr1}\\ \hline
     plots/plot\_err\_by\_weight\_m1\_obs500.pdf                 &   Lower-right plot of Figure~\ref{figureserr1}\\ \hline
     plots/art\_rates\_by\_t\_w1\_r1.pdf                        &   Upper-left plot of Figure~\ref{figureserrth} \\ \hline
     plots/art\_rates\_by\_t\_w1\_r6.pdf                        &   Upper-right plot of Figure~\ref{figureserrth} \\ \hline
     plots/art\_rates\_by\_t\_w10\_r1.pdf                       &   Lower-left plot of Figure~\ref{figureserrth} \\ \hline
     plots/art\_rates\_by\_t\_w10\_r6.pdf                       &   Lower-right plot of Figure~\ref{figureserrth} \\ \hline
     tables/tables\_no\_covariates\_results.txt                 &   Table~\ref{tabnoreg} \\ \hline
     plots/plot\_err\_by\_weight\_m2\_obs1.pdf                  &   Upper-left plot of Figure~\ref{figureserr2pl} \\ \hline
     plots/plot\_err\_by\_weight\_m2\_obs10.pdf                 &   Upper-right plot of Figure~\ref{figureserr2pl} \\ \hline
     plots/plot\_err\_by\_weight\_m2\_obs50.pdf                 &   Lower-left plot of Figure~\ref{figureserr2pl} \\ \hline
     plots/plot\_err\_by\_weight\_m2\_obs500.pdf                 &   Lower-right plot of Figure~\ref{figureserr2pl} \\ \hline
     tables/tables\_covariates\_results.txt                     &   Table~\ref{tablecompwomodels1} \\ \hline
     plots/plot\_rates\_by\_t\_R76\_total.pdf                   &   Upper-left plot of Figure~\ref{fig:allstepsrect} \\ \hline        
     plots/plot\_rates\_by\_t\_R76\_c0.pdf                      &   Upper-right plot of Figure~\ref{fig:allstepsrect} \\ \hline
     plots/plot\_rates\_by\_t\_R76\_c1.pdf                      &   Lower-left plot of Figure~\ref{fig:allstepsrect} \\ \hline
     plots/plot\_rates\_by\_t\_R76\_c2.pdf                      &   Lower-right plot of Figure~\ref{fig:allstepsrect} \\ \hline
     plots/plot\_rates\_by\_t\_R160\_total.pdf                  &   Upper-left plot of Figure~\ref{fig:allstepsdist} \\ \hline        
     plots/plot\_rates\_by\_t\_R160\_c0.pdf                     &   Upper-right plot of Figure~\ref{fig:allstepsdist} \\ \hline
     plots/plot\_rates\_by\_t\_R160\_c1.pdf                     &   Lower-left plot of Figure~\ref{fig:allstepsdist} \\ \hline
     plots/plot\_rates\_by\_t\_R160\_c2.pdf                     &   Lower-right plot of Figure~\ref{fig:allstepsdist} \\ \hline
     plots/plot\_rates\_by\_t\_R226\_total.pdf                  &   Upper-left plot of Figure~\ref{fig:allstepshexa} \\ \hline        
     plots/plot\_rates\_by\_t\_R226\_c0.pdf                     &   Upper-right plot of Figure~\ref{fig:allstepshexa} \\ \hline
     plots/plot\_rates\_by\_t\_R226\_c1.pdf                     &   Lower-left plot of Figure~\ref{fig:allstepshexa} \\ \hline
     plots/plot\_rates\_by\_t\_R226\_c2.pdf                     &   Lower-right plot of Figure~\ref{fig:allstepshexa} \\ \hline
    \end{tabular}
    
    \caption{\label{table_script} Correspondence between outputs generated by the replication script and figures and tables from Section \ref{sec:examples}.}
\end{table}

The outputs
generated by C++
are written in 
directory Replication/cpp\_tests/results 
with four subdirectories. Subdirectory ex1 contains the results for the first model without covariates from Section~\ref{sec:simulated_example}. Each file, named err\_by\_weight\_wA\_gG\_aN\_m1.txt (for several values of A, G, and N, see below), contains the coordinates (abscissas and ordinates) of the points
of a given plot in Figure~\ref{figureserr1} (the plots we refer to here are the individual plots of Figure ~\ref{figureserr1}, for instance the upper left figure of Figure ~\ref{figureserr1}
contains four different plots). 
In the file name
err\_by\_weight\_wA\_gG\_aN\_m1.txt, the letter 
A denotes the number of observations \(N_{it}\), letter G denotes the number of time groups, and letter N is a flag that indicates whether space regularization is applied or not. Subdirectory ex2 contains similar files err\_by\_weight\_wA\_gG\_aN\_m2.txt
for the results of example 2. They are used to generate Figure~\ref{figureserr2pl}. Table~\ref{tabnoreg} is generated by the Python script and saved in file replication\_results/tables/table\_no\_covariates\_results.txt. Subdirectory ex3 contains the results for the model with covariates (Table~\ref{tablecompwomodels1} of the paper) and subdirectory real\_data contains the results for the experiments with real data for Rio de Janeiro Emergency Health Service. Figures~\ref{fig:allstepsrect}, \ref{fig:allstepsdist}, and~\ref{fig:allstepshexa} are built with output  files rates\_by\_t76.txt, rates\_by\_t160.txt, and rates\_by\_t226.txt, respectively.

\section{Conclusion}

We described Poisson models for spatio-temporal data and a software package called LASPATED for discretization of space and time and for the calibration of such models.

Planned extensions of LASPATED include the following:
\begin{itemize}
\item
Include discretization methods that adapt to the available data.
\item
Develop calibration methods for combinations of non-parametric (low bias, high variance) models with parametric (high bias, low variance) models that use covariate data.
\item
Include methods that test the Poisson model, that is, tests of the complete spatial randomness property.
\item
Include methods to deal with missing data (for instance missing emergency location data).
\item
This paper considers spatio-temporal Poisson processes, which has the property that the number of points in disjoint subsets are independent, with an intensity that is a function of space and time.
Important generalizations are spatio-temporal point processes in which the number of points depend on the (time-)history of the process, with a conditional intensity that is a function of space and time as well as the history of the process; spatio-temporal point processes with clustered points, also called self-exciting spatio-temporal point processes, such as Poisson cluster processes; as well as inhibition processes \citep{hawkes71,moller03,scha:05,gelf:10,cres:11,digg:14,du2016,mei:17,chenr:21,zhus:22}.
Applications in which clustering plays an important role include earthquakes that result in aftershocks, crime with possible crime sprees, and processes with contagion such as epidemics \citep{rein:18}.
Poisson cluster processes are modeled with a Poisson background process of ``parents'' or ``cluster centers'', and an i.i.d.\ triggering process that results in additional offspring points, each of which may also trigger more offspring.
LASPATED does not estimate such triggering processes, but does estimate nonhomogeneous Poisson processes such as those used as background processes.
LASPATED can be extended to include the estimation of triggering processes.
Existing software for self-exciting point processes includes PyHawkes (\url{https://github.com/slinderman/pyhawkes}) that implements Bayesian inference algorithms to use point process observations for estimating network structures.
\item
Other extensions include the estimation of Cox processes and Markov point processes.
\end{itemize}

\bibliographystyle{plainnat}
\bibliography{Biblio}

\end{document}